\documentclass{aa}

\usepackage{txfonts}
\usepackage{graphicx}
\begin{document}

\title{MHD Wave Propagation in the Neighbourhood of Two Null Points}
\author{J.~A. McLaughlin \and A.~W. Hood}
\institute{School of Mathematics and Statistics, University of St Andrews, KY16 9SS, UK}
\offprints{J.~A. McLaughlin,
\email{\emph{james@mcs.st-and.ac.uk}}}
\date{Received 12 November 2004 / Accepted 24 January 2005}

\authorrunning{McLaughlin \& Hood}

\abstract{The nature of fast magnetoacoustic and Alfv\'en waves is investigated in a zero $\beta$ plasma in the neighbourhood of a pair of two-dimensional null points. This gives an indication of wave propagation in the low $\beta$ solar corona, for a more complicated magnetic configuration than that looked at by McLaughlin \& Hood (2004). It is found that the fast wave is attracted to the null points and that the front of the wave slows down as it approaches the null point pair, with the wave splitting and part of the wave accumulating at one null and the rest at the other. Current density will then accumulate at these points and ohmic dissipation will then extract the energy in the wave at these points. This suggests locations where wave heating will occur in the corona. The Alfv\'en wave behaves in a different manner in that the wave accumulates along the separatrices. Hence, the current density will accumulate at this part of the topology and this is where wave heating will occur. However, the phenomenon of wave accumulation at a specific place is a feature of both wave types, and illustrates the importance of studying the topology of the corona when considering MHD wave propagation.

\keywords{MHD -- Waves -- Sun:~corona}
}

\maketitle


\section{Introduction}\label{sec:0}

The solar corona is dominated by the magnetic field. In order to understand the myriad of phenomena that occur on the Sun, it is important to understand the structure (topology) of the magnetic field itself. Potential field extrapolations of the coronal magnetic field can be made from photospheric magnetograms. Such extrapolations show the existence of an important feature of the topology; \emph{null points}. Null points are points in the field where the Alfv\'en speed is zero. Detailed investigations of the coronal magnetic field, using such potential field calculations, can be found in \cite{Beveridge2002} and \cite{Brown2001}.

Another key feature of the magnetic topology is the \emph{separator}; a special field line that connects two null points. A stable magnetic separator is the intersection of two separatrix surfaces. These surfaces divide the magnetic field into regions of different connectivity. A state is topologically stable if any small perturbation in the parameters does not change the state (\cite{Brown1999a}). Separators have an important role in magnetic reconnection (\cite{PriestTitov1996}) and current sheets can form along them (\cite{Longcope1996}).

McLaughlin \& Hood (2004) found that for a single 2D null point, the fast magnetoacoustic wave was attracted to the null and the wave energy accumulated there. In addition, they found that the Alfv\'en wave energy accumulated along the separatrices of the topology. The aim of this paper is to see if their ideas  carry through to magnetic configurations of more than one null point. It can be argued that null points appear in pairs, (e.g. by a local bifurcation) which makes it relevant to investigate multiple null point topologies. A double null point may arise as a bifurcation of a single 2D null point (\cite{Galsgaarddynamic}; \cite{Brown1998}).

Waves in the neighbourhood of a single 2D null point have been investigated
by various authors. \cite{Bulanov1980} provided a
detailed discussion of the propagation of fast and Alfv\'en waves
using cylindrical symmetry. In their paper, harmonic fast waves are generated and
these propagate towards the null point. However, the assumed
cylindrical symmetry means that the disturbances can only propagate
either towards or away from the null point. \cite{CraigWatson1992}
mainly consider the radial propagation of the $m=0$ mode (where $m$ is the
azimuthal wavenumber) using a mixture of analytical and numerical
solutions. In their investigation, the outer radial boundary is held fixed so that any
outgoing waves will be reflected back towards the null point. This
means that all the energy in the wave motions is contained within a
fixed region. They show that the propagation of the $m=0$ wave
towards the null point generates an exponentially large increase in
the current density and that magnetic resistivity dissipates this current in
a time related to $\log { \eta }$. Their initial disturbance is given as a
function of radius. In this paper, we are interested in generating
the disturbances at the boundary rather than internally. Craig and
McClymont (1991, 1993) investigate the normal mode solutions for both
$m=0$ and $m\ne 0$ modes with resistivity included. Again they
emphasise that the current builds up as the inverse square of the
radial distance from the null point. However, attention was restricted to a circular reflecting boundary.

Multiple null point topologies have also been investigated by various authors. \cite{Galsgaarddynamic} looked at the dynamic reconnection properties of a 3D double null. They investigated responses of the magnetic field to specific perturbations on the boundaries of a 3D box; they tilted the spine axis to form a pure $m=1$ mode. They found that for a perturbation parallel or orthogonal to the separator (between the two nulls), current accumulation occured in the separator plane perpendicular to the direction in which the spine was moved. For any other orientation, the current was focussed along the separator line. In the model, the boundary motions move the field lines but do not return them to their original positions. Thus, the Poynting flux induced by the imposed motion (and then fixing the field after the motion is complete) accumulates at the resulting current sheet and provides the energy released in the reconnection. However, if the boundary motions are simply due to the passing of incoming waves
through the boundary, then it is not clear that the current sheet will form. If this is the case, will the energy in the wave, again due to the Poynting flux through the boundary, dissipate or simply pass
through one of the other boundaries?

Galsgaard \& Nordlund (1997) found that when magnetic structures containing many null points are perturbed, current accumulates along separator lines. The perturbations used were random shear motions on two opposite boundaries. Again, the field lines were not returned to their original position. \cite{Klausdouble} looked at shearing a 3D potential null point pair, with continuous (opposite) shear on two opposite boundaries (parallel to the separator). This generated a wave pulse that travelled towards the interior of the domain from both directions, and resulted in current accumulation along the separator line with maximum value at the null points. From these experiments (which share a lot with \cite{Galsgaarddynamic}) it was concluded that to drive current along the full length of the separator line, the perturbation wavelength had to be longer than the length of the separator line. \cite{Galsgaardenergy} looked at perturbations in 3D magnetic configurations containing a double null point pair connected by a separator. The boundary motions used were very similar to those described above (i.e. shear the boundary and fix). Their experiments showed that the nulls can either accumulate current individually or act together to produce a singular current collapse along the separator. However, in all these previous works the boundary conditions used have tried to mimic the effect of photspheric footpoint motions by moving the boundary and holding it fixed. This paper will look at the 2D null point pair and investigate the propagation and transient behaviour of an individual wave pulse entering the magnetic structure.

The propagation of fast magnetoacoustic waves in an inhomogeneous
coronal plasma has been investigated by \cite{Nakariakov1995},
who showed how the waves are refracted into regions of low Alfv\'en
speed. In the case of null points, it is the aim of this paper to
see how this refraction proceeds when the Alfv\'en speed actually
drops to zero.

The paper has the following outline. In Section \ref{sec:1}, the basic equations are described. The results for an uncoupled fast magnetoacoustic wave are presented in Section \ref{sec:2}. This section discusses fast wave propagation with a pulse coming in from the top and side boundaries, and numerical and anaytical results are presented. Section \ref{sec:3} discusses the propagation of Alfv\'en waves and the conclusions are given in Section \ref{sec:4}

\section{Basic Equations}\label{sec:1}

The usual MHD equations for an ideal, zero $\beta$ plasma appropriate to
the solar corona are used. Hence,
\begin{eqnarray}
\qquad  \rho \left( {\partial {\bf{v}}\over \partial t} + \left( {\bf{v}}\cdot\nabla \right) {\bf{v}} \right) &=& {1\over \mu}\left(\nabla \times {\bf{B}}\right)\times {\bf{B}}\; ,\label{eq:2.1a} \\
  {\partial {\bf{B}}\over \partial t} &=& \nabla \times \left ({\bf{v}}\times {\bf{B}}\right ) \; ,\label{eq:2.1b} \\
{\partial \rho\over \partial t} + \nabla \cdot \left (\rho {\bf{v}}\right ) &=& 0\; , \label{eq:2.1c}
\end{eqnarray}
where $\rho$ is the mass density, ${\bf{v}}$ is the plasma
velocity, ${\bf{B}}$ the magnetic induction (usually called the
magnetic field), $ \mu = 4 \pi \times 10^{-7} \/\mathrm{Hm^{-1}}$  the magnetic
permeability, and $\sigma$ the electrical conductivity. The gas pressure and the
adiabatic energy equation are neglected in the zero $\beta$ approximation. The magnetic diffusivity is neglected in the ideal approximation.

\subsection{Basic equilibrium}\label{sec:1.1}

\begin{figure*}[ht]
\begin{center}
\includegraphics[width=2.8in]{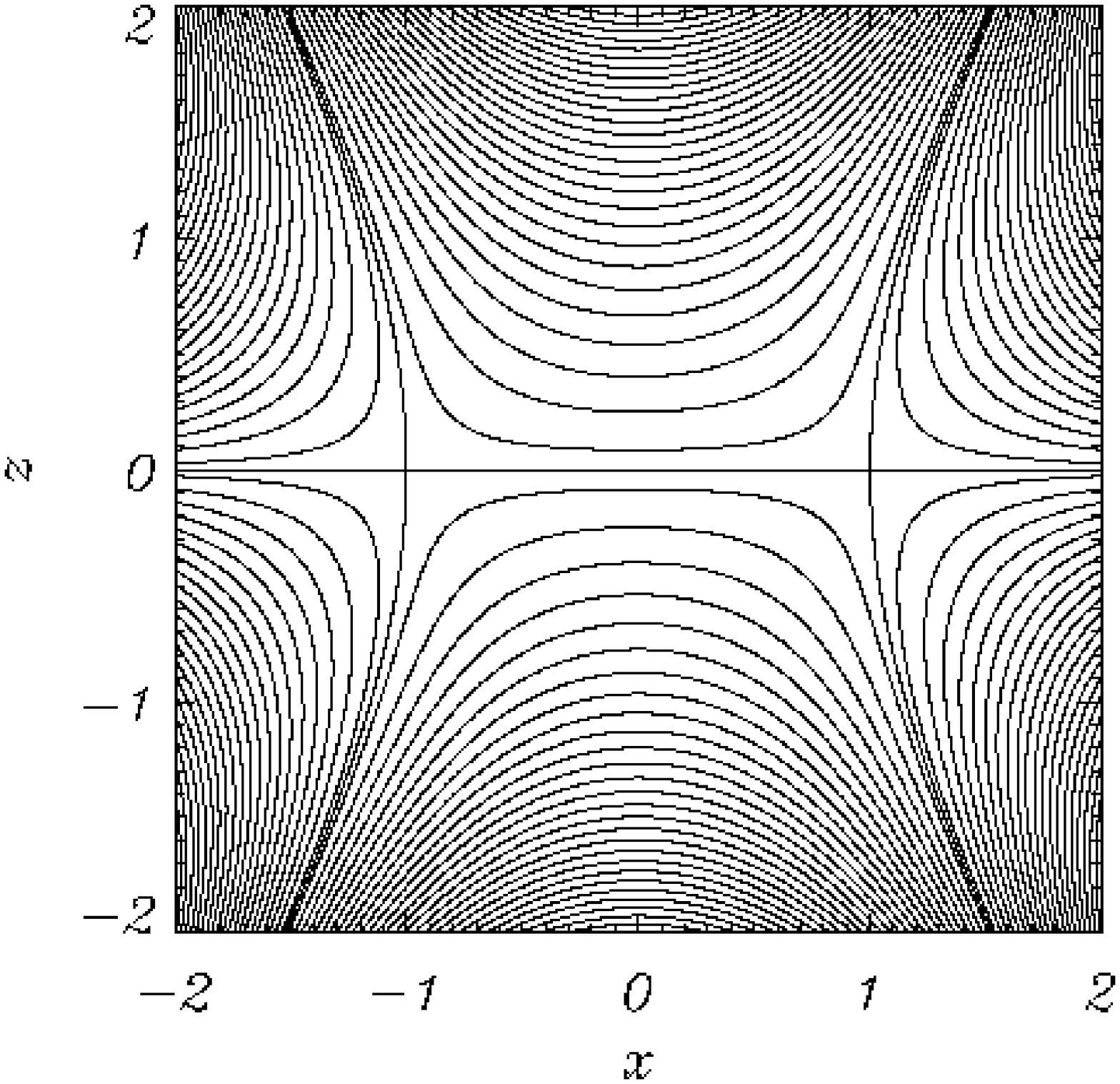}
\hspace{0.2in}
\includegraphics[width=2.8in]{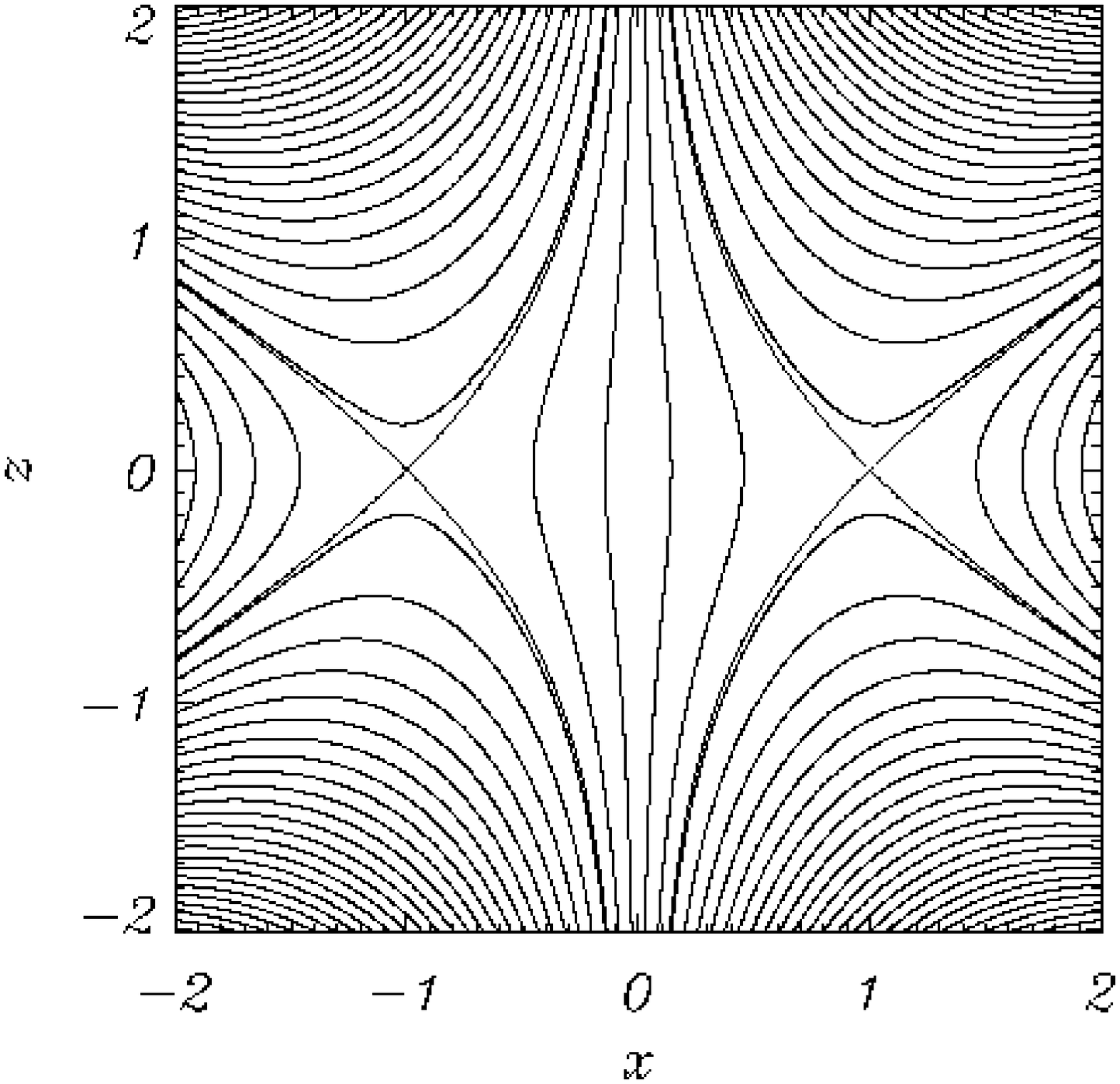}
\caption{(left) Magnetic field configuration containing two nulls joined by a separator. (right) Magnetic field configuration containing two nulls not joined by a separator.}
\label{figureone}
\end{center}
\end{figure*}

The basic magnetic structure is taken as a simple 2D two null points configuration. There are two such configurations to consider; one containing a separator and one that does not. The aim of studying waves in a 2D configuration is one of simplicity; the individual effects are much easier to identify when there is no coupling between the fast and Alfv\'en modes. The magnetic field with a separator is taken as:
\begin{equation}\label{eq:left}
\qquad {\bf{B}}_0 = {B_0 \over a^2} \left({x^2-z^2-a^2}, 0, {-2xz}\right),
\end{equation}
where $B_0$ is a characteristic field strength and $a$ is the length scale for magnetic field variations. This configuration can be seen in Figure \ref{figureone} (left). The other magnetic configuration considered takes the form:
\begin{equation}\label{eq:right}
\qquad {\bf{B}}_0 = {B_0 \over a^2} \left({2xz}, 0, {x^2-z^2-a^2}\right),
\end{equation}
and can be seen in Figure \ref{figureone} (right). Obviously, these magnetic configurations are no longer valid far away from the null points, as the field strength tends to infinity.

As in \cite{McLaughlin2004}, the linearised MHD equations are used to study the nature of wave propagation near the two null points. Using subscripts of $0$ for equilibrium quantities and $1$ for
perturbed quantities, the linearised equation of motion becomes
\begin{equation}\label{eq:2.3}
\qquad  \rho_0 \frac{\partial \mathbf{v}_1}{\partial t} = \left(\frac{ \nabla \times \mathbf{B}_1}{\mu} \right) \times \mathbf{B}_0 \; ,
\end{equation}
the linearised induction equation
\begin{equation}\label{eq:2.5}
\qquad    {\partial {\bf{B}}_1\over \partial t} = \nabla \times
    ({\bf{v}}_1 \times {\bf{B}}_0) \; ,
\end{equation}
and the linearised equation of mass continuity
\begin{equation}\label{eq:2.4}
\qquad \frac{\partial \rho_1} {\partial t} + \nabla\cdot\left( \rho_0 \mathbf{v} _1 \right) =0 \; .
\end{equation}
We will not discuss equation (\ref{eq:2.4}) further as it can be solved once we know $\mathbf{v} _1$. In fact, it has no influence on the momentum equation (in the zero $\beta$ approximation) and so in effect the plasma is arbitrarily compressible (Craig \& Watson, 1992). We assume the background gas density is uniform and label it as $\rho_0$. A spatial variation in $\rho _0$ can cause phase mixing (\cite{Heyvaerts1983}).

We now consider a change of scale to non-dimensionalise; let ${\mathbf{\mathrm{v}}}_1 = \bar{\rm{v}} {\mathbf{v}}_1^*$, ${\mathbf{B}}_0 = B_0 {\mathbf{B}}_0^*$, ${\mathbf{B}}_1 = B_0 {\mathbf{B}}_1^*$, $x = a x^*$, $z=az^*$, $\nabla = \frac{1}{a}\nabla^*$ and $t=\bar{t}t^*$, where we let * denote a dimensionless quantity and $\bar{\rm{v}}$, $B_0$, $a$ and $\bar{t}$ are constants with the dimensions of the variable they are scaling. We then set $\frac {B_0}{\sqrt{\mu \rho _0 } } =\bar{\rm{v}}$ and $\bar{\rm{v}} =  {a} / {\bar{t}}$ (this sets $\bar{\rm{v}}$ as a sort of constant background Alfv\'{e}n speed). This process non-dimensionalises equations (\ref{eq:2.3}) and (\ref{eq:2.5}), and under these scalings, $t^*=1$ (for example) refers to $t=\bar{t}=  {a} / {\bar{\rm{v}}}$; i.e. the time taken to travel a distance $a$ at the background Alfv\'en speed. For the rest of this paper, we drop the star indices; the fact that they are now non-dimensionalised is understood.

The ideal linearised MHD equations naturally decouple into two equations for the fast
MHD wave and the Alfv\'en wave. The slow MHD wave is absent in
this limit and there is no velocity component along the background
magnetic field (as can be seen by taking the scalar product of equation (\ref{eq:2.3}) with ${\mathbf{B}} _0$.

The linearised equations for the fast magnetoacoustic wave are:
\begin{eqnarray}
\qquad \frac{\partial V}{\partial t} &=& v_A^2 \left( x,z \right) \left( \frac{\partial b_z}{\partial x} - \frac{\partial b_x}{\partial z}  \right) \nonumber \\
\frac{\partial b_x}{\partial t} &=& -\frac{\partial V}{\partial z} \; , \; \frac{\partial b_z}{\partial t} =   \frac{\partial V}{\partial x} \; \label{fastalpha},
\end{eqnarray}
where the Alfv\'{e}n speed, $v_A \left( x,z \right)$, is equal to $ \sqrt{B_x^2+B_z^2}$ $=\left[ \left( x^2+z^2 \right) ^2 - 2 \left( x^2-z^2 \right) +1 \right] ^ { 1 \over 2 } $, $ { \mathbf{B} } _1 = \left( b_x,0,b_z \right) $ and the variable $ V $ is related to the perpendicular velocity; $ V = \left[ \left( \mathbf{v} _1 \times { \mathbf{B} } _0 \right) \cdot {\hat{\mathbf{e}} }_y \right] $.
These equations can be combined to form a single wave equation:
\begin{eqnarray}
\qquad \frac{\partial ^2 V}{\partial t^2} = v_A^2 \left( x,z \right) \left( \frac{\partial^2 V}{\partial x^2} + \frac{\partial ^2 V}{\partial z^2}  \right) \; \label{fastbeta}.
\end{eqnarray}

The linearised equations for the Alfv\'en wave, with ${\mathbf{v}} _1 = \left( 0,v_y,0 \right)$ and ${\mathbf{B}} _1 = \left( 0,b_y,0 \right)$ are:
\begin{eqnarray}
\qquad \frac {\partial v_y }{\partial t} = B_x \frac {\partial b_y }{\partial x} + B_z\frac {\partial b_y }{\partial z} \; , \quad \frac {\partial b_y }{\partial t} = B_x \frac {\partial v_y }{\partial x} + B_z\frac {\partial v_y }{\partial z} \; \label{alfvenalpha} ,
\end{eqnarray}
which can be combined to form a single wave equation:
\begin{eqnarray}
\qquad \frac {\partial^2 v_y }{\partial t^2} = \left(B_x \frac {\partial }{\partial x} +B_z\frac {\partial }{\partial z} \right) ^2 v_y \; \label{alfvenbeta}.
\end{eqnarray}

It is worth noting that the zero $\beta$ approximation will be invalid near the null points, where the pressure terms become important. In fact, a finite $\beta$ model would introduce slow waves, which would be driven by the excited fast waves. Also, the model is linear but the inhomogenity of the medium would lead to nonlinear coupling of the modes (\cite{Nakariakov1997}).

\section{Fast wave}\label{sec:2}

We consider the effect of sending a fast magnetoacoustic wave into these two magnetic configurations from the top and side boundaries. Unlike \cite{McLaughlin2004}, we also consider the effect of sending in a wave pulse from a side boundary because the magnetic configuration is not symmetric. In fact, since the fast magnetoacoustic wave can cross fieldlines, it behaves identically across both magnetic configurations (Figure \ref{figureone} left and right) and so we only have two cases to investigate, i.e. (\ref{fastbeta}) is true for both magnetic configurations (unlike (\ref{alfvenbeta}), which depends upon the form of $\mathbf{B}$).

\subsection{Upper boundary}

We solve the linearised MHD equations (\ref{fastalpha}) for the fast wave numerically using a two-step Lax-Wendroff scheme. The numerical scheme is run in a box with $-3 \le x \le 3$ and $-4 \le z \le 2$, with our attention focused on  $-2 \le x \le 2$ and $-2 \le z \le 2$. For a single wave pulse coming in from the top boundary, the boundary conditions are taken as:
\begin{eqnarray*}
\qquad V(x, 2) &=& \left\{\begin{array}{cl}
{\sin { \omega t } } & {\mathrm{for} \; \;0 \leq t \leq \frac {\pi}{\omega} } \\
{0} & { \mathrm{otherwise} }
\end{array} \right. \; , \\
\; \left. \frac {\partial V } {\partial x } \right|  _{x=-3} &=& 0 \; , \quad \left.\frac {\partial V } {\partial x } \right| _{x=3} = 0 \; , \quad \left.\frac {\partial V } {\partial z } \right| _{z=-4}  = 0 \; .
\end{eqnarray*}

Tests show that the central behaviour is largely unaffected by these choices of side and bottom boundary conditions. The other boundary conditions on the perturbed magnetic field follow from the remaining equations and the solenodial condition, $\nabla \cdot {\mathbf{B} _1} =0 $.

We find that the linear, fast magnetoacoustic wave travels towards the neighbourhood of the two null points and begins to wrap around them.  Since the Alfv\'{e}n speed, $v_A \left( x,z \right)$, is spatially varying, different parts of the wave travel at different speeds, and it travels faster the further it is away from the null points ($v_A \left( x,z \right)=0$ at the null points). This is a similar effect to that described it the case of a single null point in \cite{McLaughlin2004}. However, in this case there is a non-zero  Alfv\'{e}n speed between the two null points, i.e. for $-1 < x < 1$, $z=0$. The fast wave slows down greatly when it approaches this area, but manages to cross this line (the separator if we were considering a setup such as that of Figure \ref{figureone} (left)). Hence the wave bends between the two null points and passes through the area between them. This part of the wave is then again suceptible to the refraction effect, and so continues to wrap around the null points, breaking into two waves along the line $x=0$ (due to symmetry). Each part of the wave then continues to wrap around its closest null point, repeatedly, eventually accumulating at the null points ($x=-1$ and $x=+1$, along $z=0$). This can be seen in the shaded contours of Figure \ref{figuresix}.

\subsection{Analytical results}\label{sec:2.2}

We can approximately solve equation (\ref{fastbeta}) for the fast wave to gain more insight into the numerical simulations. Substituting $V = e^{i \phi (x,z) } \cdot e^{-i \omega t}$ into (\ref{fastbeta}) and assuming that $\omega \gg 1 $ (WKB approximation), leads to a first order PDE of the form $\mathcal{F} \left( x,z,\phi,\frac {\partial \phi } {\partial x}, \frac {\partial \phi } {\partial z} \right)=0$. Applying the method of characteristics, we generate the equations:
\begin{eqnarray*}
\qquad \frac {d \phi }{ds} &=& - \omega ^2  \nonumber\\
\qquad \frac {dp}{ds} &=& 2x \left( x^2+z^2-1\right) \left( p^2+q^2 \right) \nonumber \\
\qquad \frac {dq}{ds} &=& 2z \left( x^2+z^2+1 \right) \left( p^2+q^2 \right)  \nonumber \\
\qquad \frac {dx}{ds} &=& - p \left[ \left( x^2 +z^2\right)^2 -2(x^2-z^2) +1 \right]  \nonumber \\
\qquad \frac {dz}{ds} &=& - q \left[ \left( x^2 +z^2\right)^2 -2(x^2-z^2) +1 \right] \label{fastcharacteristics}
\end{eqnarray*}
where $p=\frac {\partial \phi } {\partial x}$, $q=\frac {\partial \phi } {\partial z}$, $\omega$ is the frequency of our wave and $s$ is some parameter along the characteristic. 

\begin{figure*}[t]
\includegraphics[width=2.1176in]{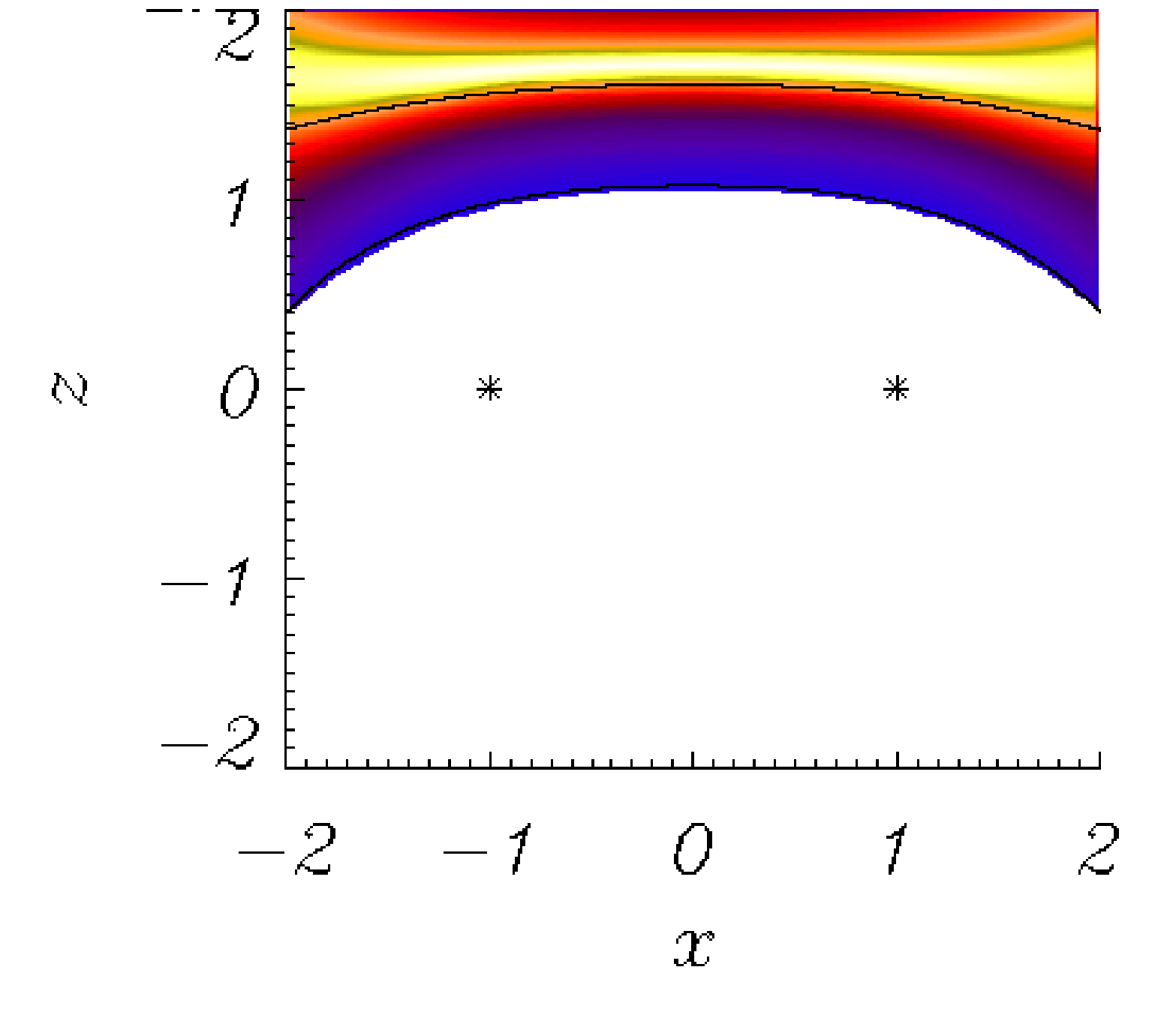}
\hspace{0.14286in}
\includegraphics[width=2.1176in]{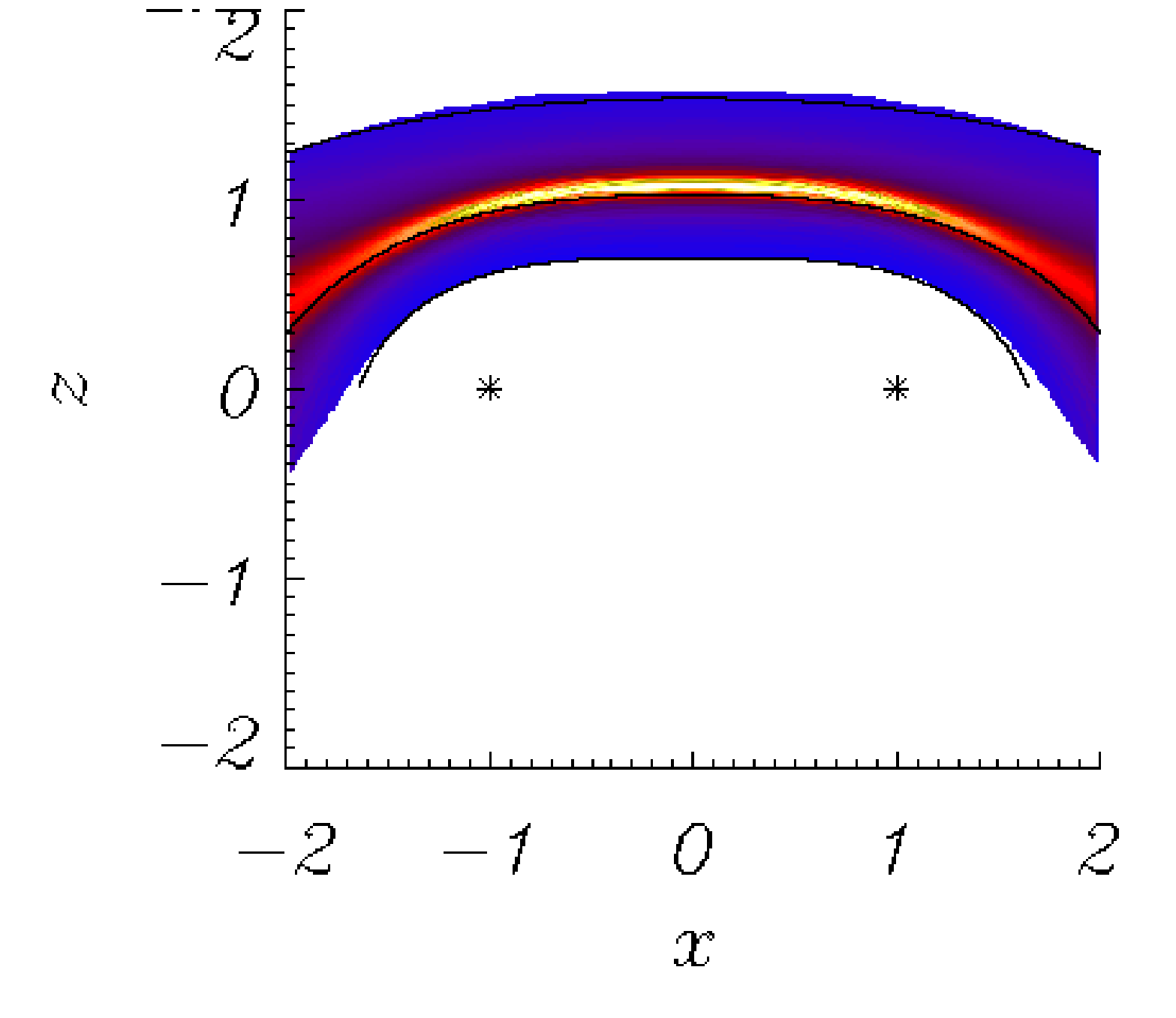}
\hspace{0.14286in}
\includegraphics[width=2.1176in]{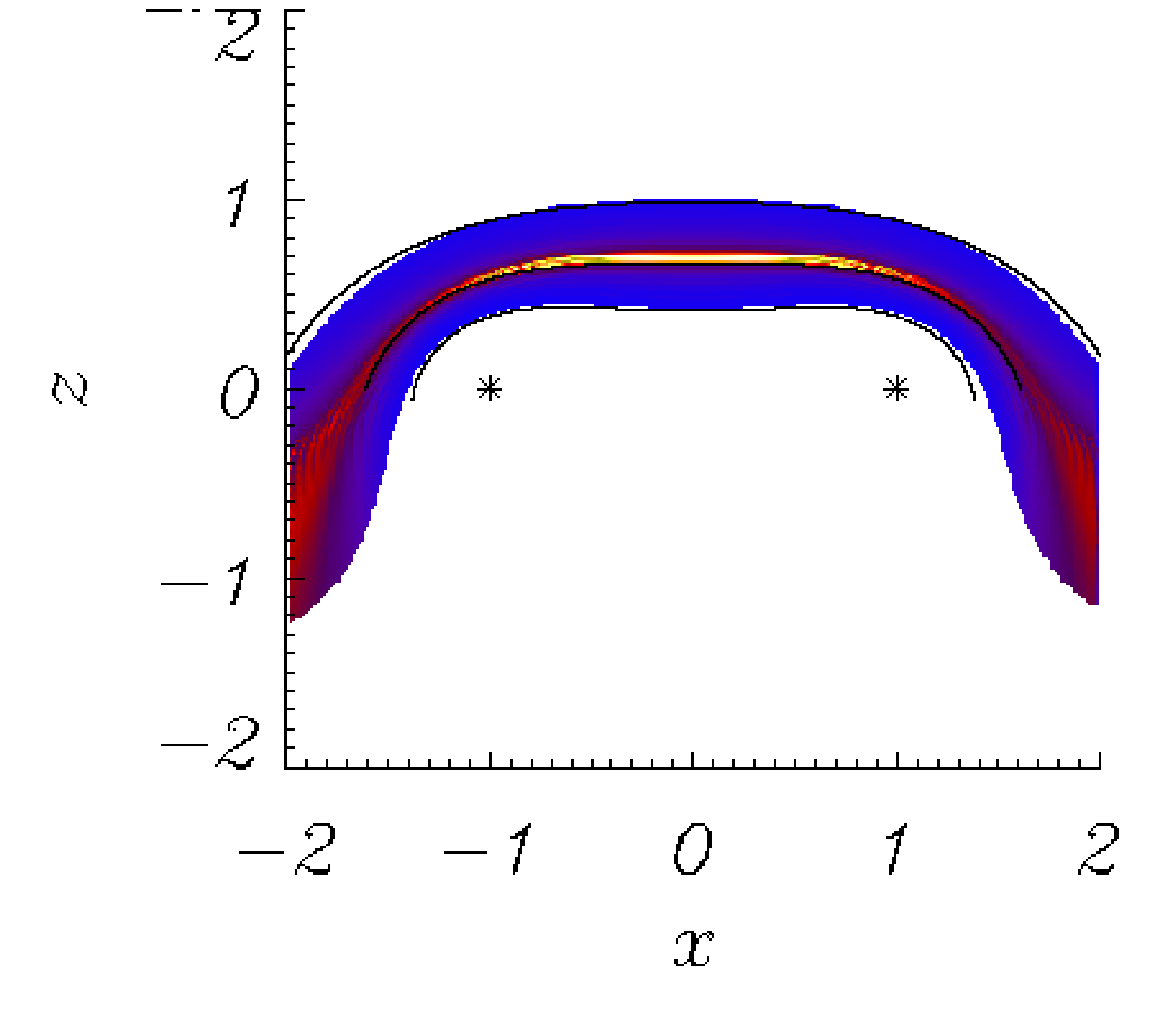}\\
\includegraphics[width=2.1176in]{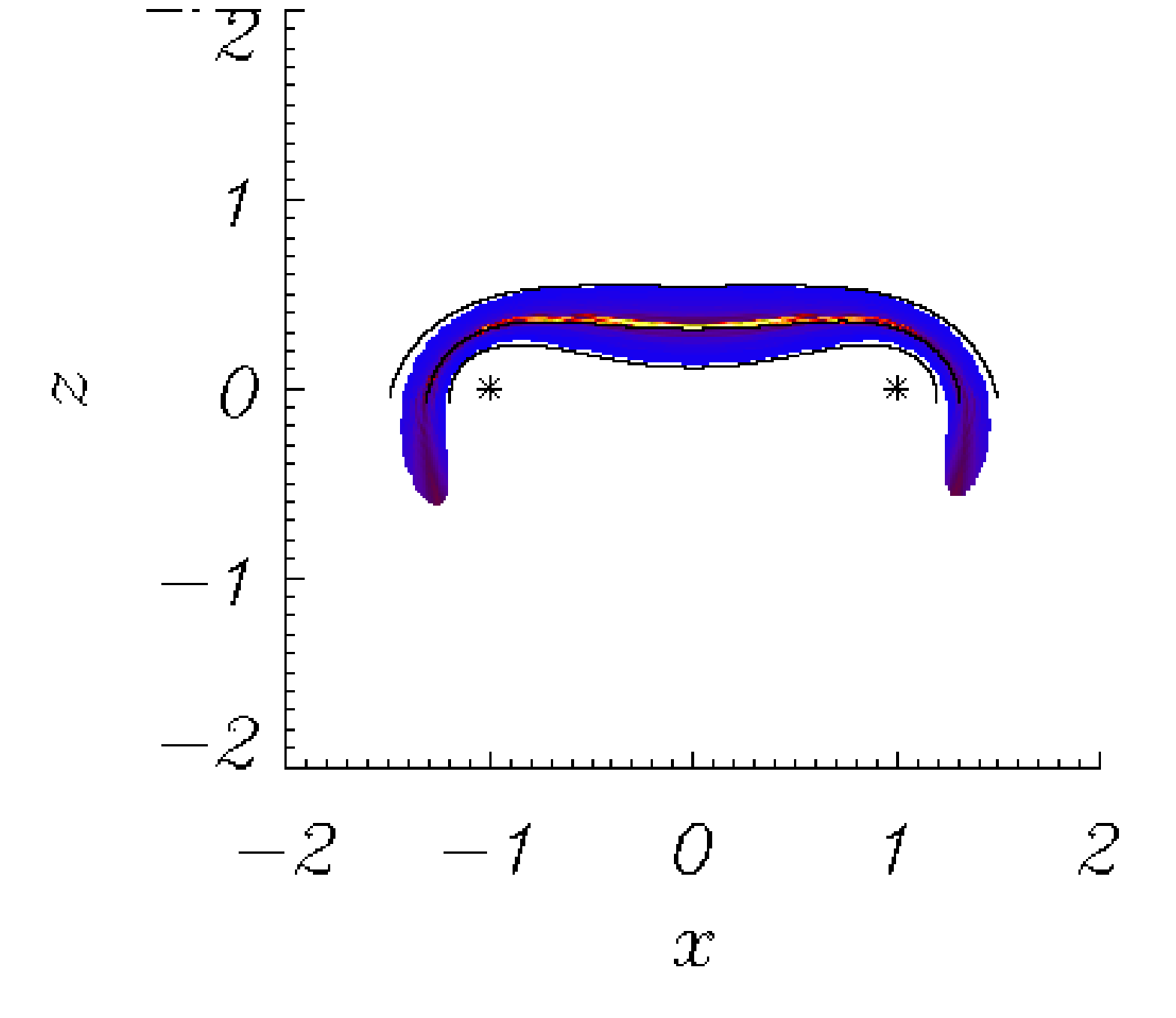}
\hspace{0.14286in}
\includegraphics[width=2.1176in]{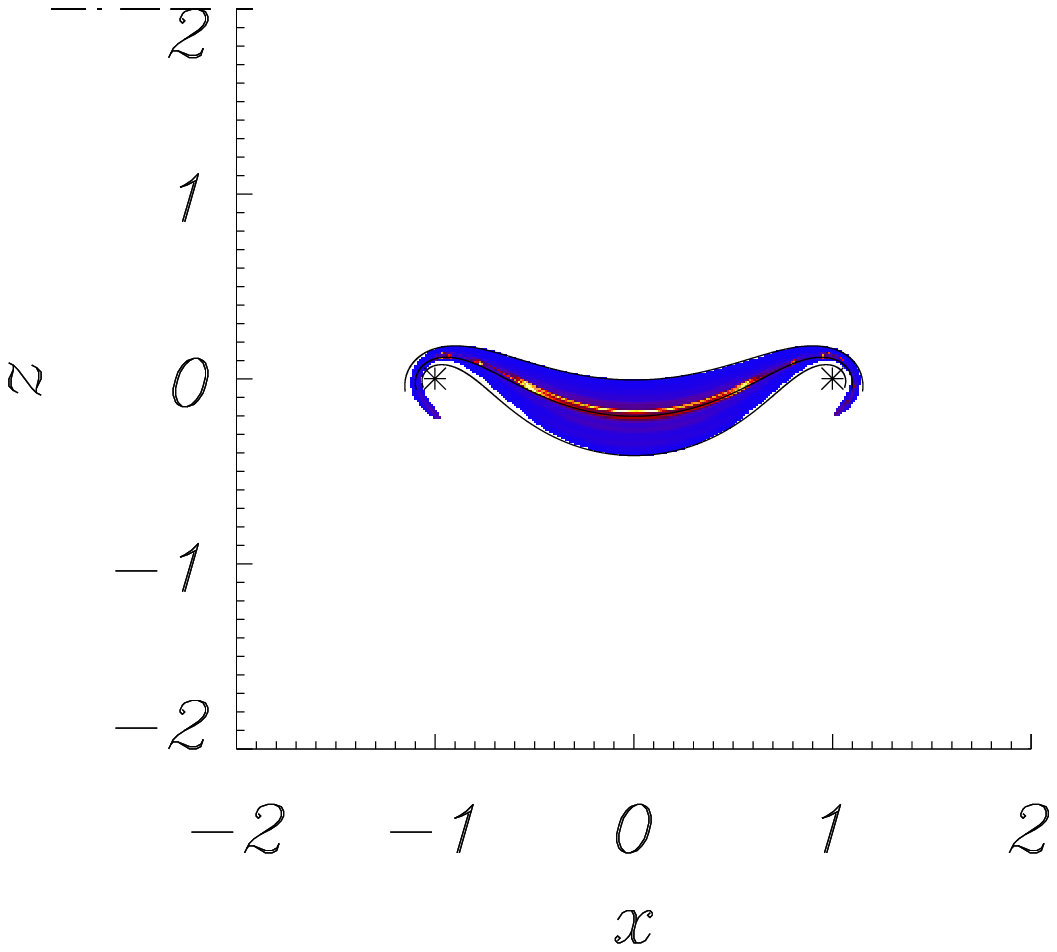}
\hspace{0.14286in}
\includegraphics[width=2.1176in]{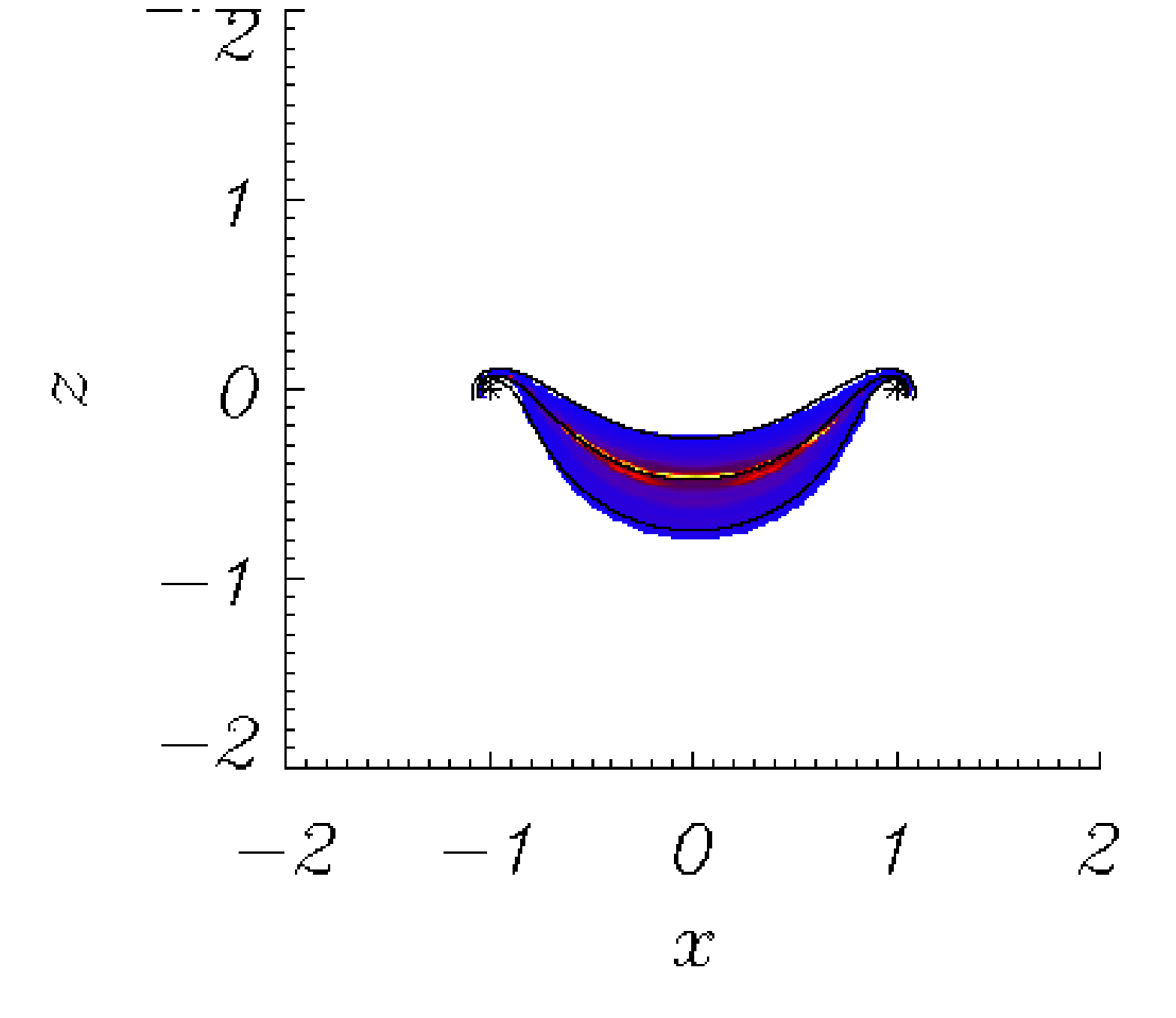}\\
\includegraphics[width=2.1176in]{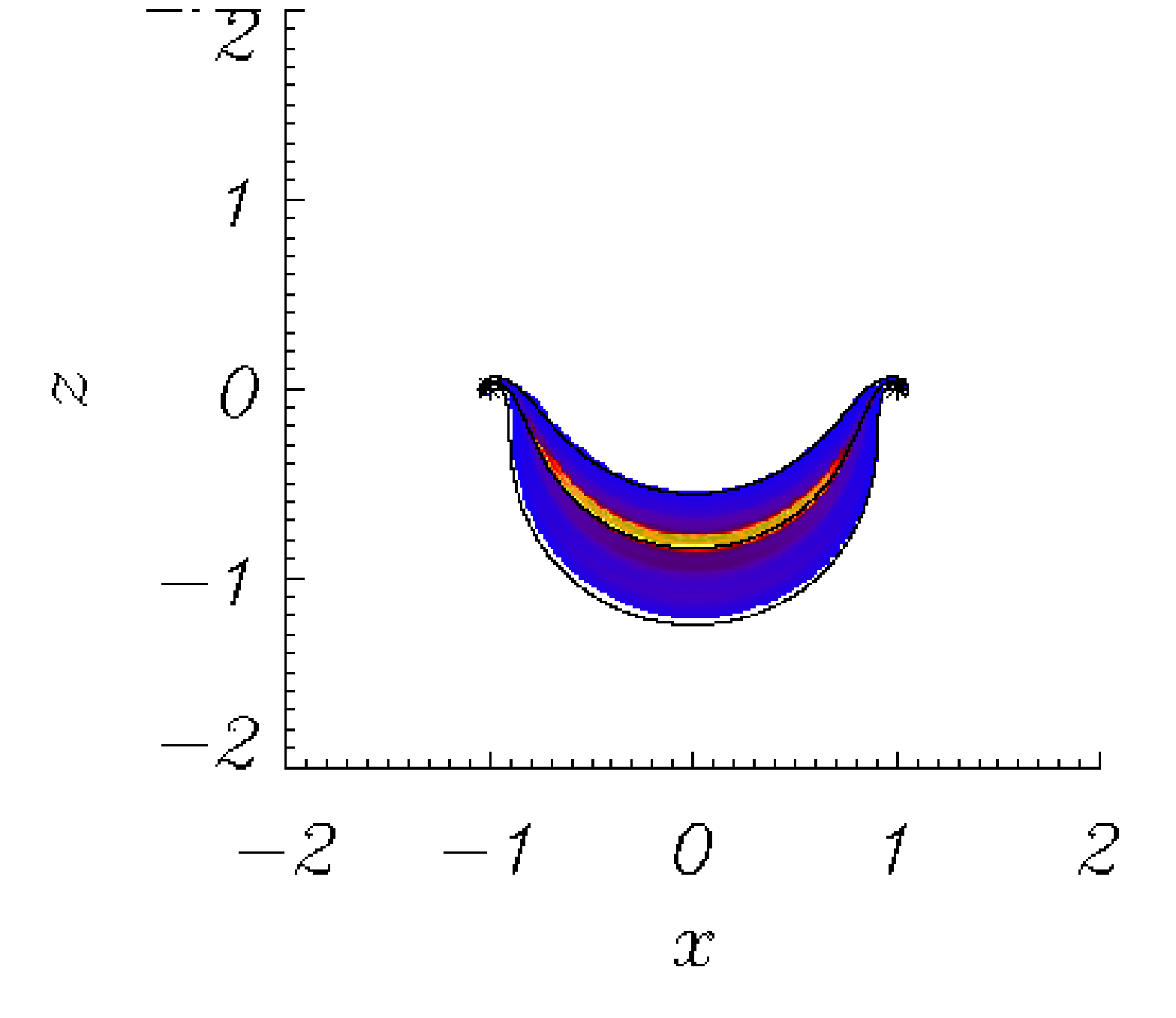}
\hspace{0.14286in}
\includegraphics[width=2.1176in]{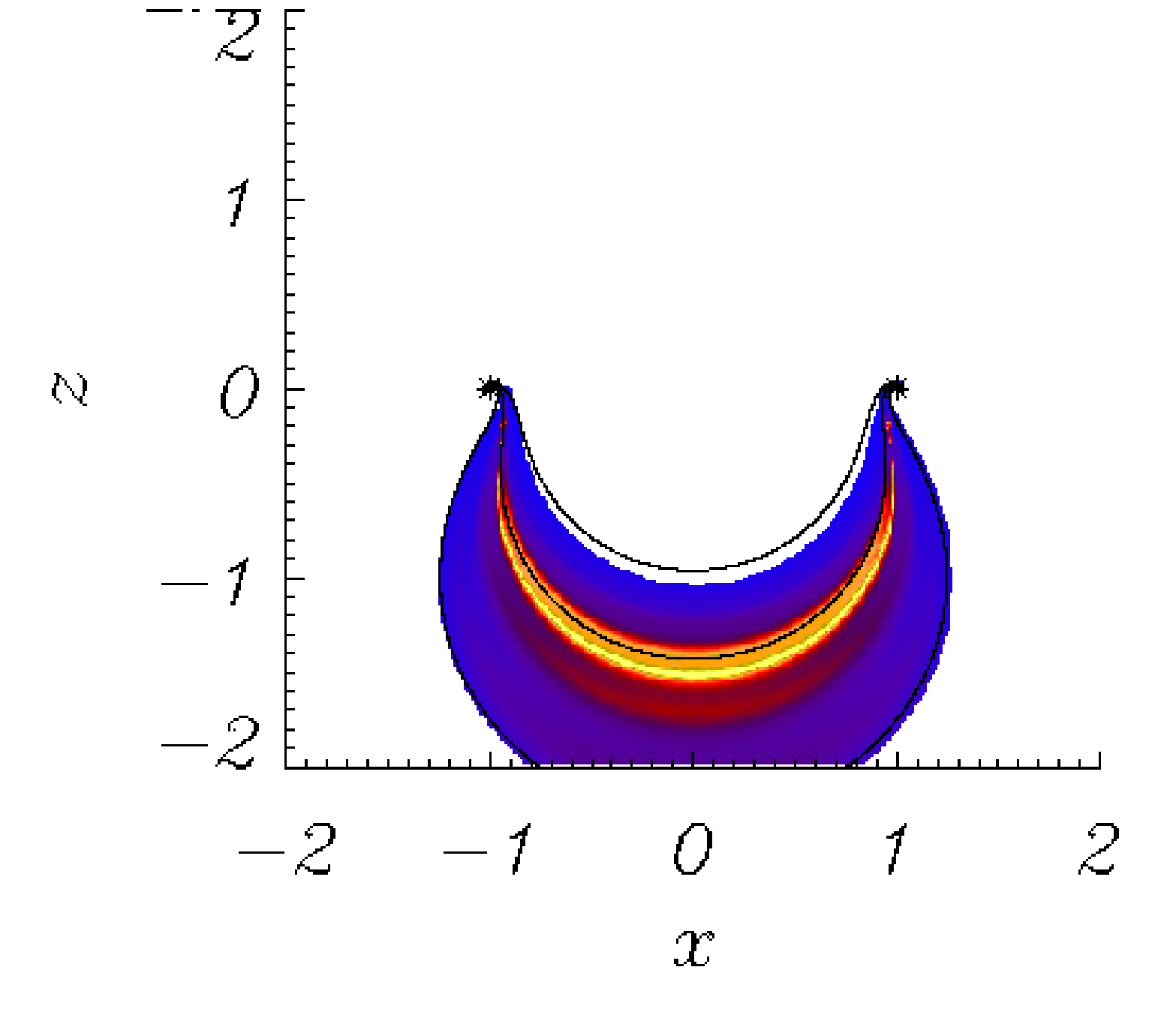}
\hspace{0.14286in}
\includegraphics[width=2.1176in]{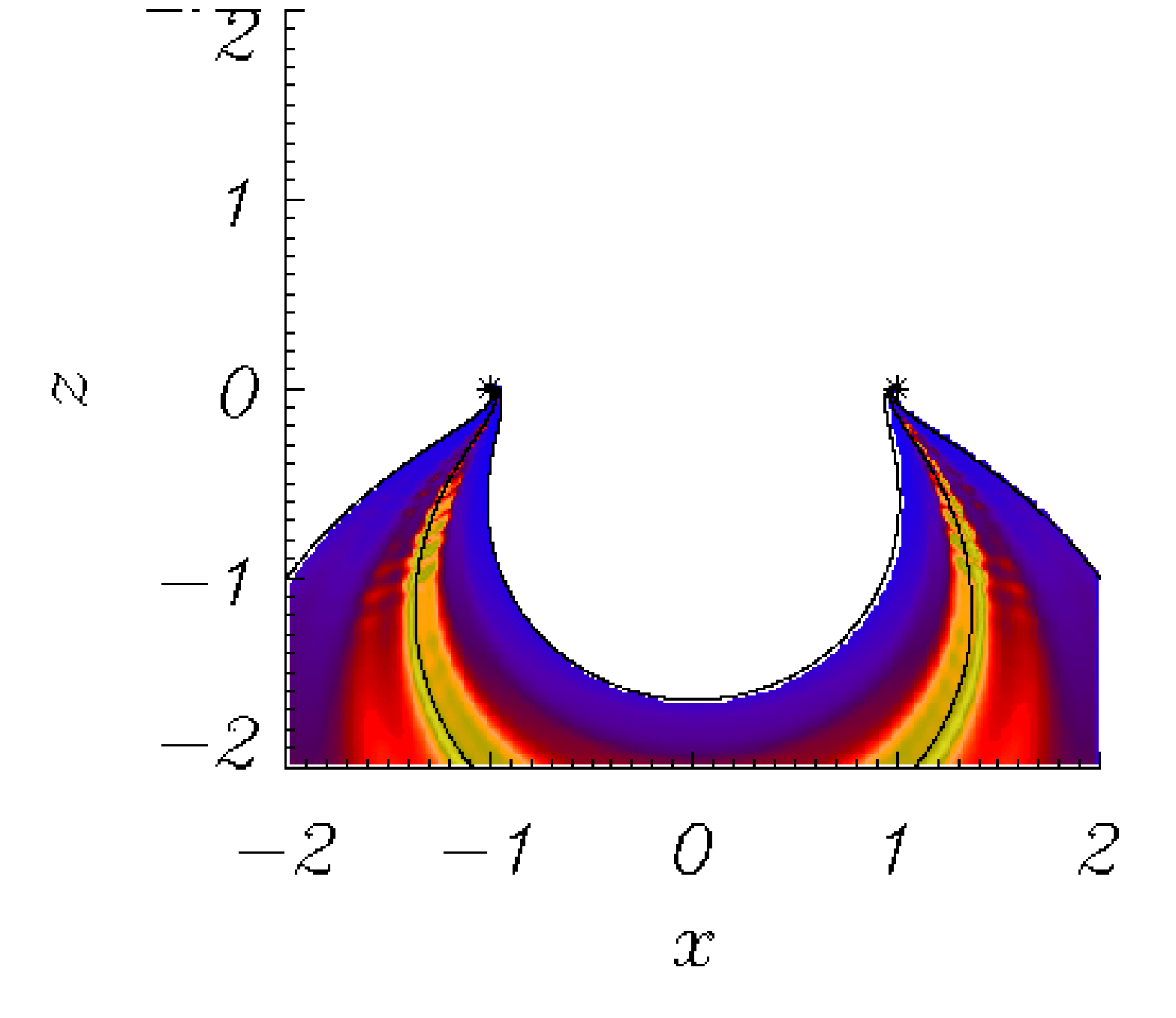}\\
\includegraphics[width=2.1176in]{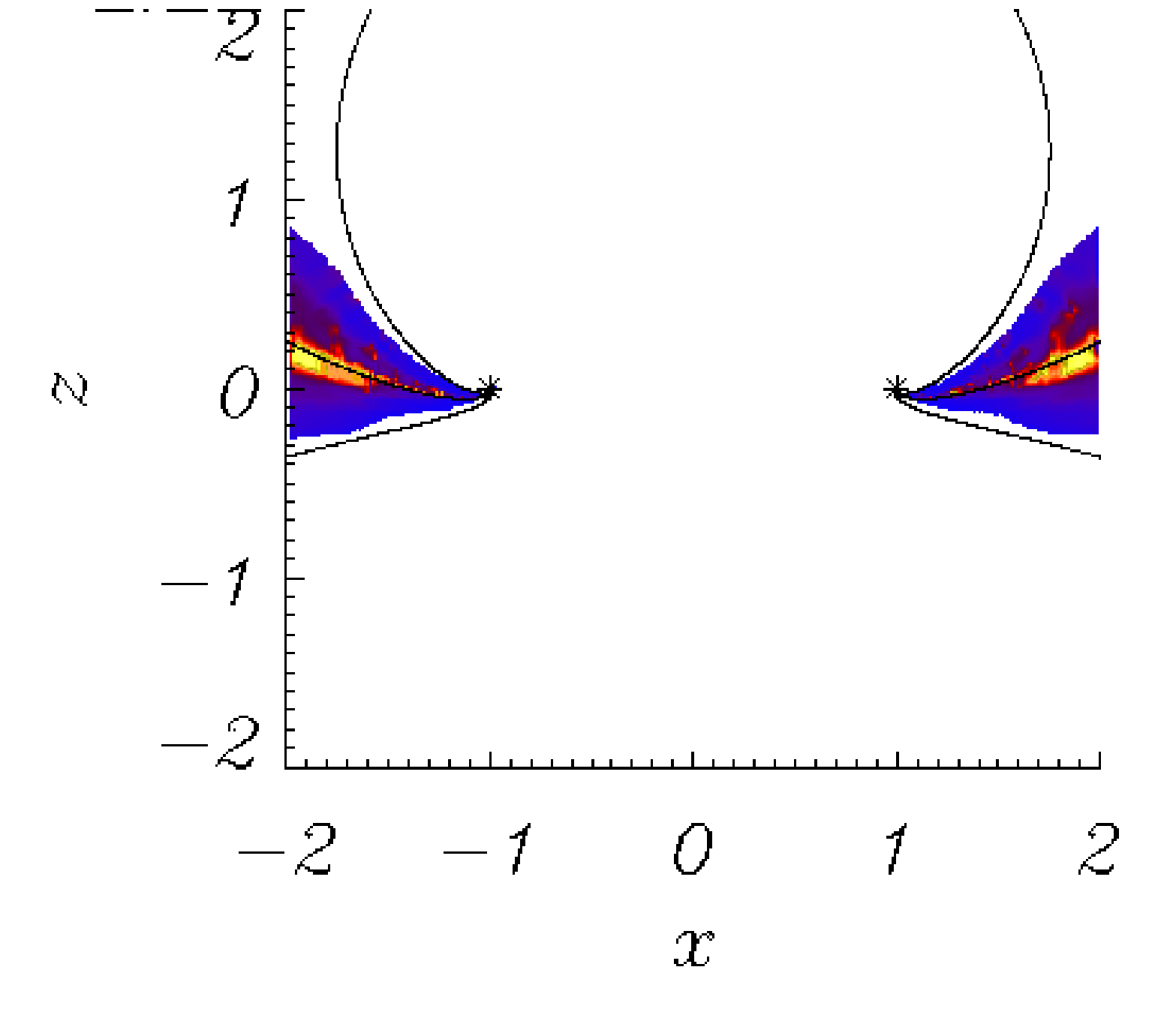}
\hspace{0.14286in}
\includegraphics[width=2.1176in]{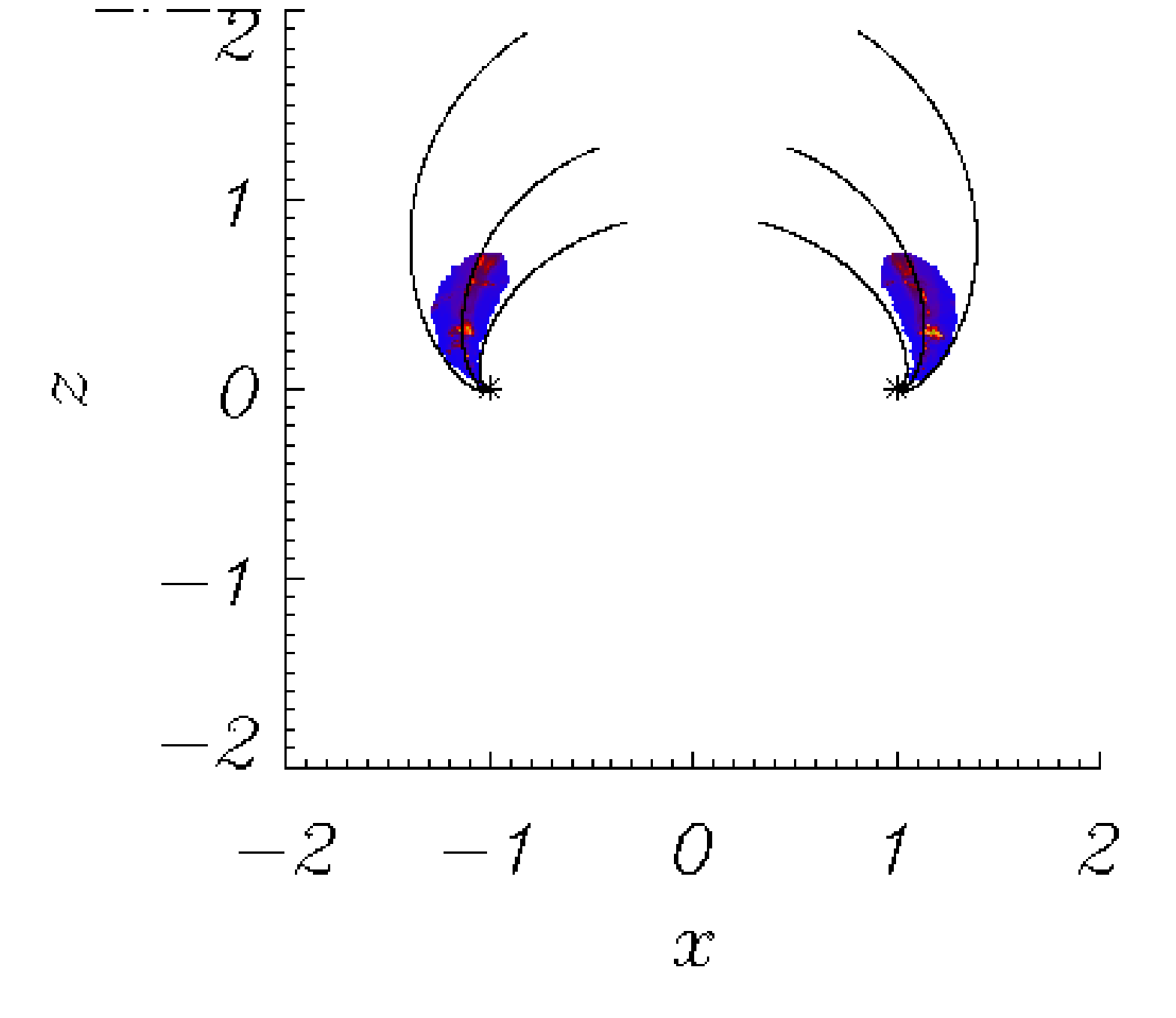}
\hspace{0.85in}
\includegraphics[width=2.056in]{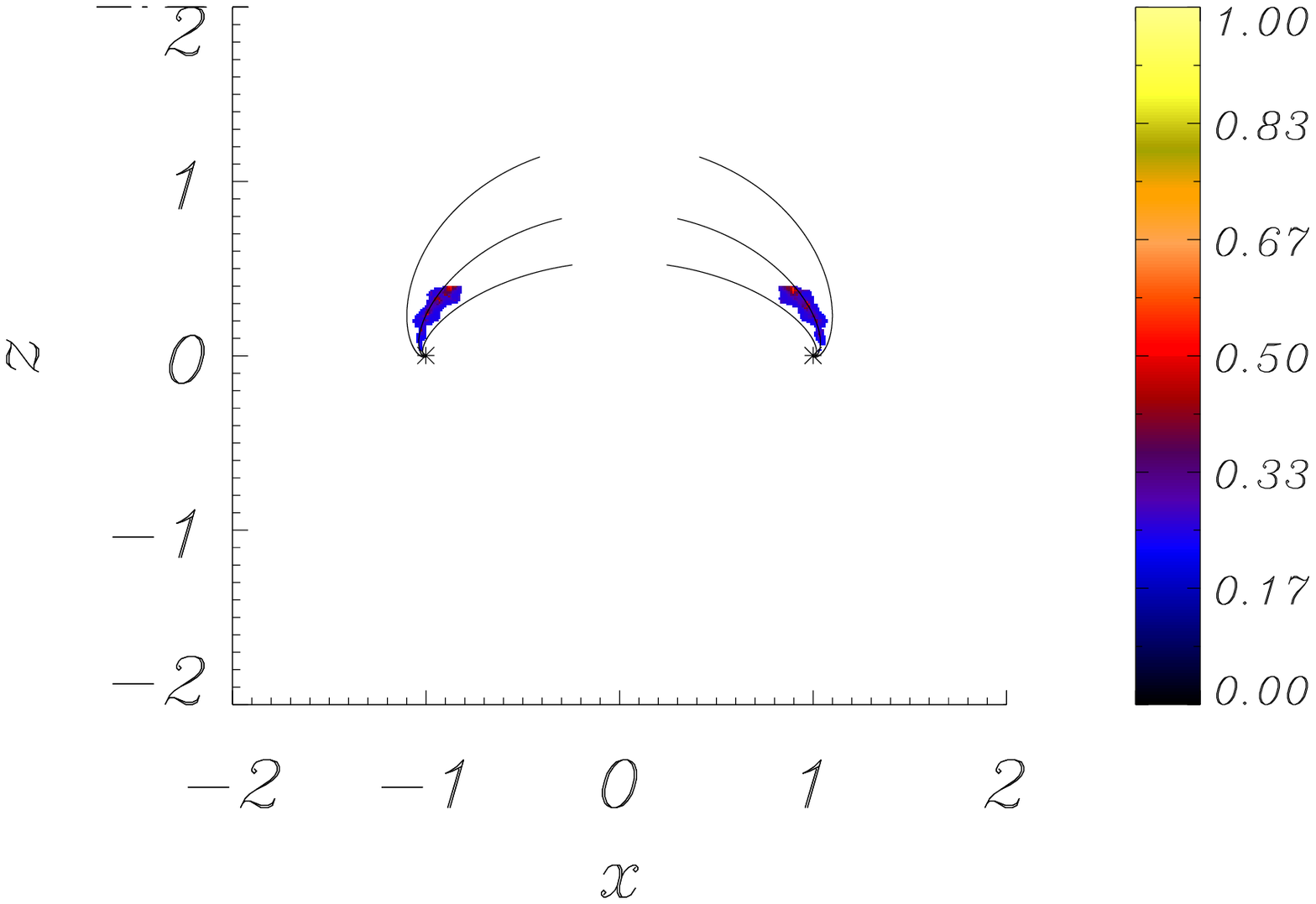}
\caption{Comparison of numerical simulation (shaded area) and analytical solution $V$ for a Fast wave sent in from upper boundary for $-3 \leq x \leq 3 $, and its resultant propagation at times $(a)$ $t$=0.25, $(b)$ $t$=0.5, $(c)$ $t$=0.75, $(d)$ $t=$1.0, $(e)$ $t$=1.5 and $(f)$ $t$=1.75, $(g)$ $t$=2.0, $(h)$ $t$=2.25, $(i)$ $t$=2.5, $(j)$ $t=$3.0, $(k)$ $t$=3.5 and $(l)$ $t$=3.75 , labelling from top left to bottom right. The lines represent the front, middle and back edges of the WKB wave solution, where the pulse enters from the top of the box.}
\label{figuresix}
\end{figure*}

These five ODEs were solved numerically using a fourth-order Runge-Kutta method. Contours of constant $\phi$ can be thought of as defining the positions of the edges of the wave pulse, i.e. with correct choices of $s$, the WKB solution represents the front, middle and back edges of the wave. $s$ is comparable to $t$ and so the numerical and analytical work can be directly compared. The agreement between the analytic model and the leading edge of the wavefront is very good, as seen in an overplot of a numerical simulation (shaded area) and our WKB solution (thick lines) in Figure {\ref{figuresix}}.

\begin{figure}[htb]
\begin{center}
\includegraphics[width=2.4in]{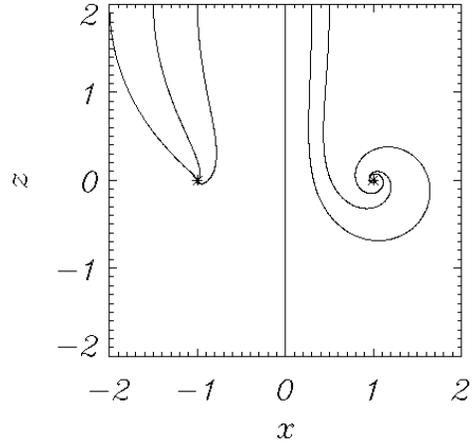}
\caption{Plots of WKB solution for a wave sent in from the upper boundary and its resultant particle paths for starting points of $x=-2$, $-1$, $0$, $0.3$ and $0.5$ along $z=2$.}
\label{figureseven}
\end{center}
\end{figure}

We can also use our WKB solution to plot the particle paths of individual elements from the initial wave. In Figure \ref{figureseven}, we see the spiral evolution of elements that begin at points $x=-2$, $-1.5$, $-1$, $0$, $0.3$ $0.5$ along $z=2$. Note that an element that begins at $x=0, z=2$ is not affected by the null points and passes between the two and off to $-\infty$. In effect, the two null points cancel out each other's effect at this point, although any perturbation to an element travelling along this path would send the element spiralling towards one of the nulls. The behaviour of such a particle can be shown:
\begin{eqnarray*}
\qquad v_A(0,z)=\left. \sqrt{ B_x^2+B_z^2} \right| _{x=0}=-(z^2+1) = \frac{\partial z}{\partial t}
\end{eqnarray*}
Hence, solving for $z$, we find $z=\tan{ \left(A -t\right)}$, where $A=\arctan{z_0}$ and $z_0$ is some starting position (which is $z_0=2$ in our simulations).

\begin{figure*}[t]
\begin{center}
\includegraphics[width=2.0in]{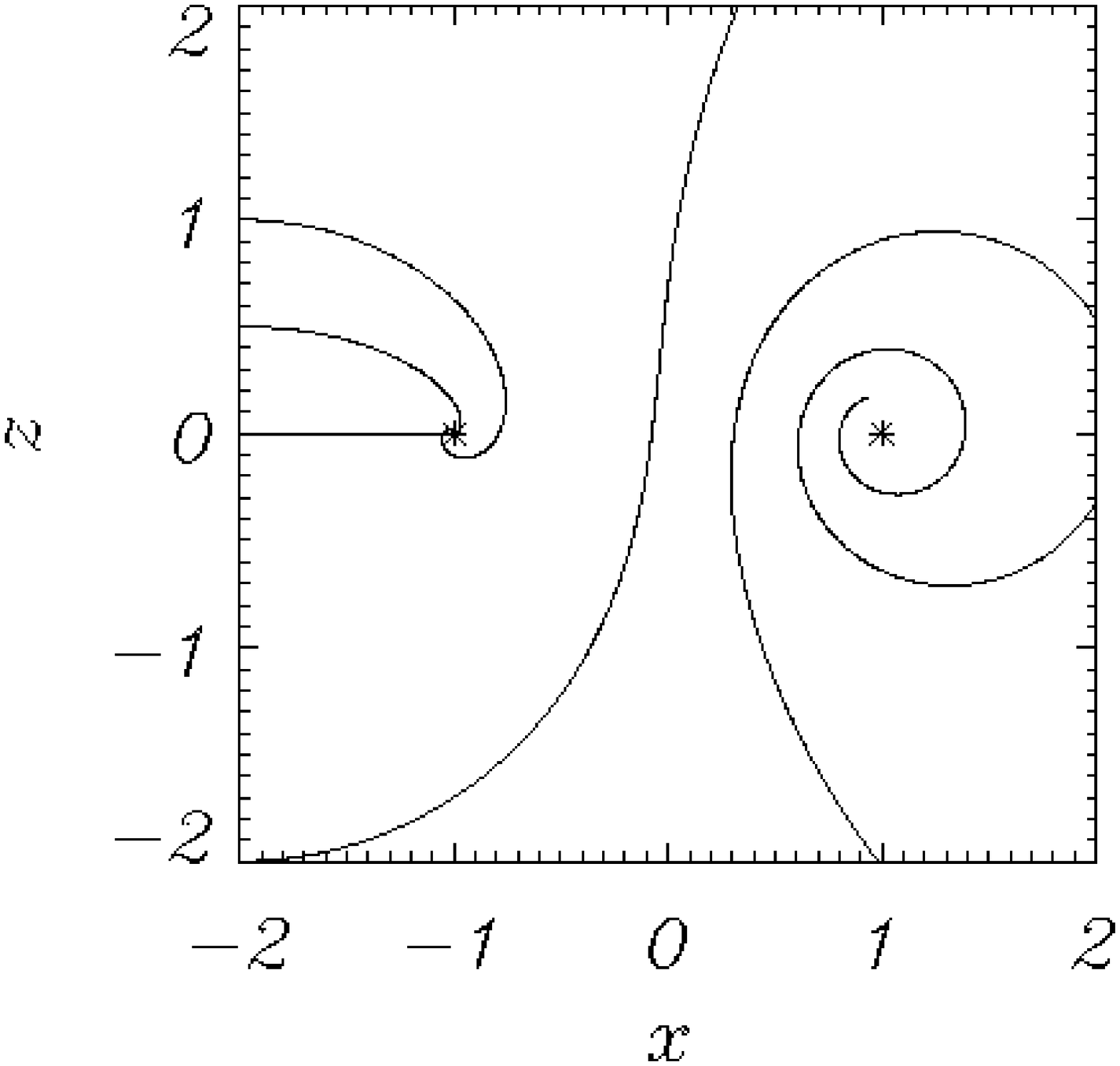}
\hspace{0.15in}
\includegraphics[width=2.0in]{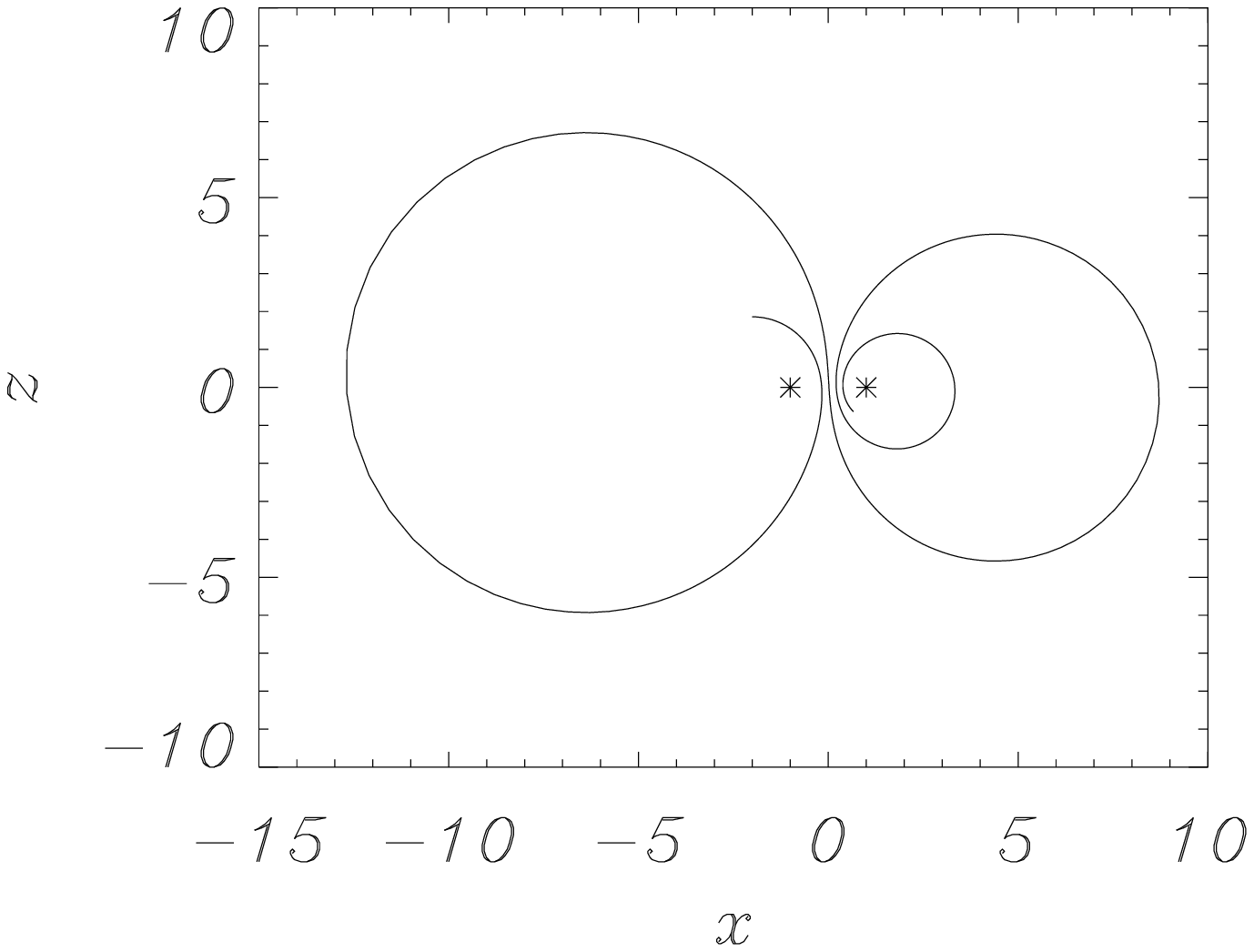}\\
\includegraphics[width=2.0in]{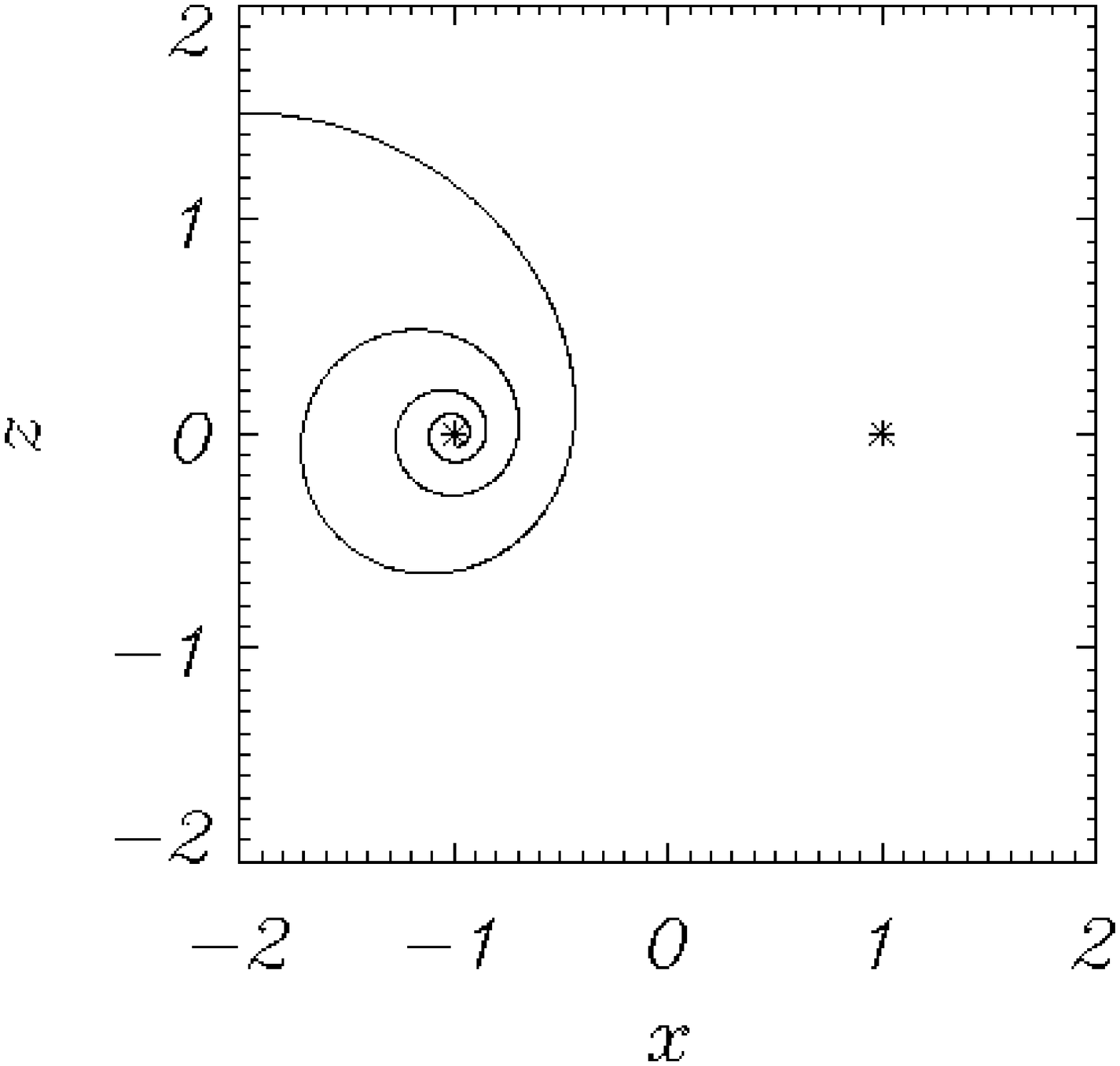}
\hspace{0.15in}
\includegraphics[width=2.0in]{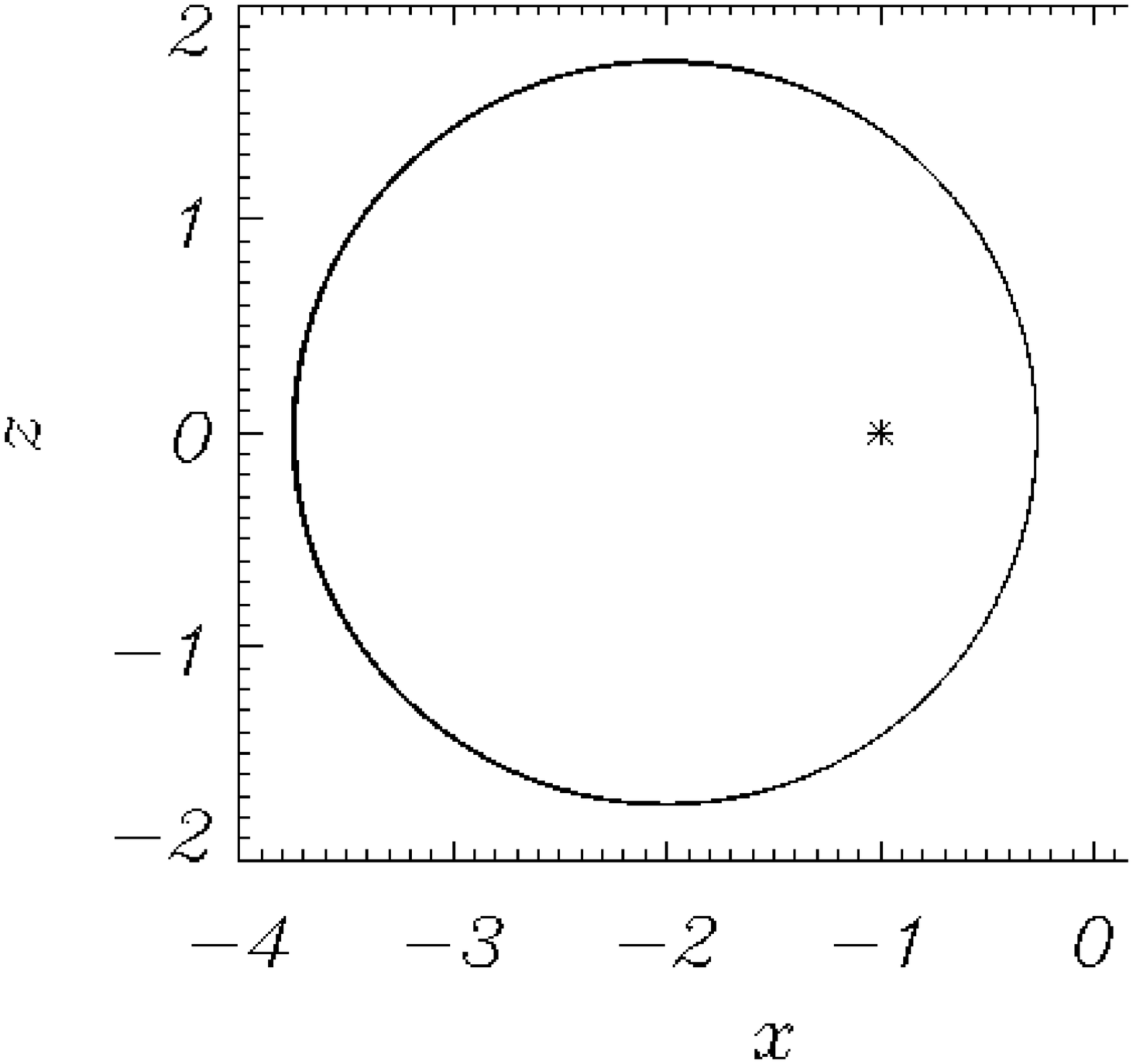}
\caption{Plots of WKB solution for a wave sent in from the side boundary and its resultant particle paths. Top left displays starting points of $z=-2$, $0$, $0.5$ and $+1$ along $x=-2$. Bottom left displays a starting point of $z=+1.5$. The top right subfigure shows a starting point of $z=1.86$, the lower right shows $z=\sqrt{3}$. Note the change of scale in the two right hand subfigures. }
\label{figureeleven}
\end{center}
\end{figure*}


\subsection{Side boundary}

We also investigated the effect of solving the linearised MHD equations for a fast wave coming in from the side boundary, along $x=-2$. Again a two-step Lax-Wendroff numerical scheme was run in a box with $-2 \le x \le 4$ and $-3 \le z \le 3$, with our attention again focused on  $-2 \le x \le 2$ and $-2 \le z \le 2$. For a single wave pulse from the side boundary, the boundary conditions were:
\begin{eqnarray*}
\qquad V(-2, z) &=& \left\{\begin{array}{cl}
{\sin { \omega t } } & {\mathrm{for} \; \;0 \leq t \leq \frac {\pi}{\omega} } \\
{0} & { \mathrm{otherwise} }
\end{array} \right. \; , \\
\;  \left.\frac {\partial V } {\partial x } \right| _{x=4} &=& 0 \; , \quad \left. \frac {\partial V } {\partial z } \right| _{z=-3}  = 0 \;  , \quad \left. \frac {\partial V } {\partial z } \right| _{z=3}  = 0 \;.
\end{eqnarray*}

Again, tests show that the central behaviour is largely unaffected by these choices of the remaining boundary conditions.

We find that the linear,  fast wave travels in from the left hand side of the box and begins to feel the effect of the left hand side null point (at $x=-1$, $z=0$). The wave thins and begins to wrap around this null point. As the ends of the wave wrap around behind the left null point, they then become influenced by the second, right hand side null point  (at $x=+1$, $z=0$). These arms of the wave then proceed to wrap around the right null point, flattening the wave. Furthermore, the two parts of the wave now travelling through the area between the null points have non-zero  Alfv\'{e}n speed, and so pop through (the wave in the lower half plan now travels up, crossing the line $-1<x<1$ and $z=0$, and visa versa for the wave in the upper half plan). These parts of the wave break along $x=0$ and then proceed to wrap around the null point closest to them. Eventually, the wave accumulates at both null points. This can be seen in the shaded contours of Figure \ref{figureten}.

\subsection{Analytical results}

\begin{figure*}[htb]
\includegraphics[width=2.1176in]{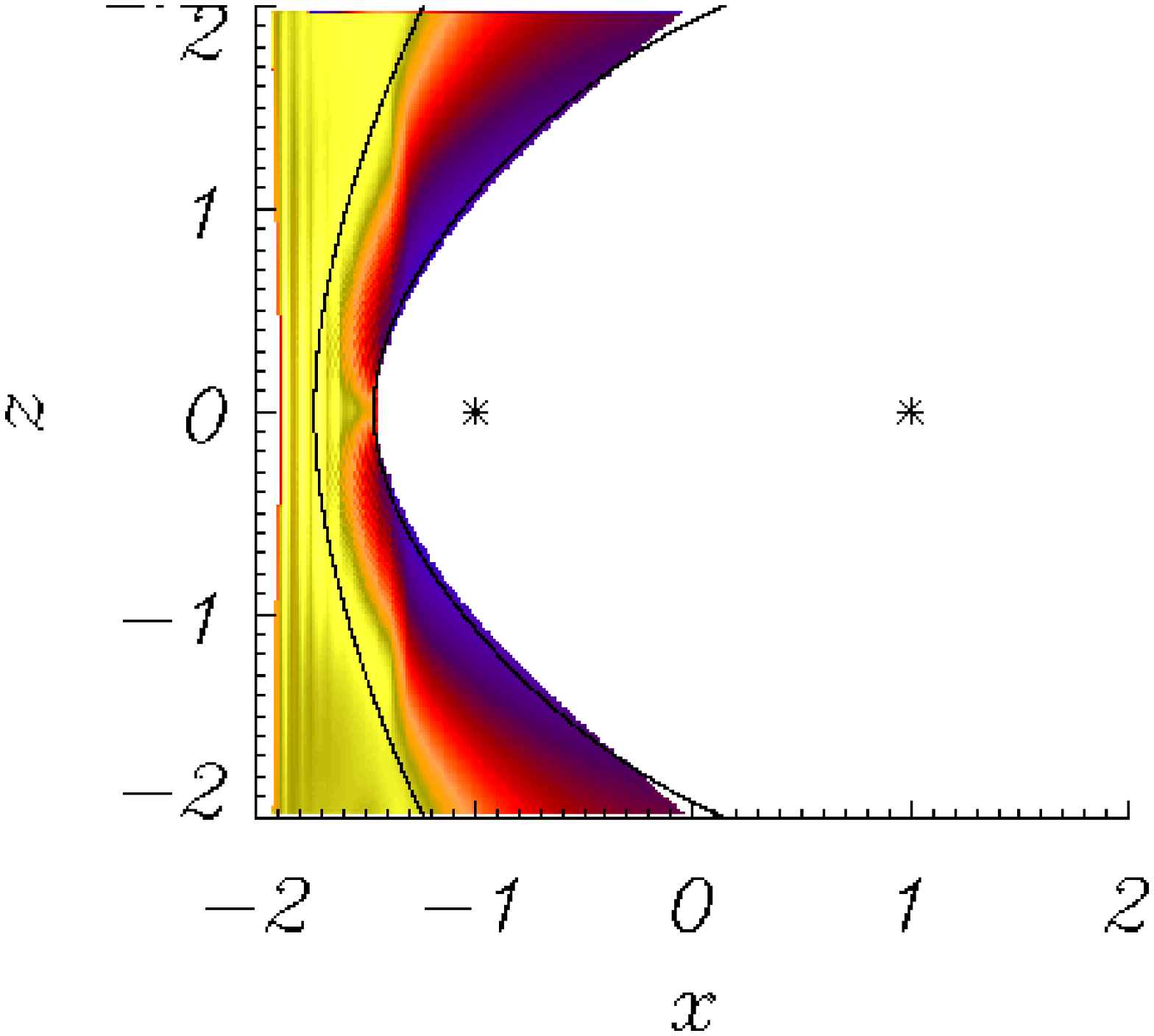}
\hspace{0.14286in}
\includegraphics[width=2.1176in]{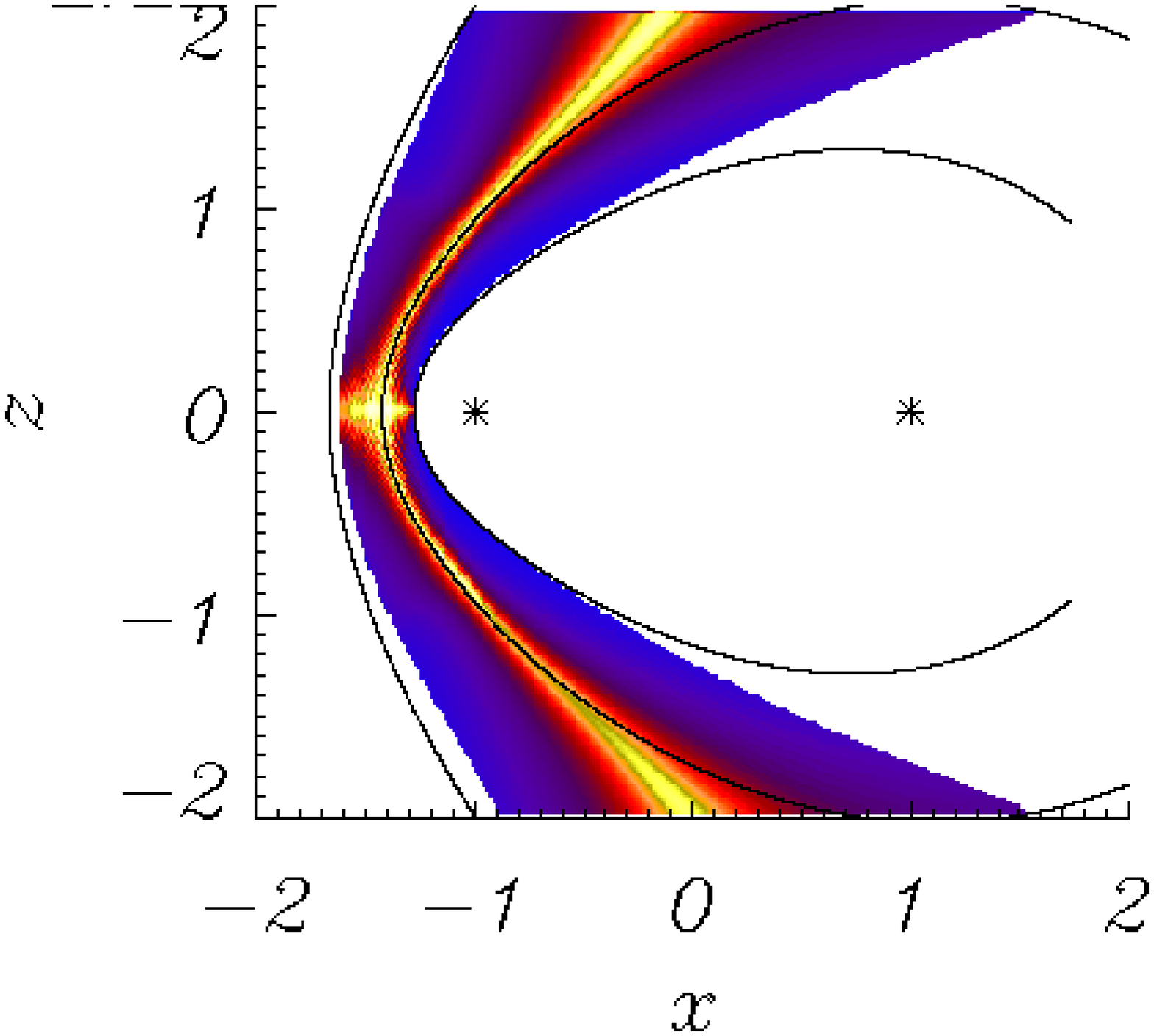}
\hspace{0.14286in}
\includegraphics[width=2.1176in]{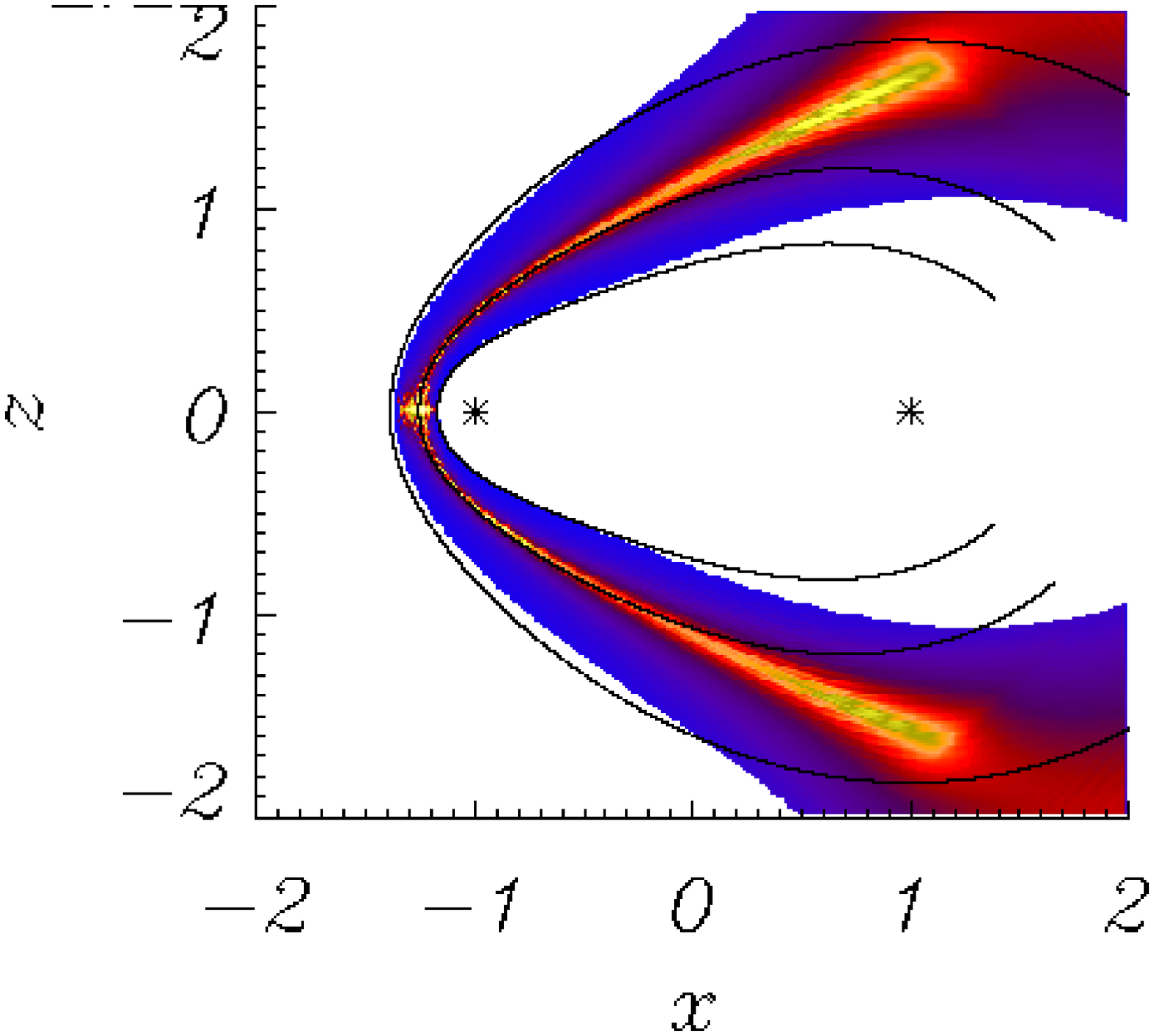}\\
\includegraphics[width=2.1176in]{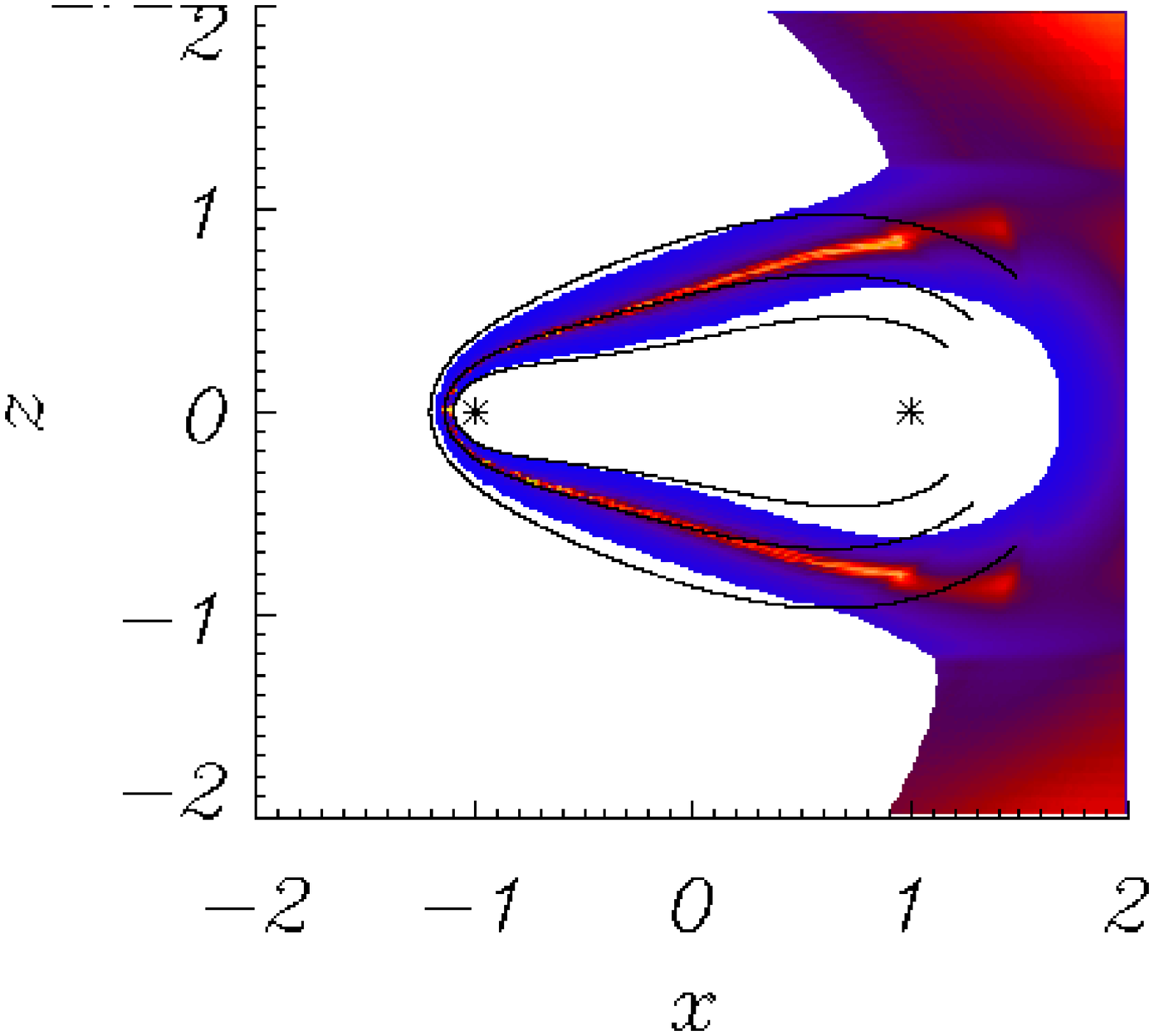}
\hspace{0.14286in}
\includegraphics[width=2.1176in]{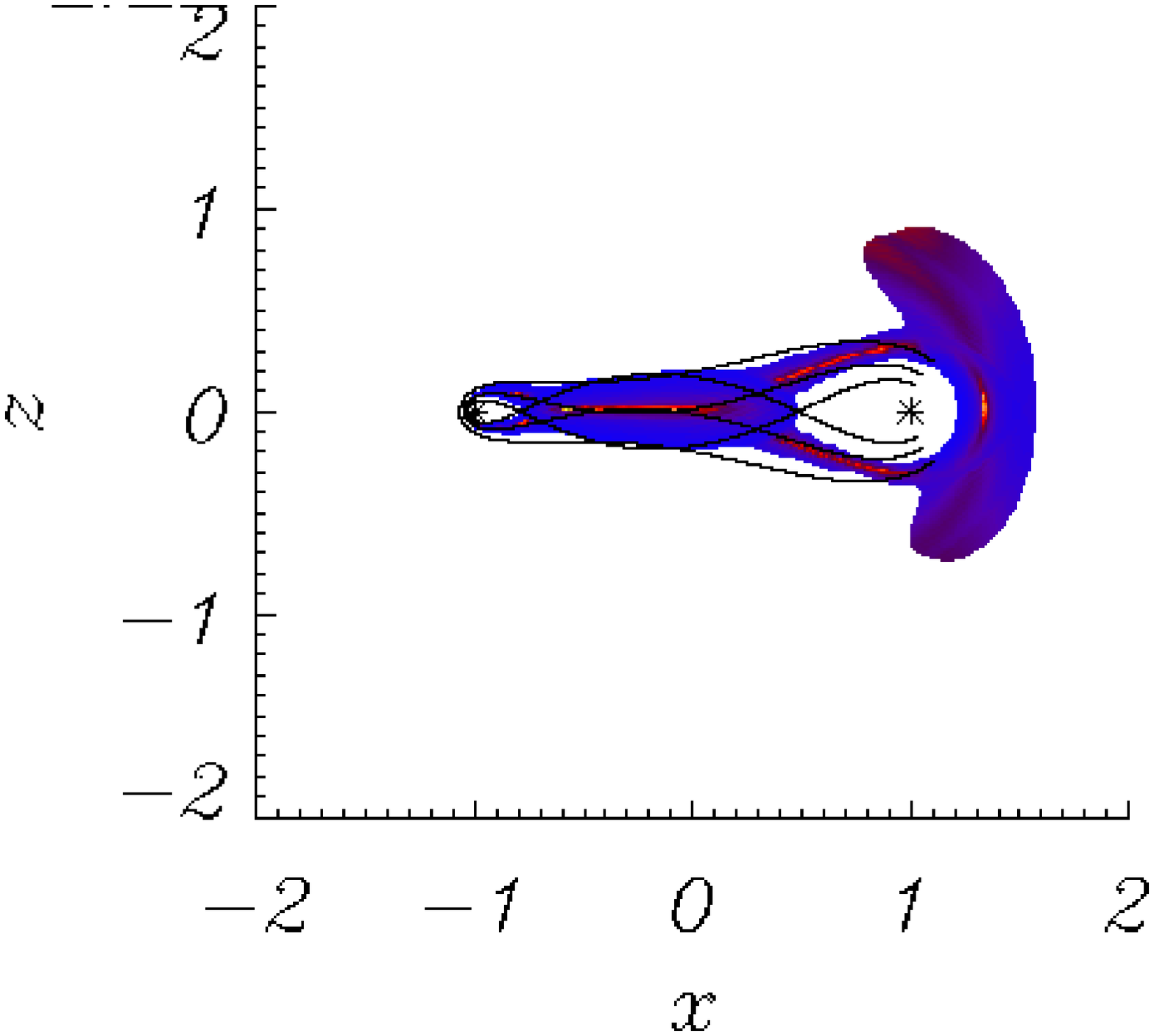}
\hspace{0.14286in}
\includegraphics[width=2.1176in]{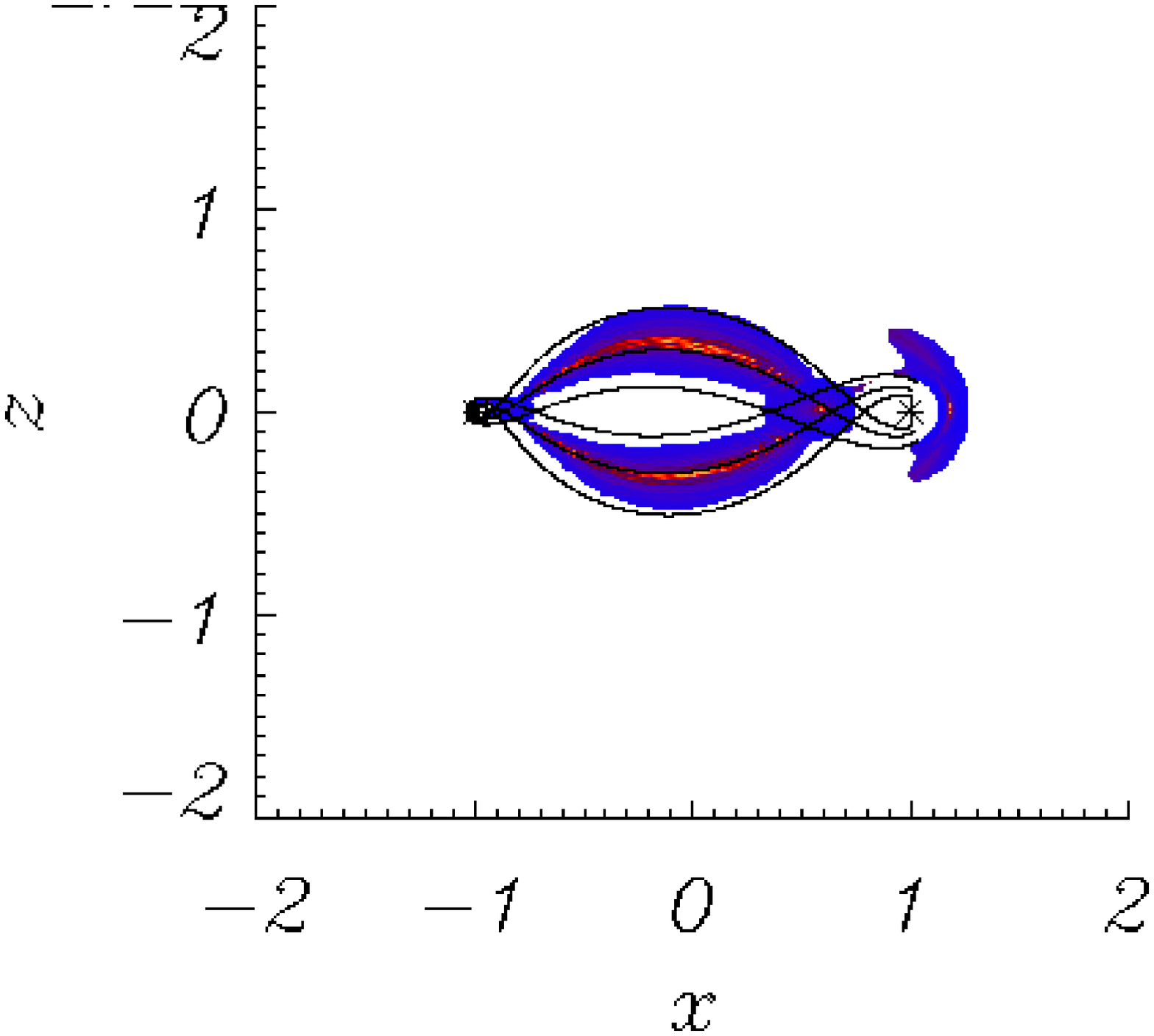}\\
\includegraphics[width=2.1176in]{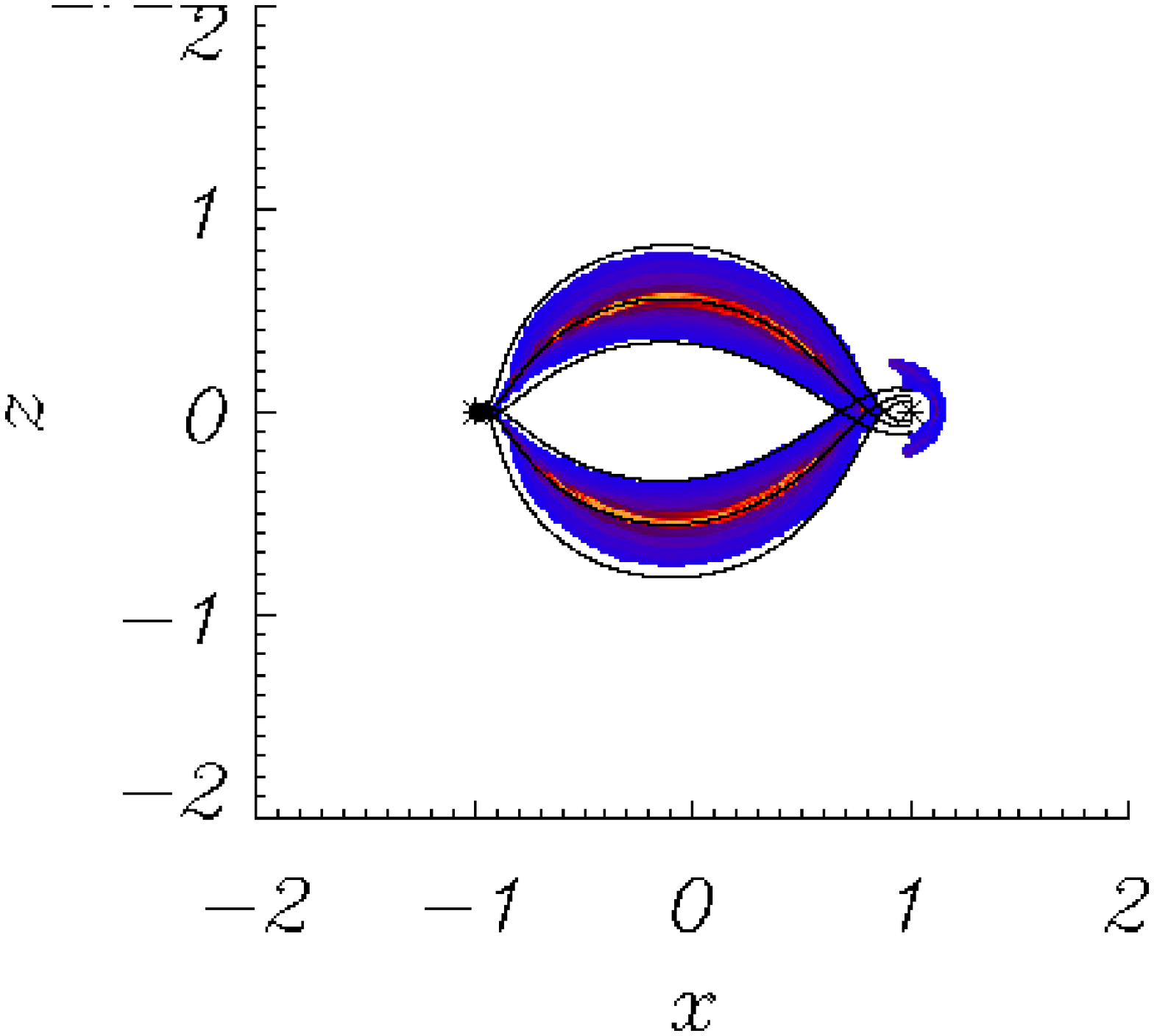}
\hspace{0.14286in}
\includegraphics[width=2.1176in]{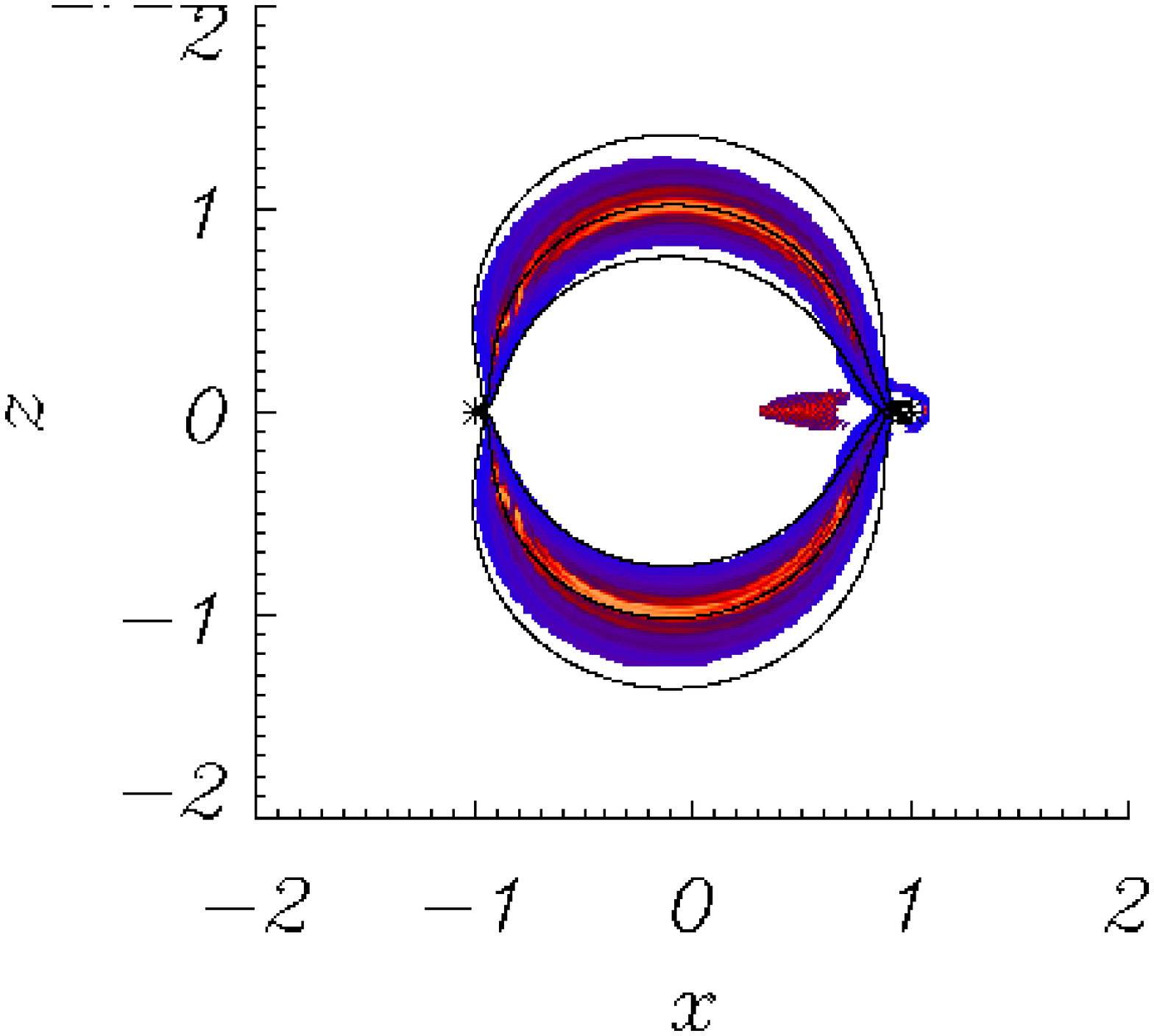}
\hspace{0.14286in}
\includegraphics[width=2.1176in]{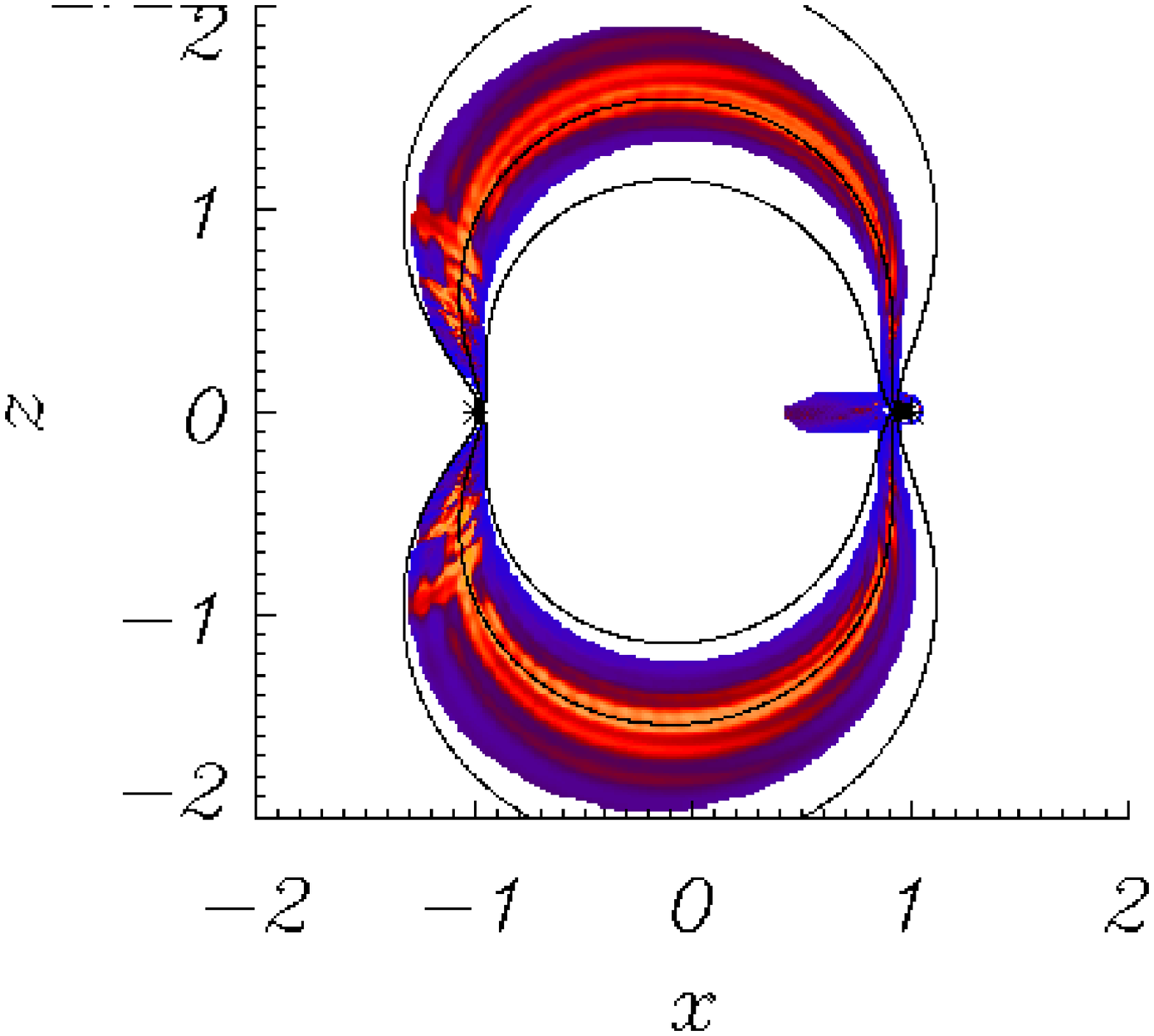}\\
\includegraphics[width=2.1176in]{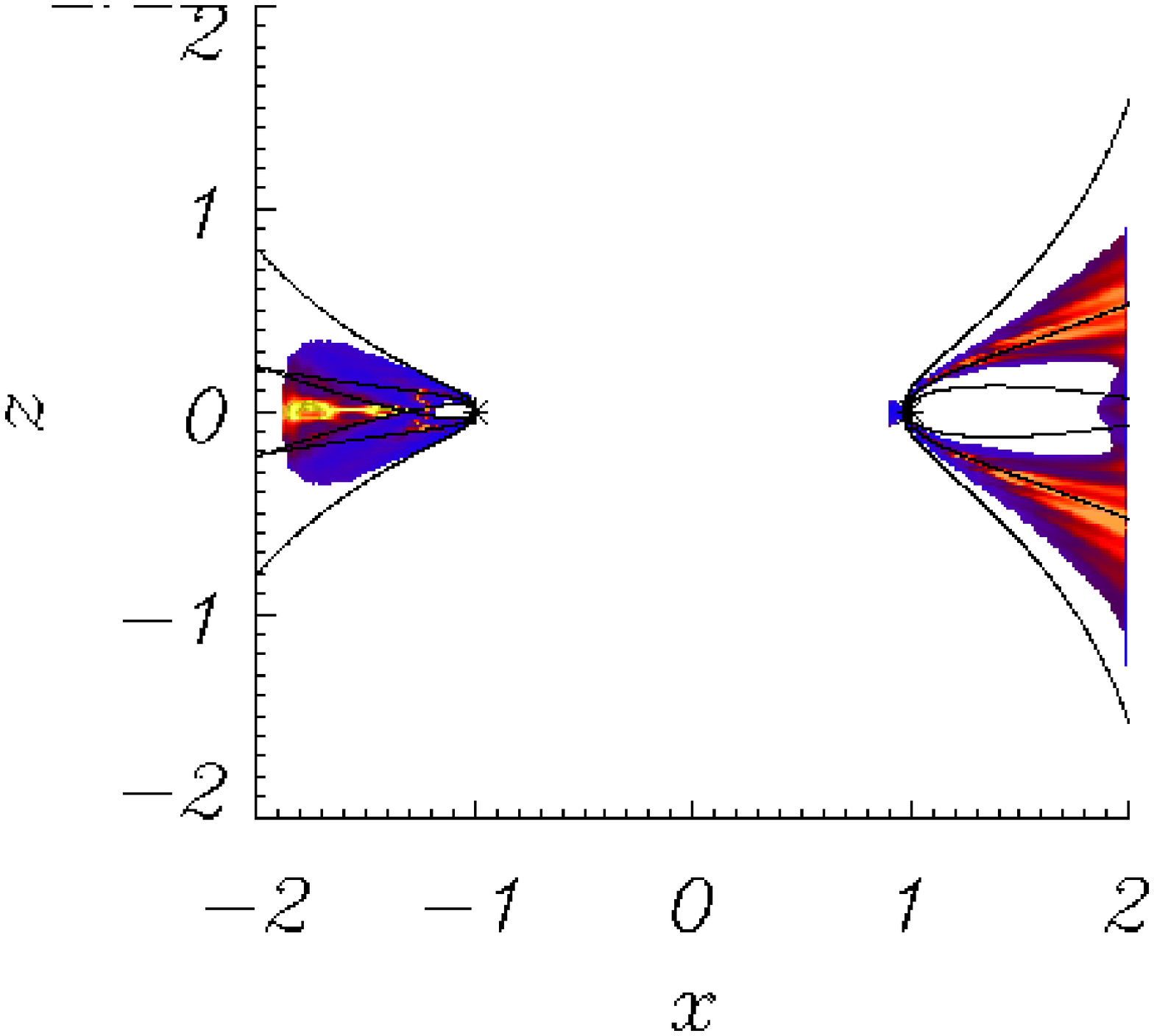}
\hspace{0.14286in}
\includegraphics[width=2.1176in]{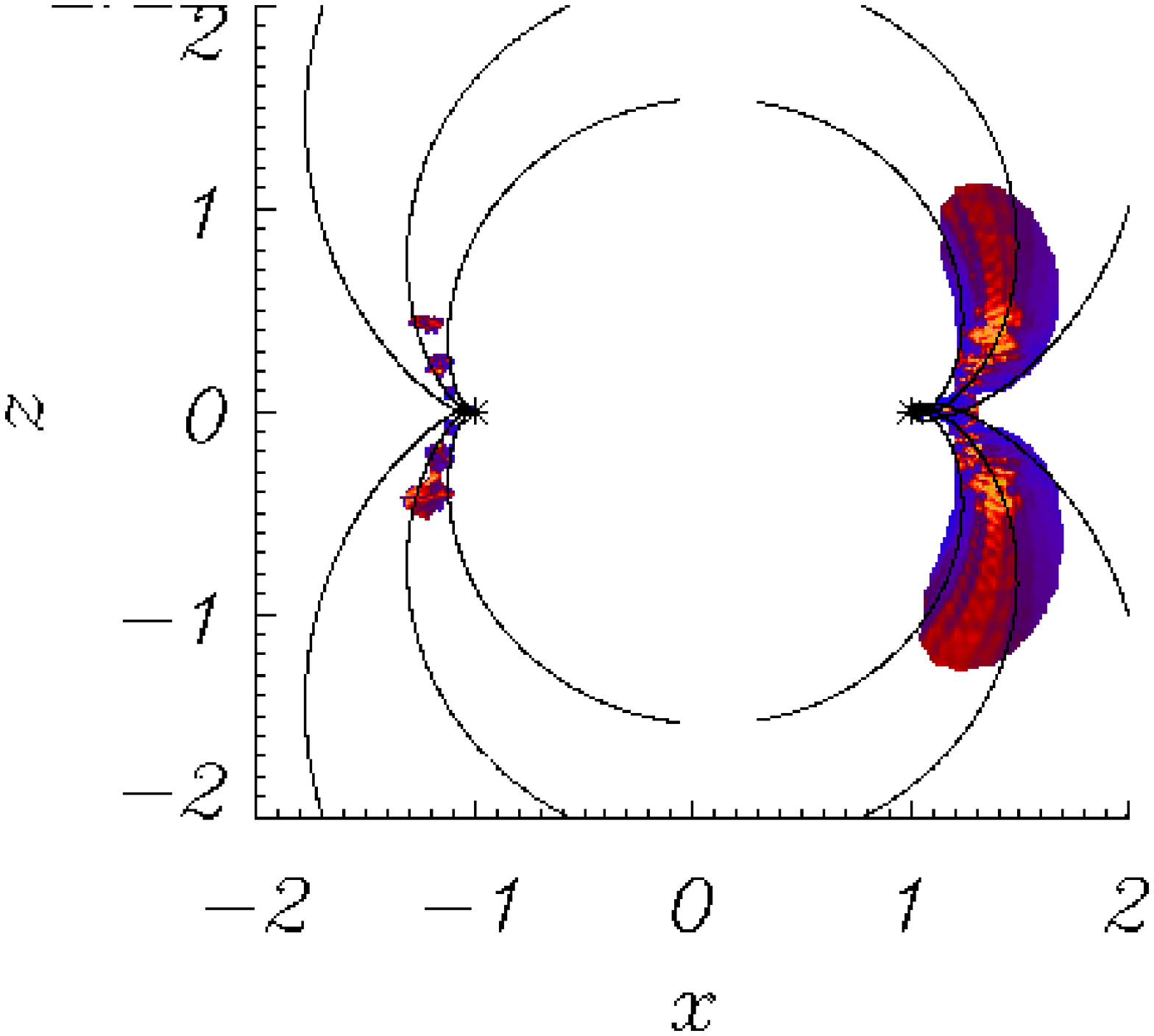}
\hspace{0.85in}
\includegraphics[width=2.056in]{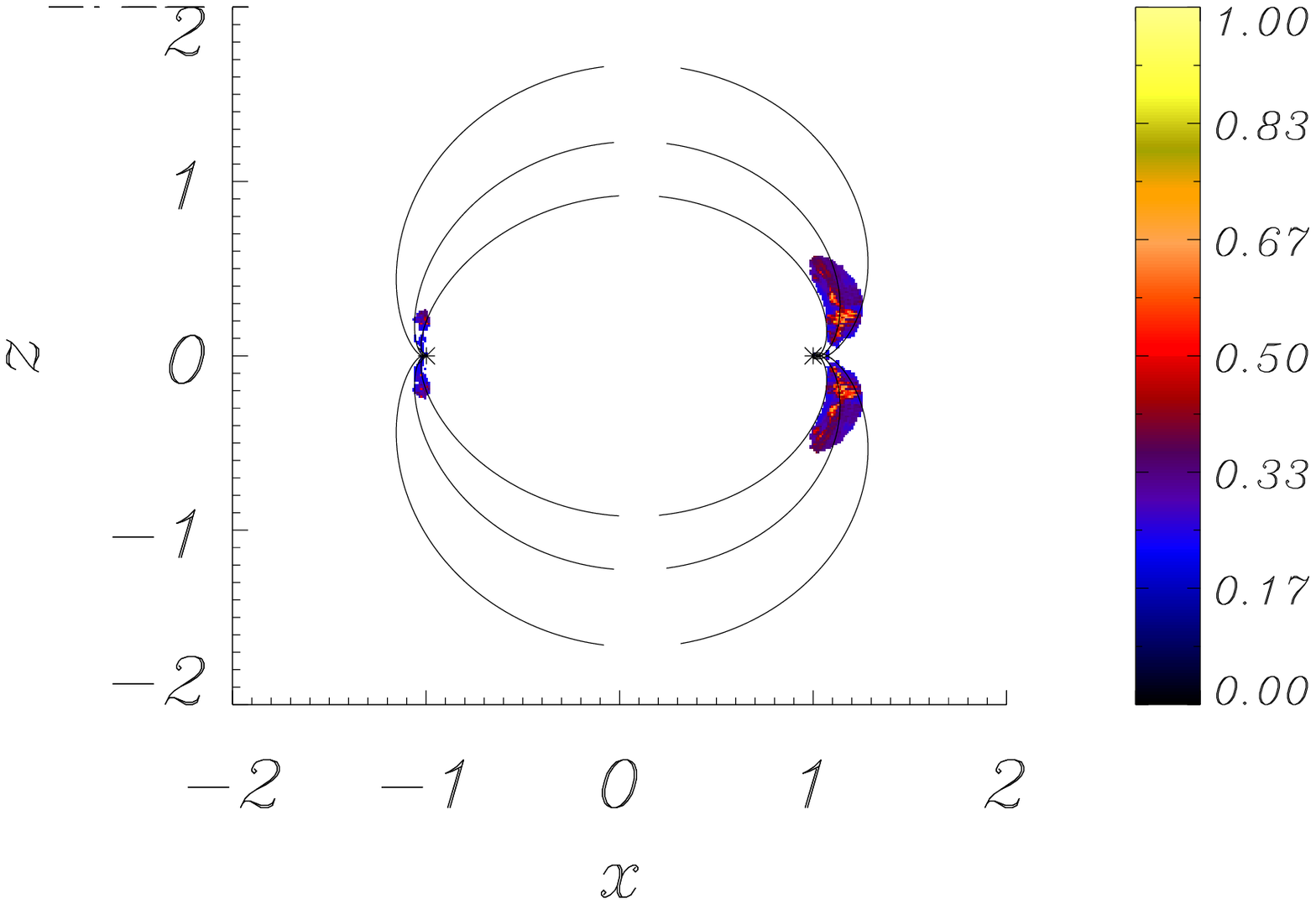}
\caption{Comparison of numerical simulation and analytical solution of $V$ for a Fast wave sent in from left hand boundary for $-3 \leq x \leq 3$, and its resultant propagation at times $(a)$ $t$=0.25, $(b)$ $t$=0.5, $(c)$ $t$=0.75, $(d)$ $t=$1.0, $(e)$ $t$=1.5 and $(f)$ $t$=1.75, $(g)$ $t$=2.0, $(h)$ $t$=2.25, $(i)$ $t$=2.5, $(j)$ $t=$3.0, $(k)$ $t$=3.5 and $(l)$ $t$=3.75, labelling from top left to bottom right. The lines represent the front, middle and back edges of the wave, where the pulse enters from the side of the box.}
\label{figureten}
\end{figure*}

As before, equations \ref{fastcharacteristics} were solved numerically using a fourth-order Runge-Kutta method, but this time using different initial conditions. The thick black lines in Figure \ref{figureten} shows constant $\phi$ at different values of the parameter $s$.  Constant $\phi$ can be thought of as defining the position of the edge of the wave pulse, i.e. with correct choices of $s$, the WKB solution represents the front, middle and back edges of the wave. As before, the agreement between the analytic model and the wavefront is very good and can be seen in an overplot of a numerical simulation and our WKB solution in Figure {\ref{figureten}}.

We can also use our WKB solution to plot the particle paths of individual elements from the initial wave. In Figure \ref{figureeleven}, we see the spiral evolution of elements that begin at points $z=-2$, $0$, $0.5$, $1.0$, $1.5$ and $1.86$ along $x=-2$. Note that for a wavefront extending between $-2 \le z \le +2$, there is a critical distance where some elements of the wave are sucked into the nearest (left hand) null point, and others spiral round it at such a great distance that they feel the influence of the second (right hand) null point and begin spiralling towards it ($x=+1$, $z=0$). Note that for a starting position of  $z=\sqrt {3}$, the wave element orbits in a circle (radius $\sqrt {3}$). This is the critical starting distance; a starting distance, say $z_0$, less than this will sprial into the left hand null point, whereas the wave particle will spiral into the right hand null point if $z_0 > \sqrt{3}$. This can be seen in Figure \ref{figureeleven}. Simulations show that this critical distance is $\sqrt {L^2-1}$, where $L$ is the distance between the origin and the side boundary (here $L= \left| x_0 \right| =2$). When $x_0 = -2$ and $z_0 =\sqrt {3}$ (i.e. critical starting distance), the equations are satisfied by: 
\begin{eqnarray*}
\qquad x &=& \frac {-1} {\left( 2+\sqrt{3}\sin{\theta} \right)}  \; , \quad p = - \frac {\pi}{\sqrt{3}} \cos{\theta} \; , \\
\qquad  z &=& \frac {\sqrt{3} \cos{\theta}} {\left( 2+\sqrt{3}\sin{\theta} \right)} \; , \quad  q = \pi + \frac {2 \pi}{\sqrt{3}} \sin{\theta} \; , 
\end{eqnarray*}
where $\theta = 4 \pi  s - {\pi} / {3}$, and confirm the periodic circular orbit.

\newpage
\clearpage

\begin{figure*}[t]
\includegraphics[width=2.1176in]{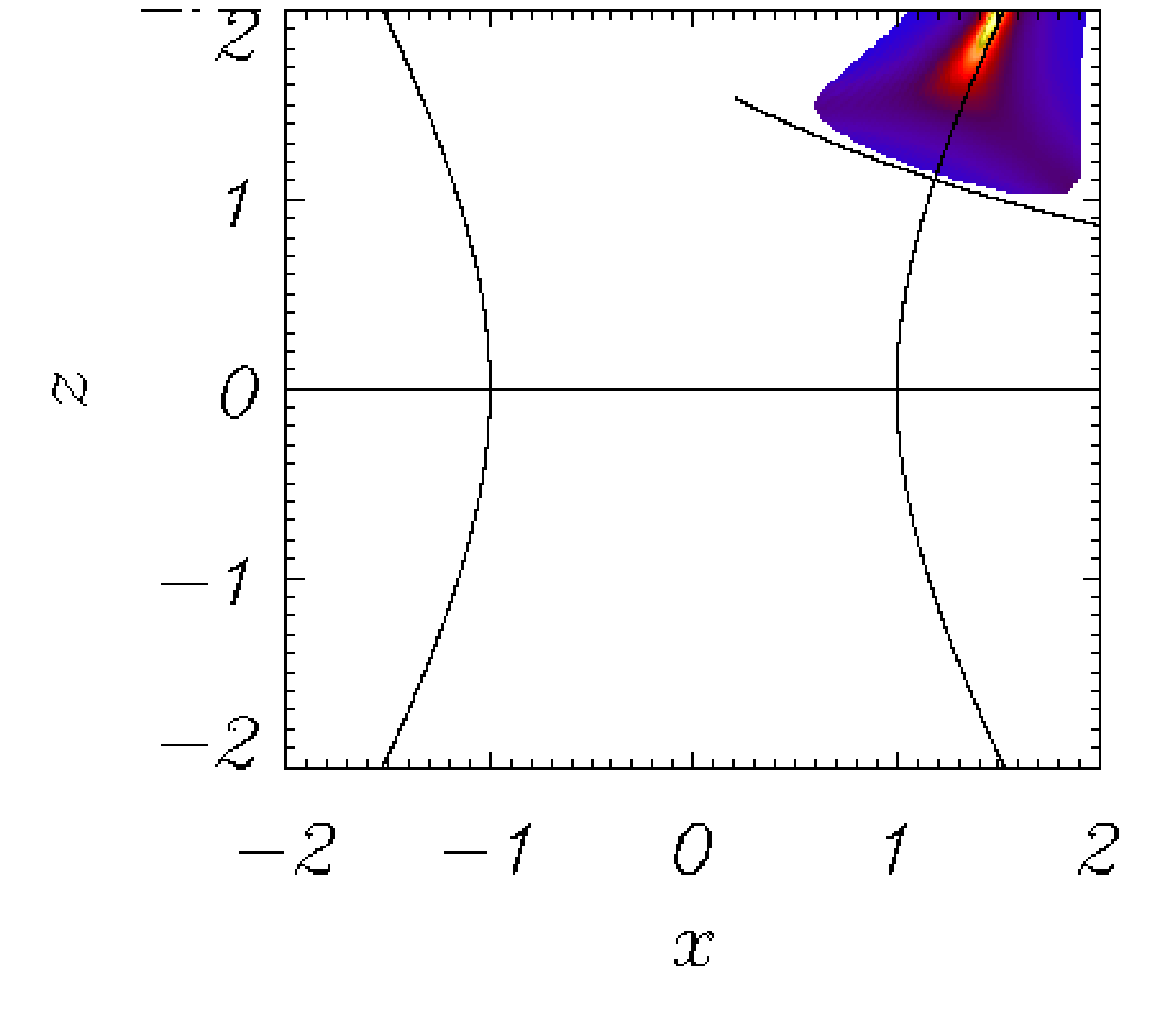}
\hspace{0.14286in}
\includegraphics[width=2.1176in]{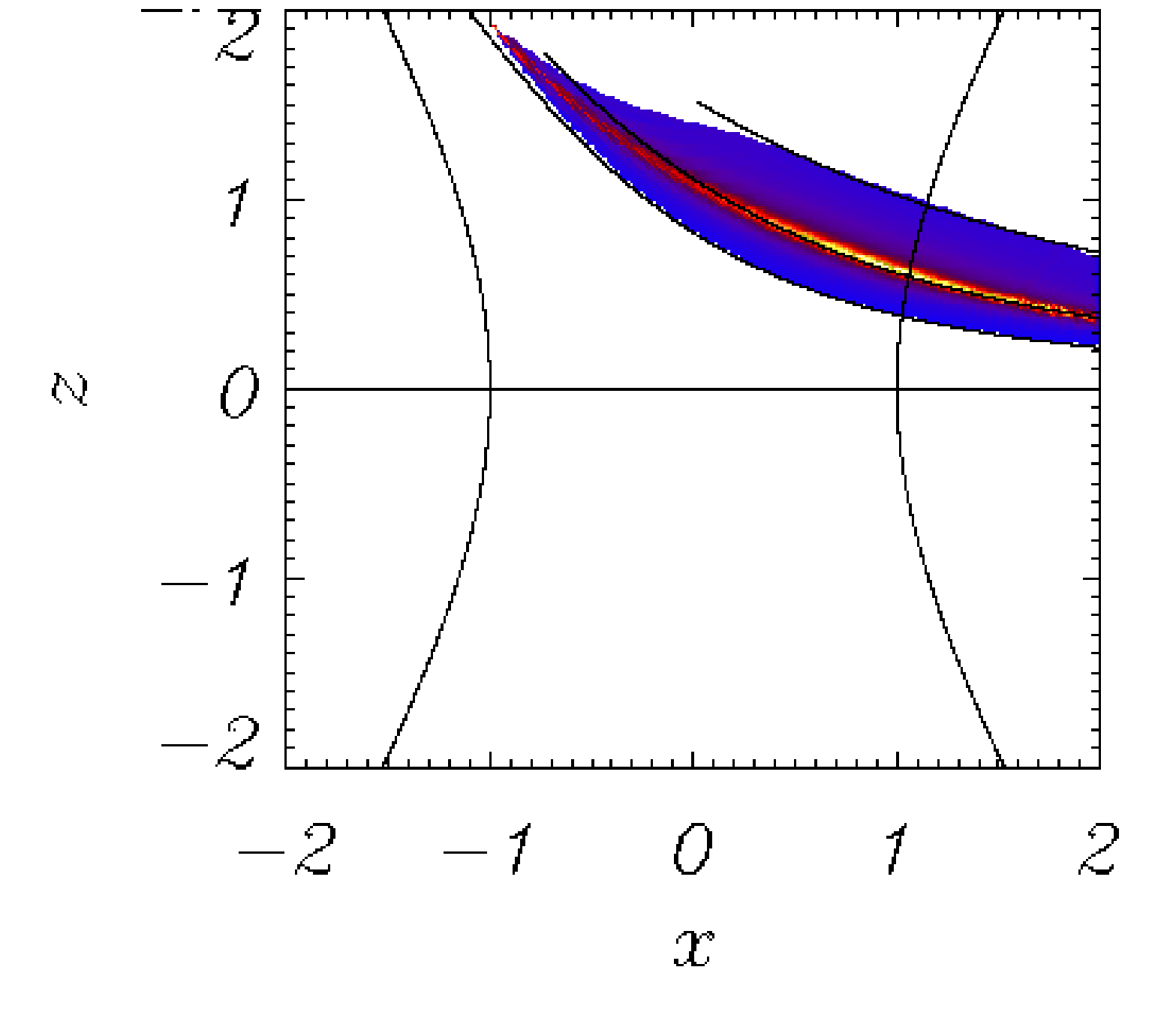}
\hspace{0.14286in}
\includegraphics[width=2.1176in]{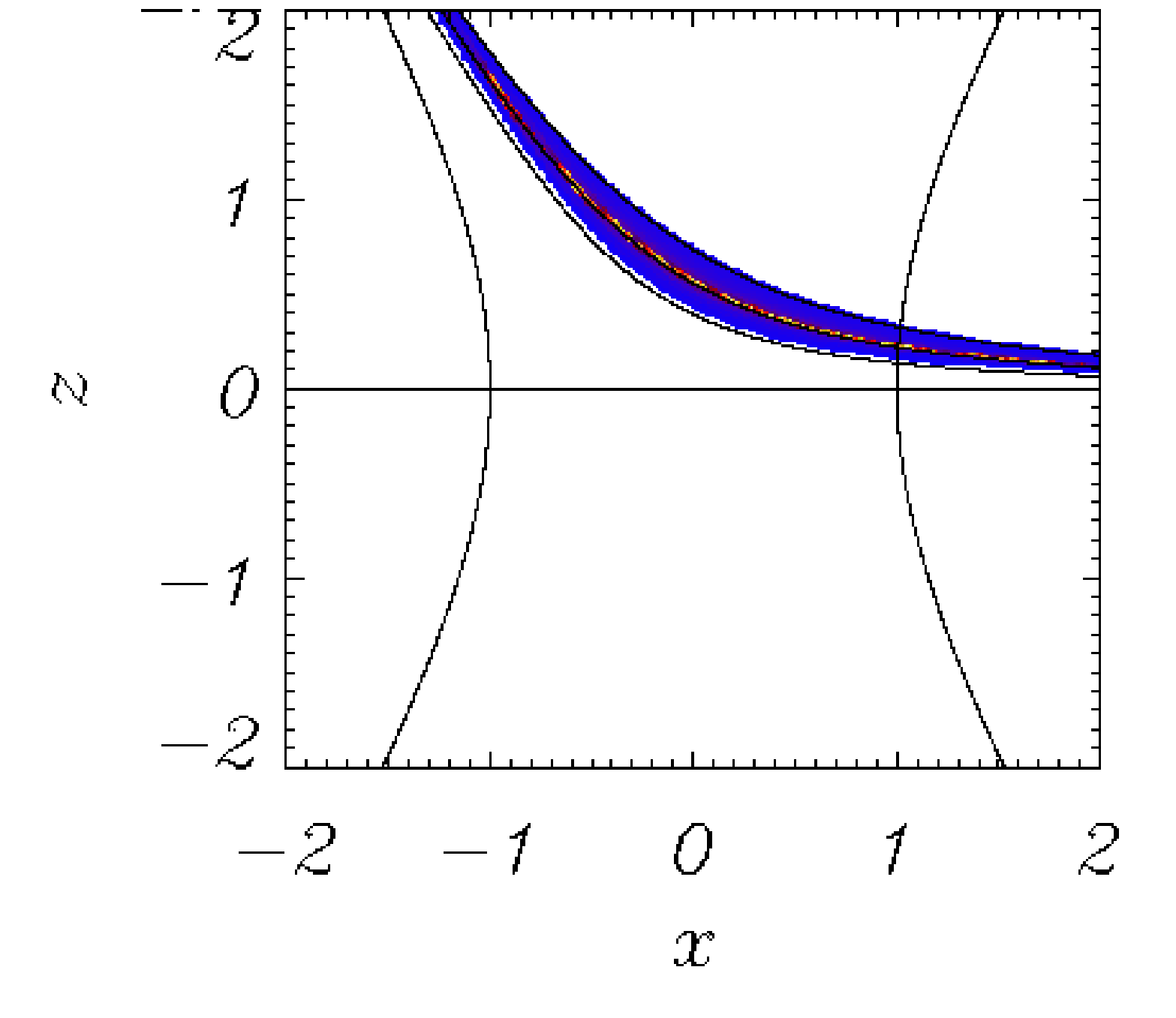}\\
\includegraphics[width=2.1176in]{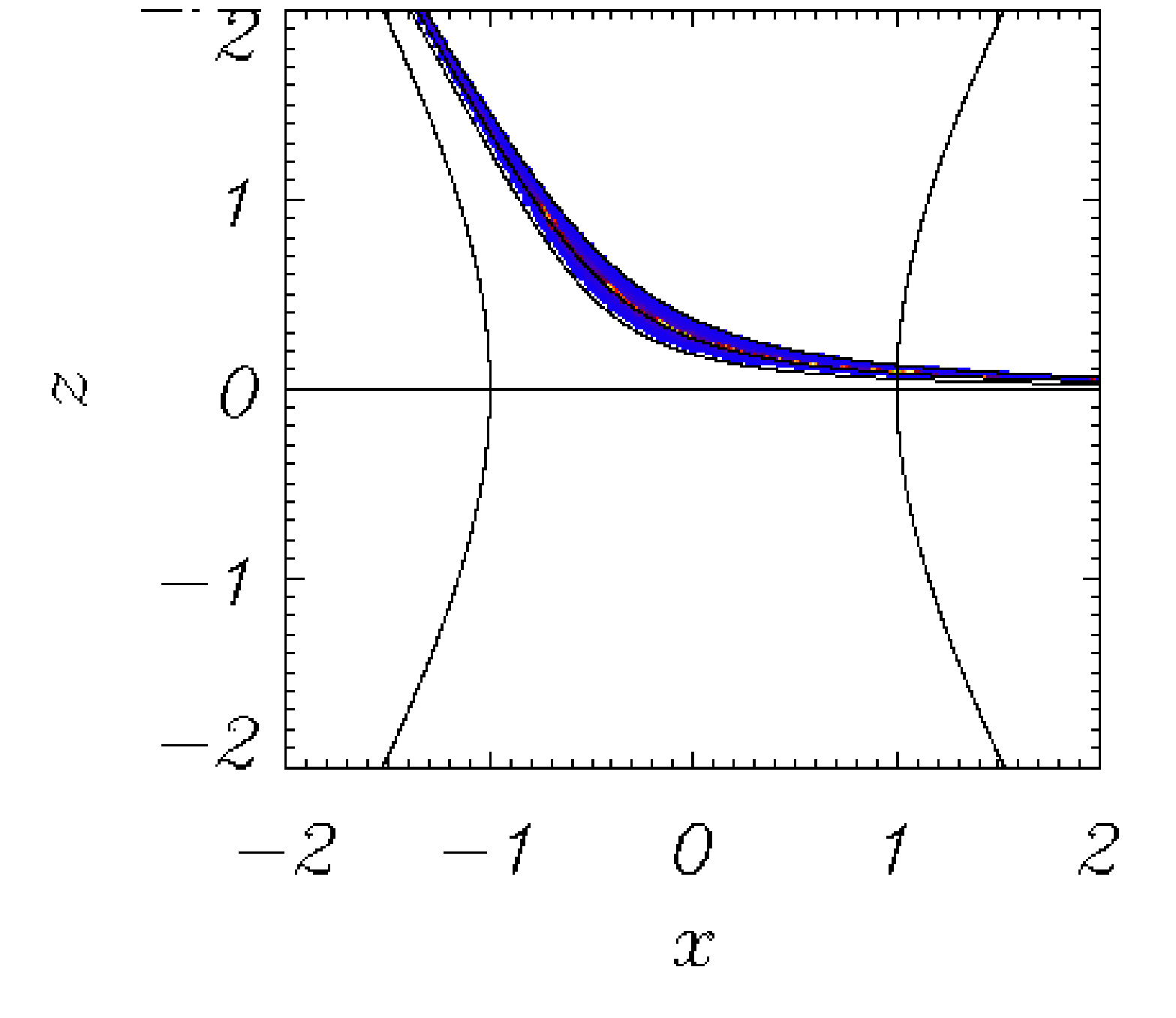}
\hspace{0.14286in}
\includegraphics[width=2.1176in]{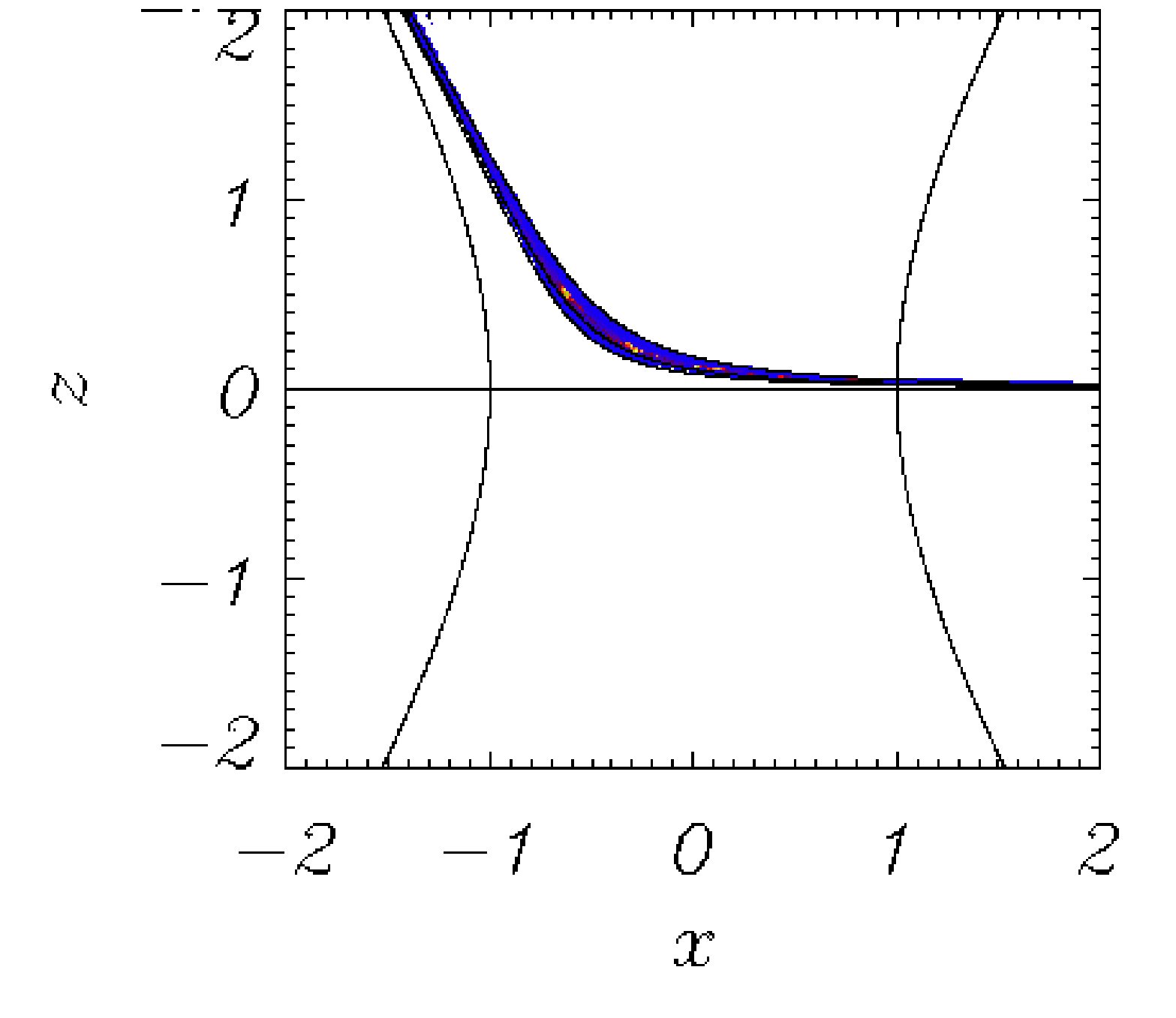}
\hspace{0.8in}
\includegraphics[width=2.05in]{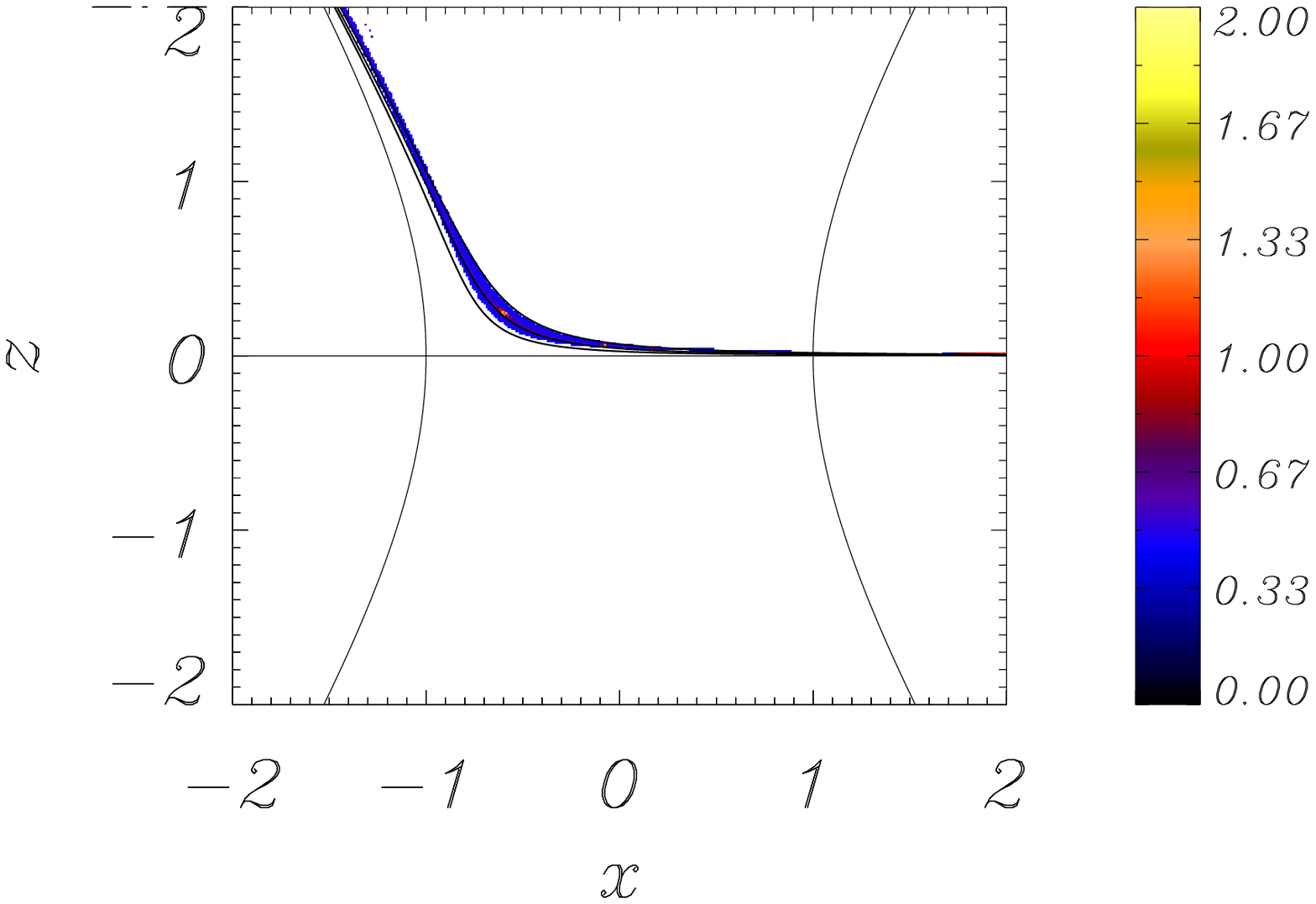}
\caption{Comparison of numerical simulation (shaded wave) and analytical (WKB) solution at times $(a)$ $t$=0.25, $(b)$ $t$=0.75, $(c)$ $t$=1.25, $(d)$ $t$=1.75, $(e)$ $t$=2.25 and $(f)$ $t$=2.75, labelling from top left to bottom right. The lines represent the front, middle and back edges of the WKB wave solution, where the pulse enters from the top of the box.}
\label{figure23}
\end{figure*}

\section{Alfv\'en Wave}\label{sec:3}

The equations describing the behaviour of the Alfv\'en Wave, equations (\ref{alfvenalpha}), were solved numerically using a two-step Lax-Wendroff scheme. \cite{McLaughlin2004} showed that in the neighbourhood of a two-dimensional X-point, the linear Alfv\'en wave spreads out along the field lines, eventually accumulating along the separatrices. Hence, for our two null points, we have \emph{four} cases to consider; an Alfv\'en wave pulse coming in from the top boundary and coming in from the side boundary for both magnetic configurations shown in Figure \ref{figureone}.

\subsection{Two null points connected by a separator}

We consider the two null point magnetic configuration containing a separator, as shown in Figure \ref{figureone} (left).

\subsubsection{top boundary}\label{sec:4.1.1}

We initially consider a box ($-2 \le x \le 2$, $-2 \le z \le 2$) with a single wave pulse coming in across part of the top boundary ($1 \le x \le 2$, $z=2$). We choose such a pulse because, as initial experiments showed, the Alfv\'en wave spreads out along the field lines as it propagates and we found that this choice of boundary condition illustrated this effect much clearer. In fact, the wave crosses the separator ($z = 0$) and so the final, high resolution run was performed in a box  ($-2.5 \le x \le 2.5$, $-0.5 \le z \le 2$). The full boundary conditions were;
\begin{eqnarray*}
\begin{array}{cl}
{v_y(x, 2) = 2\left(\sin { \omega t } \right)\left( \sin {\pi x }/{2} \right) } & {\mathrm{for} \; \; \left\{\begin{array}{c}  {1 \leq x \leq 2} \\ {0 \leq t \leq \frac {\pi}{\omega} } \end{array}\right. } \\
{\left. \frac {\partial v_y }{\partial z} \right| _{z=2} = 0 } & { \mathrm{otherwise} }
\end{array} \; ,  \\
\qquad \left. \frac {\partial v_y } {\partial x } \right| _{x=2.5} =0 \; , \qquad \left. \frac {\partial v_y } {\partial x } \right| _{x=-2.5} = 0 \; , \qquad \left. \frac {\partial v_y } {\partial z } \right| _{z=-0.5}  = 0 \; .
\end{eqnarray*}
Tests show that the central behaviour is unaffected by these choices. The other boundary conditions follow from the remaining equations and the solenodial condition, $\nabla \cdot {\mathbf{B} _1} =0$.

We found that the linear Alfv\'en wave travels down from the top boundary and begins to spread out, following the field lines. As the wave approaches the separator along $z=0$ and the separatrix that passes through $x=-1$, the wave thins but keeps its original amplitude. The wave eventually accumulates very near the separator and separatrix. This can be seen in Figure \ref{figure23}.

\subsubsection{analytical results}\label{sec:WKB}

As in all four cases, the Alfv\'en equations we have to solve are:
\begin{eqnarray}
\qquad \frac {\partial ^2 v_y } {\partial t^2 } = \left( B_x\frac {\partial }{\partial x} +B_z \frac {\partial }{\partial z } \right) ^2 v_y \; \label{alphaone},
\end{eqnarray}
In this experiment, $B_x = x^2-z^2-1$, $B_z = -2xz$ and we can apply the WKB approximation. As in Section \ref{sec:2.2}, substituting $v_y = e^{i \phi (x,z) } \cdot e^{-i \omega t}$ into (\ref{alphaone}) and assuming that $\omega \gg 1 $, leads to a first order PDE of the form $\mathcal{F} \left( x,z,\phi,\frac {\partial \phi } {\partial x}, \frac {\partial \phi } {\partial z} \right)=0$. Applying the method of characteristics, we generate the equations:
\begin{eqnarray}
\qquad \frac {d \phi }{ds} &=& - \omega ^2  \nonumber \\
\qquad \frac {dp}{ds} &=& \left( 2zq -2xp  \right) \xi \nonumber \\
\qquad \frac {dq}{ds} &=& \left( 2zp+2xq   \right) \xi \nonumber \\
\qquad \frac {dx}{ds} &=& \left( x^2-z^2-1 \right) \xi\nonumber \\
\qquad \frac {dz}{ds} &=& \left(-2xz       \right) \xi \label{C1_characteristics}
\end{eqnarray}
where $\xi = \left[ \left( x^2-z^2-1 \right) p -2xzq \right]$, $p=\frac {\partial \phi } {\partial x}$, $q=\frac {\partial \phi } {\partial z}$, $\omega$ is the frequency of our wave and $s$ is some parameter along the characteristic.

\begin{figure}[htb]
\begin{center}
\includegraphics[width=2.4in]{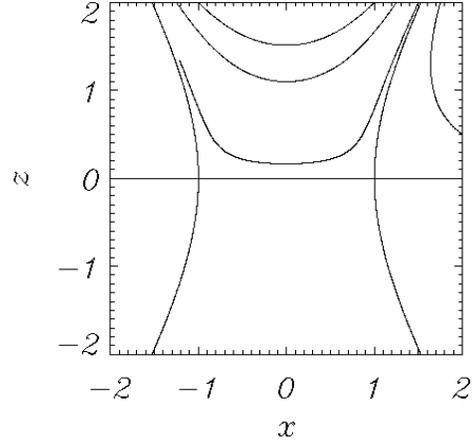}
\caption{Plots of WKB solution for a wave sent in from the upper boundary and its resultant particle paths (thick lines). Starting points of $x$=1, 1.25, 1.5, $\sqrt{7/3}$ and 1.75 are plotted.} 
\label{figure50}
\end{center}
\end{figure}

These five ODEs were solved numerically using a fourth-order Runge-Kutta method. Constant $\phi$ can be thought of as defining the position of the edge of the wave pulse, so with correct choices of $s$ the solution can be directly compared to our numerical solution (seen in Figure {\ref{figure23}} as the dark overplot lines). Figures {\ref{figure23}}, {\ref{figure17}}, {\ref{figure30}} and {\ref{figure37}} all show comparisons of the numerical simulation and corresponding WKB solutions. In each case, the agreement between the analytic model and numerical results is very good.

We can also use our WKB solution to plot the particle paths of individual elements from the initial wave. In Figure \ref{figure50}, starting points of $x$=1, 1.25, 1.5, $\sqrt{ 7 / 3 }$ (\emph{separatrix}) and 1.75 are plotted.

\subsubsection{side boundary}

We now consider a box ($-2\le x\le 2$, $-2\le z\le 2$) with a single wave pulse coming in across part of the side boundary ($x=-2$, $-1\le z\le 1$). In fact, the wave never passes the (left) separatrix and so the final, high resolution run was performed in a box ($-2 \le x \le 0.5$, $-2.5 \le z \le 2.5$). The full boundary conditions were;
\begin{eqnarray*}
\qquad v_y(-2, z) = \left\{\begin{array}{cl} {2\left(\sin { \omega t }\right) \left[\sin \frac {\pi \left( 1+z \right) }{2} \right]} & {\mathrm{for} \; \; \left\{\begin{array}{c}  {-1 \leq z \leq 1} \\ {0 \leq t \leq \frac {\pi}{\omega} } \end{array}\right. } \\
{0} & { \mathrm{otherwise} }
\end{array} \right. \; ,  \\
\left. \frac {\partial v_y } {\partial x } \right| _{x=0.5} =0 \; , \qquad \left. \frac {\partial v_y } {\partial z } \right| _{z=-2.5} = 0 \; , \qquad \left. \frac {\partial v_y } {\partial z } \right| _{z=2.5}  = 0 \; .
\end{eqnarray*}
Tests show that the central behaviour is unaffected by these choices. The other boundary conditions follow from the remaining equations and the solenodial condition, $\nabla \cdot {\mathbf{B} _1} =0$.

We found that the linear Alfv\'en wave travels in from the side boundary and begins to spread out, following the field lines. As the wave approaches the (left) separatix (that passes through $x=-1$), the wave thins (but keeps its original amplitude). The wave eventually accumulates very near this separatrix. This can be seen in Figure \ref{figure17}.

\subsubsection{analytical results}

\begin{figure*}[th]
\includegraphics[width=2.0in]{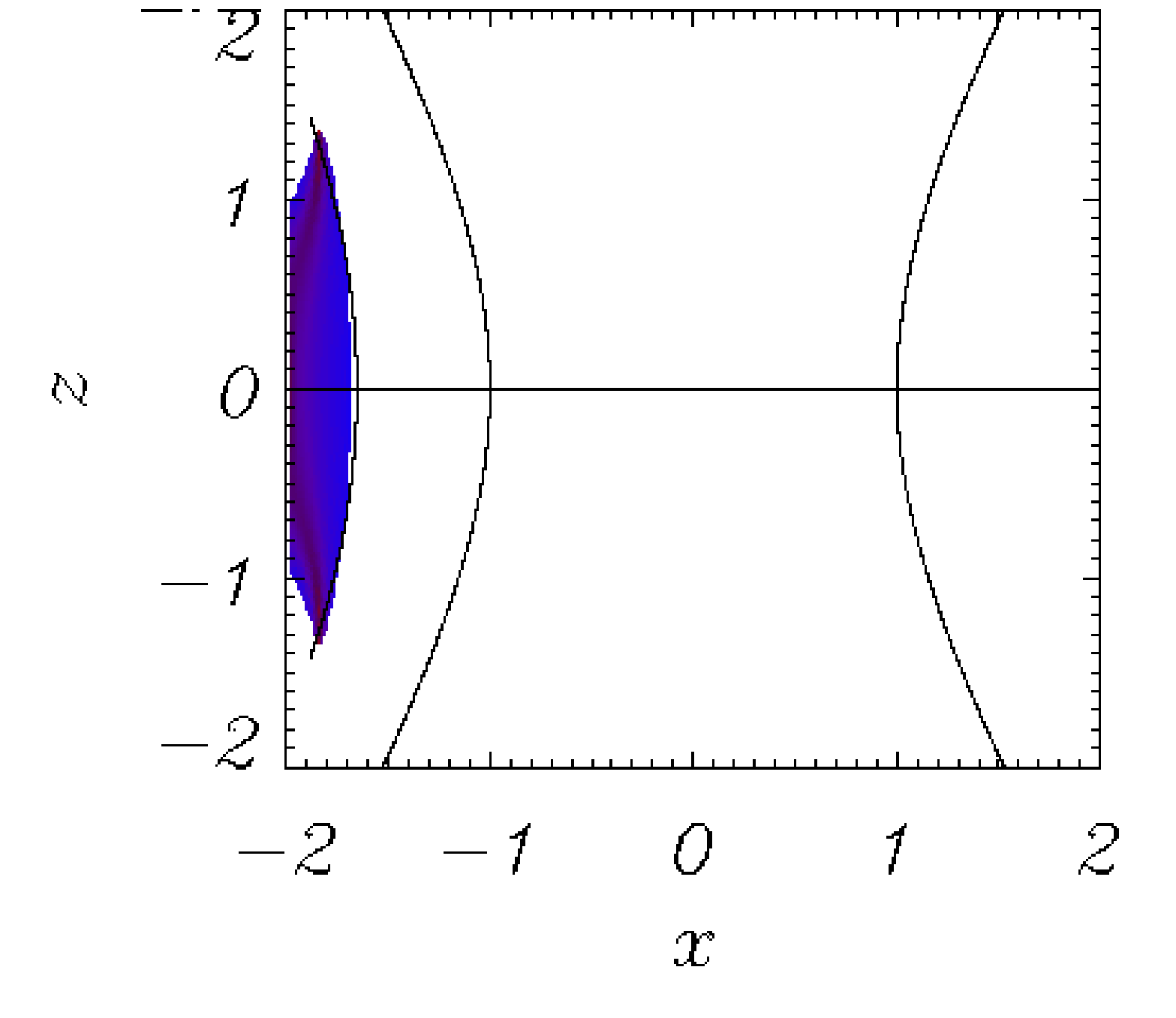}
\hspace{0.15in}
\includegraphics[width=2.0in]{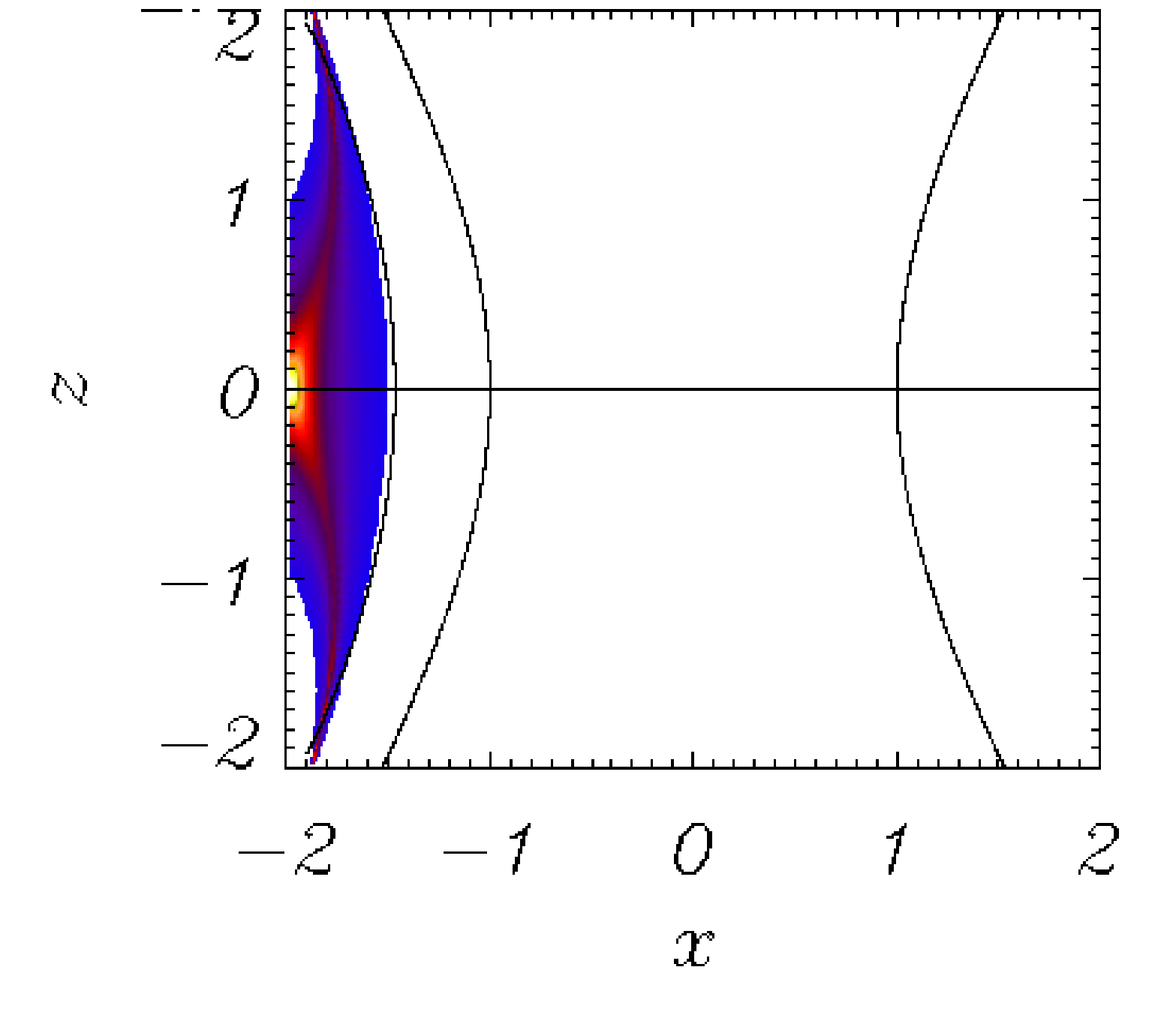}
\hspace{0.15in}
\includegraphics[width=2.0in]{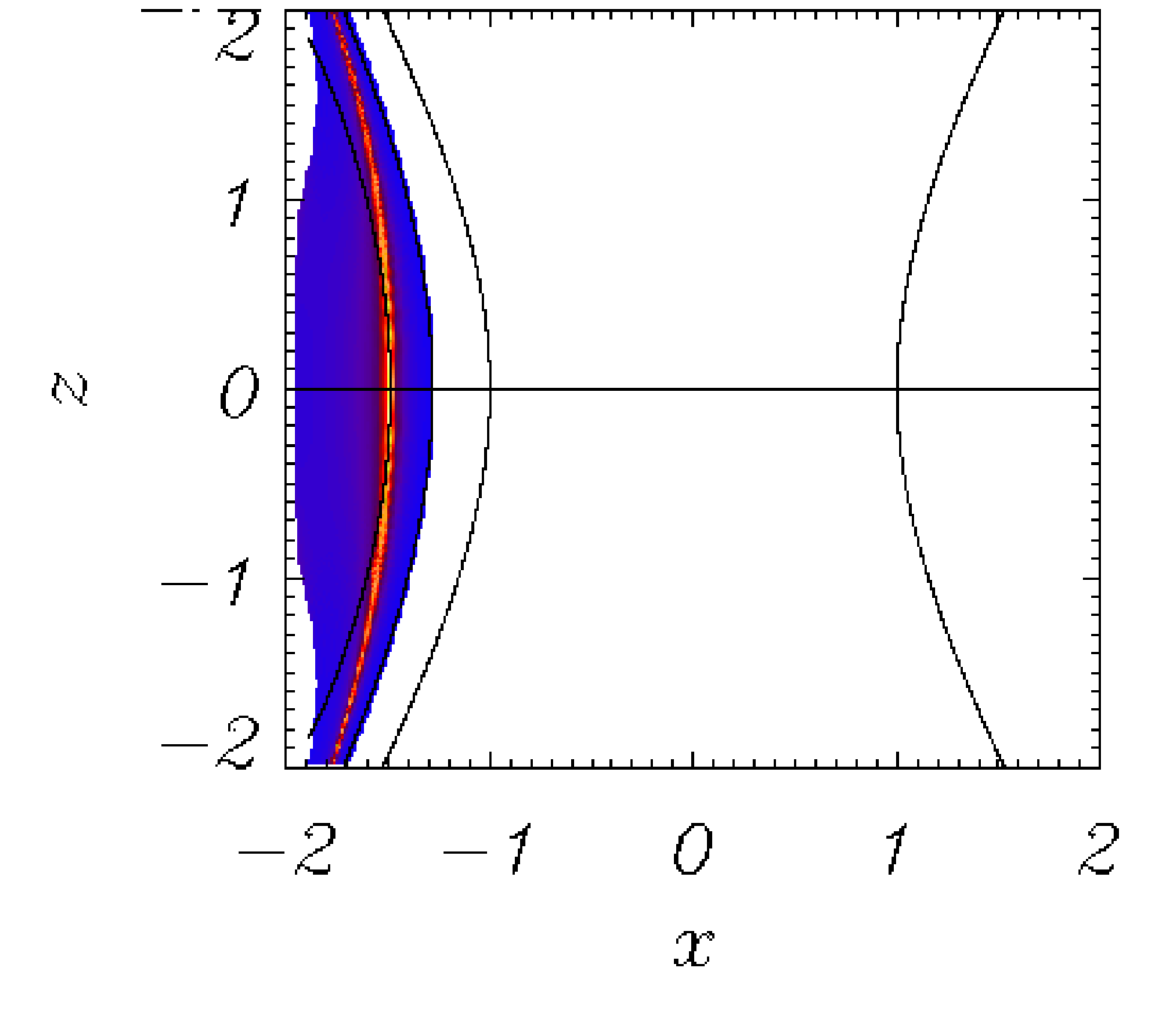}\\
\includegraphics[width=2.0in]{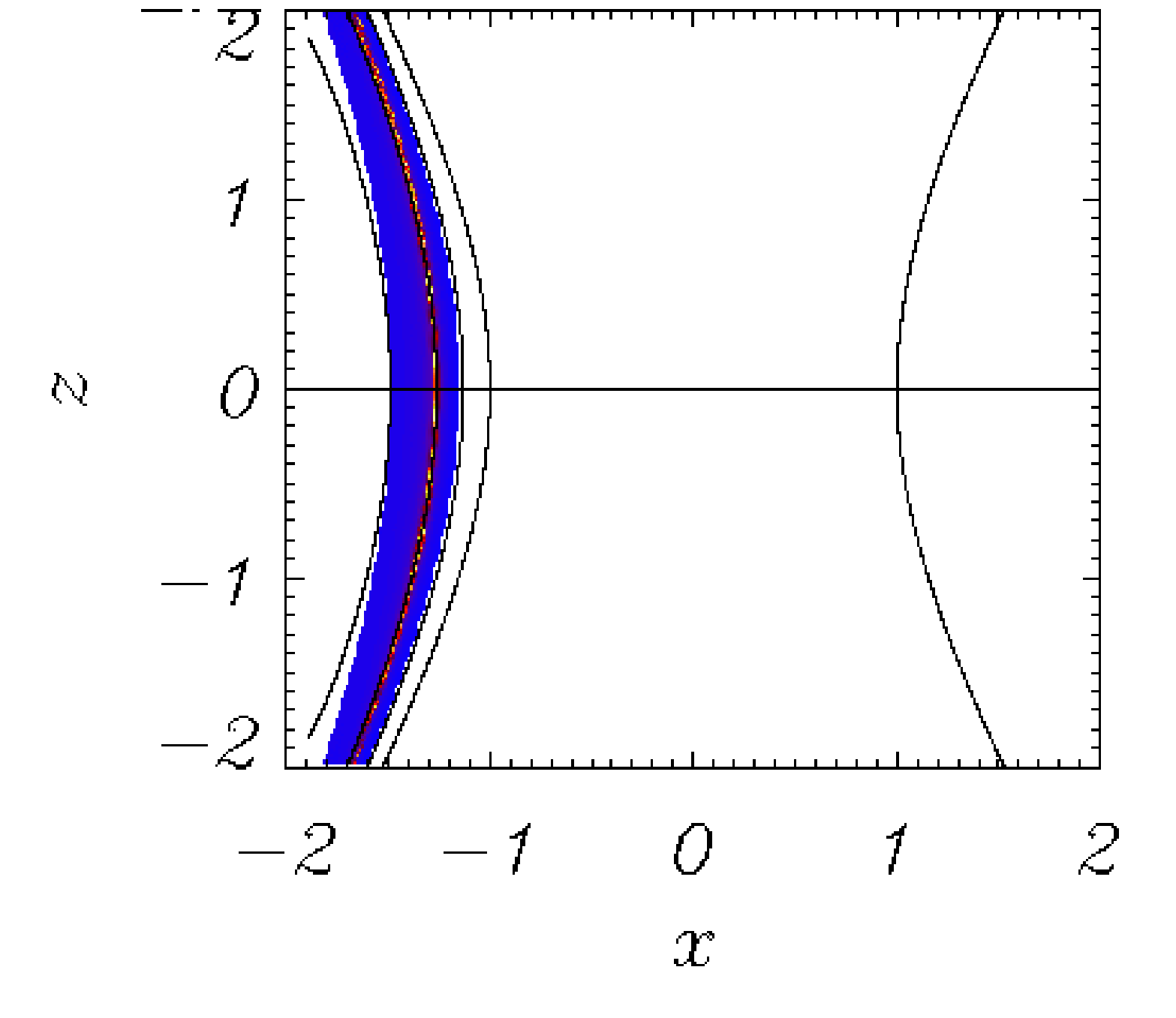}
\hspace{0.15in}
\includegraphics[width=2.0in]{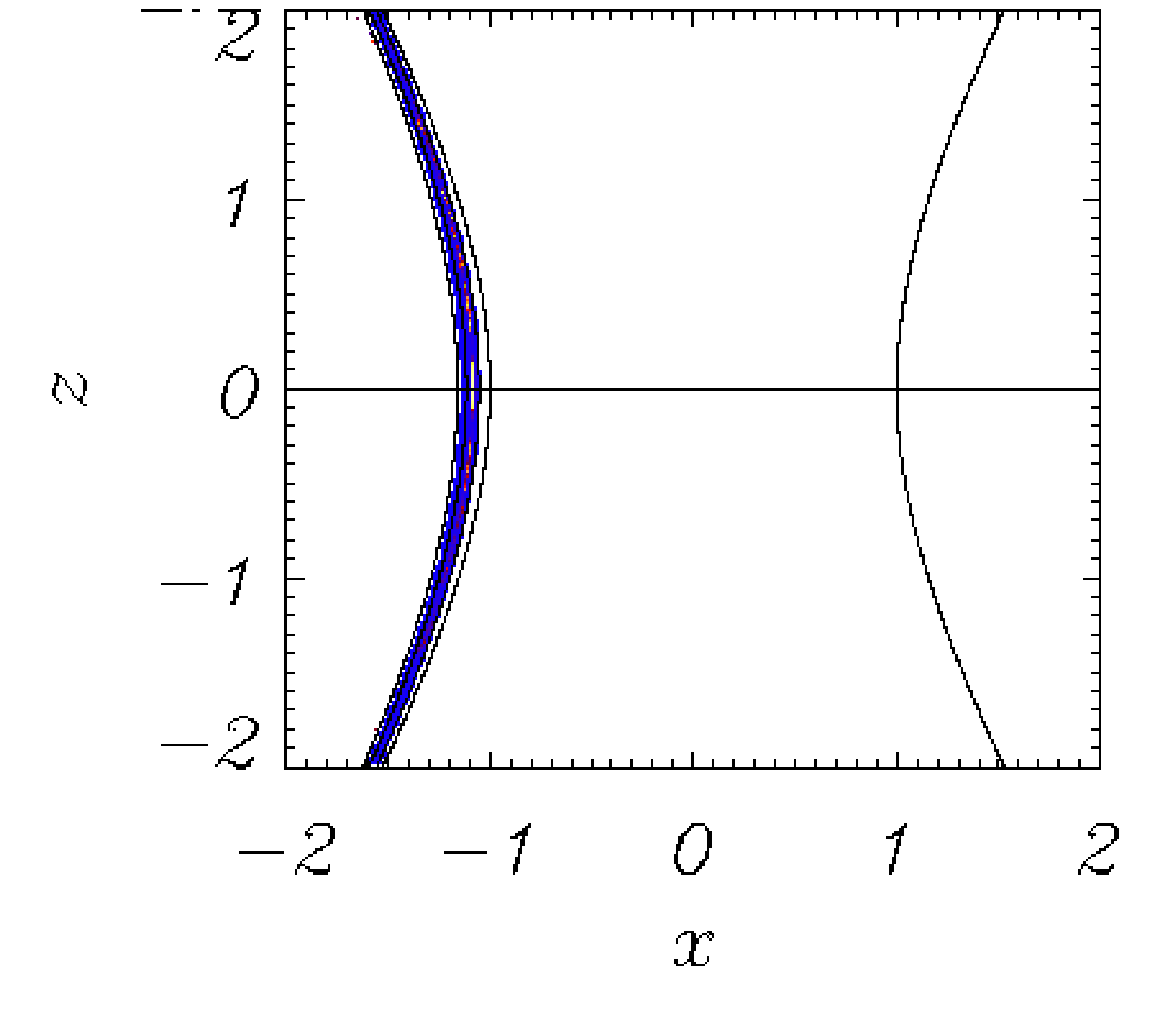}
\hspace{0.7666in}
\includegraphics[width=1.93in]{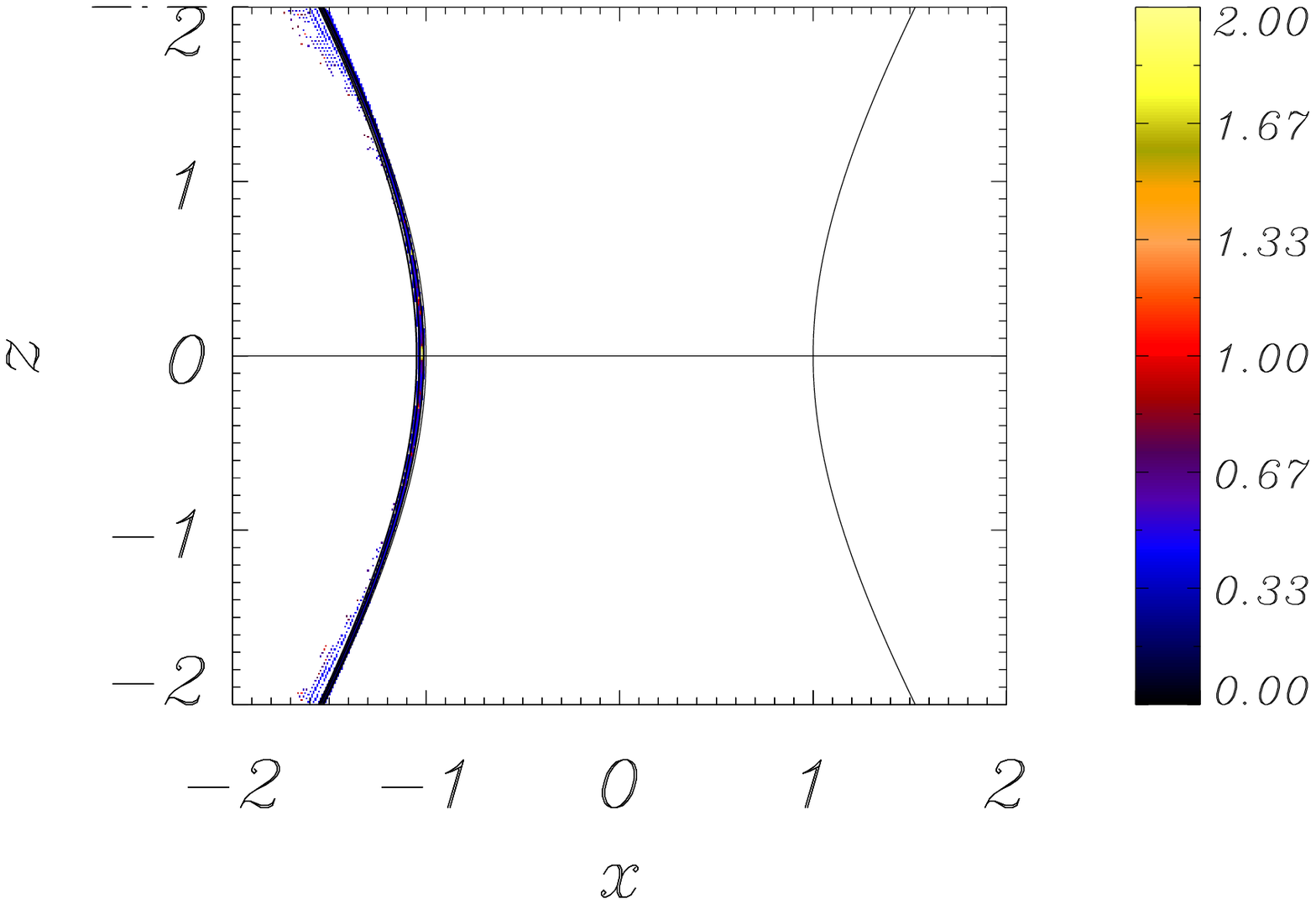}
\caption{Comparison of numerical simulation (shaded) and analytical solution at times $(a)$ $t$=0.14, $(b)$ $t$=0.25, $(c)$ $t=$0.5, $(d)$ $t$=0.75,  $(e)$ $t$=1.25 and $(f)$ $t$=1.75, labelling from top left to bottom right. The lines represent the front, middle and back edges of the wave, where the pulse enters from the side of the box}
\label{figure17}
\end{figure*}

By solving the same characteristic equations, equations \ref{C1_characteristics}, but with different choices of initial condition (i.e. modelling the wave coming in from the side), we can obtain an analytical result to match our numerical simulation. Again, we use a fourth-order Runge-Kutta method. The comparison of the numerical simulation and our WKB solution can be seen in Figure {\ref{figure17}}.

\subsection{Two null points not connected by a separator}

We now consider the two null point magnetic configuration that does not contain a separator, as shown in Figure \ref{figureone} (right).

\subsubsection{top boundary}

We consider a box ($-2 \le x \le 2$, $-2 \le z \le 2$) with a single wave pulse coming in across part of the top boundary ($-1 \le x \le 1$, $z=2$), i.e. crossing the separatrices. The final, high resolution run was performed in a box  ($-2.5 \le x \le 2.5$, $-3 \le z \le 2$). The full boundary conditions were;
\begin{eqnarray*}
\qquad v_y(x, 2) = \left\{\begin{array}{cl} {2\left(\sin { \omega t }\right)\left[ \sin \frac {\pi \left( 1+x \right) }{2} \right]} & {\mathrm{for} \; \; \left\{\begin{array}{c}  {-1 \leq x \leq 1} \\ {0 \leq t \leq \frac {\pi}{\omega} } \end{array}\right. } \\
{0} & { \mathrm{otherwise} }
\end{array} \right. \; ,  \\
\left. \frac {\partial v_y } {\partial x } \right| _{x=2.5} =0 \; , \qquad \left. \frac {\partial v_y } {\partial x }  \right| _{x=-2.5} = 0 \; , \qquad \left. \frac {\partial v_y } {\partial z }  \right| _{z=-3}  = 0 \; .
\end{eqnarray*}
The other boundary conditions are chosen in the same way as before.

We found that the linear Alfv\'en wave travels down from the top boundary and begins to spread out, following the field lines. The wave thins as it descends, but keeps its original amplitude. The wave takes the shape of the separatrices that pass through $x=\pm{2}$, $z=+\sqrt{2/3}$. The wave accumulates along these separatrices for $z > 0$. The rest of the wave (the part between $ -1 < x < +1 $) continues descending and eventually accumulates along the separatrices in the lower half of the box.  This can be seen in Figure \ref{figure30}. The part of wave in the small region about $x=0$ descends indefinitely, since the separatrices converge along $x=0$ at infinity. This can be shown:
\begin{eqnarray*}
\qquad \frac {\partial ^2 v_y } {\partial t^2 } = \left( B_x\frac {\partial }{\partial x} +B_z \frac {\partial }{\partial z } \right) ^2 v_y = \left(B_z \frac {\partial }{\partial z } \right) ^2 v_y | _{x=0}\; ,
\end{eqnarray*}
Let $\frac {\partial }{\partial s} =  \left( B_z \frac {\partial }{\partial z } \right) $ and comparing the original equation with $\frac {\partial v_y}{\partial s} =   \frac {\partial x}{\partial s} \frac {\partial v_y}{\partial x} +  \frac {\partial z}{\partial s} \frac {\partial v_y}{\partial z }$ leads to:
\begin{eqnarray*}
\quad z = \tan{(A-t)} | _{x=0} \; ,
\end{eqnarray*}
where $A=\arctan{z_0}$ and $z_0$ is the starting postion of our characteristic. Thus $z$ starts at $x=0$, $z=2$ and descends, following the line $x=0$ but becoming closer to zero. The particle passes through $z=0$ at $t=$ $arctan{2}$ and becomes more and more negative (i.e. continues to descend) thereafter.

\subsubsection{analytical results}

\begin{figure*}[t]
\includegraphics[width=2.0in]{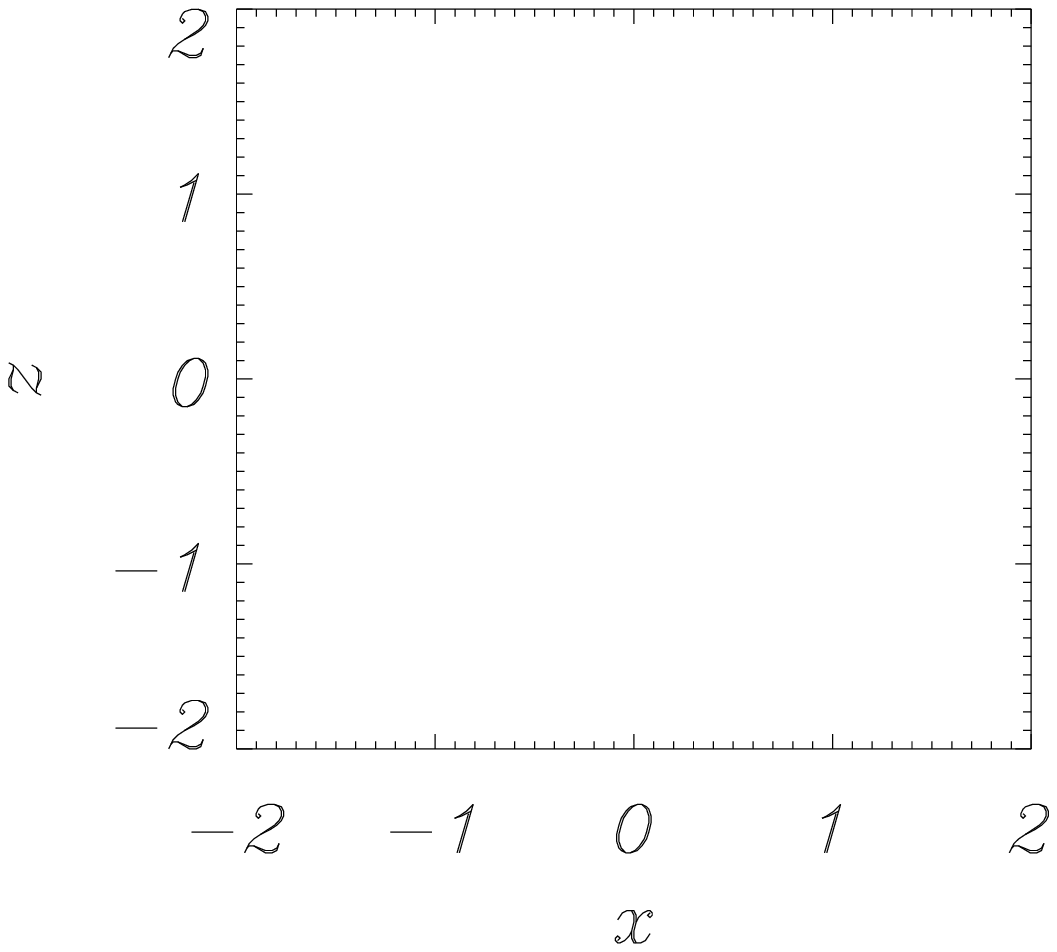}
\hspace{0.15in}
\includegraphics[width=2.0in]{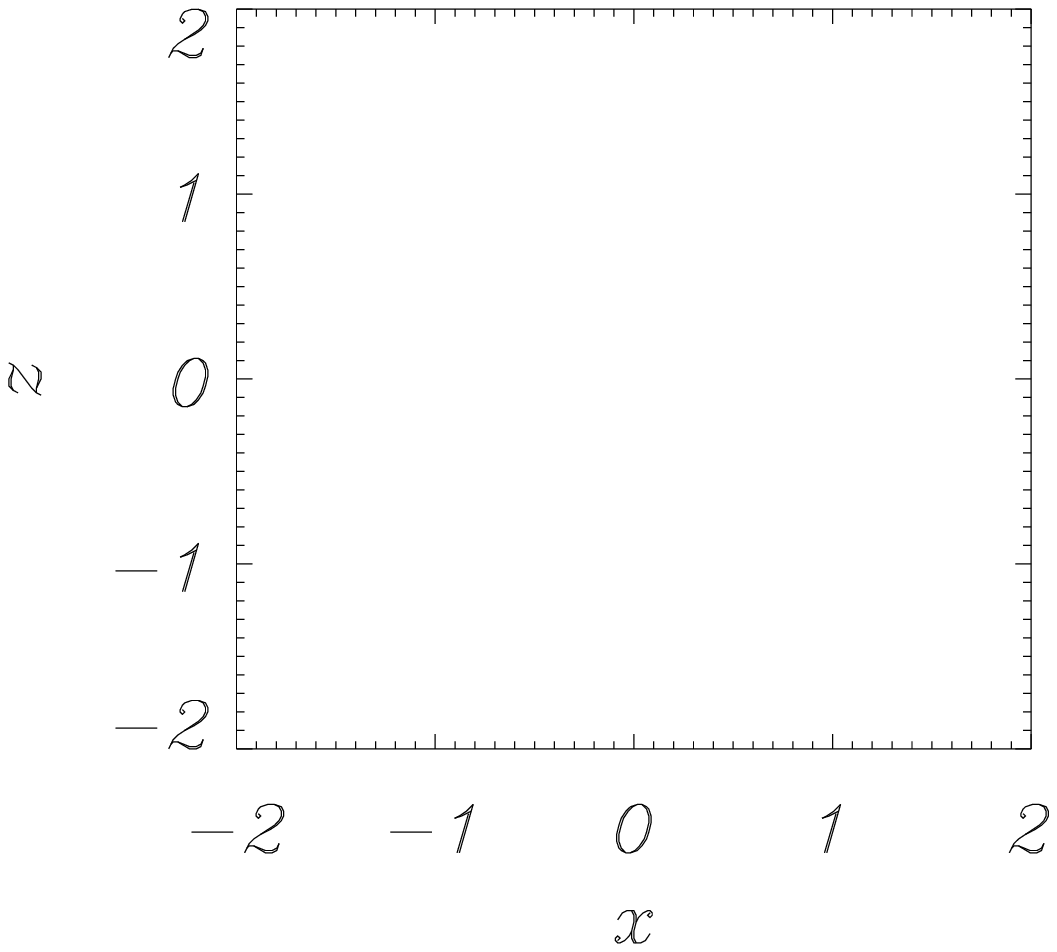}
\hspace{0.15in}
\includegraphics[width=2.0in]{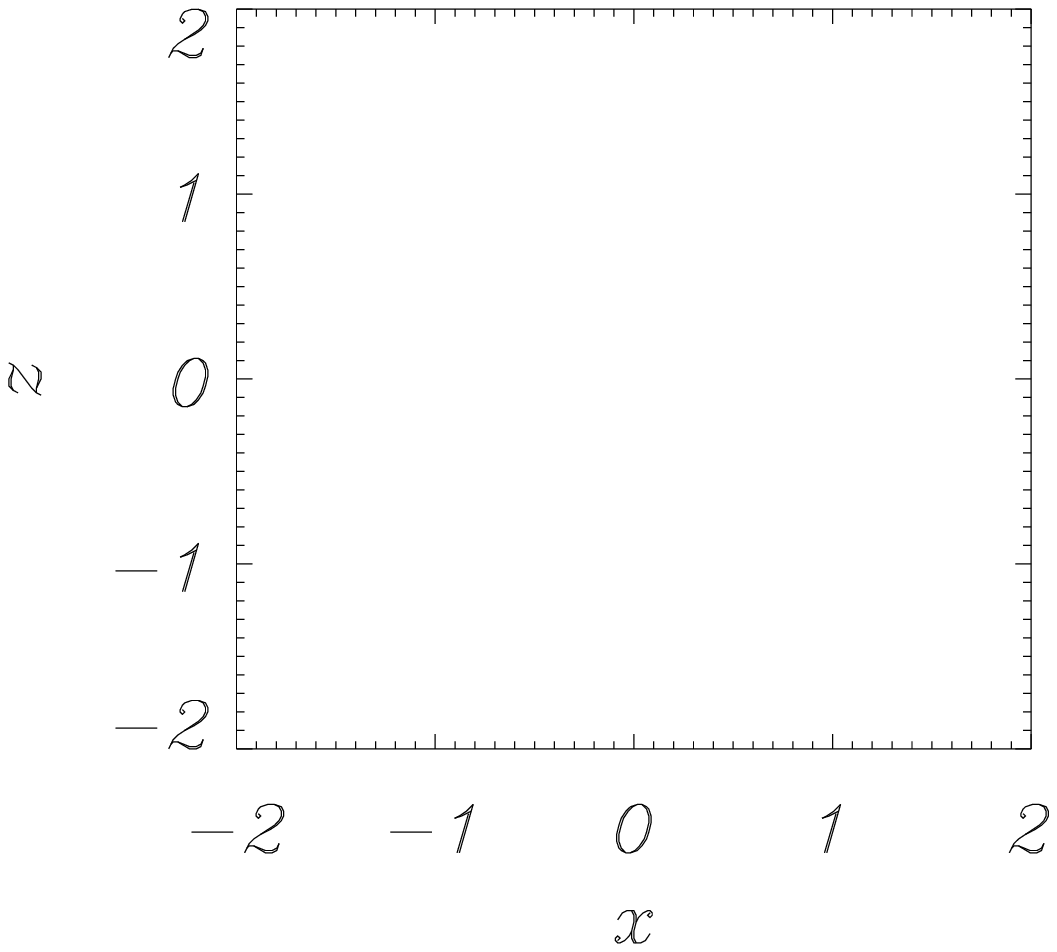}\\
\includegraphics[width=2.0in]{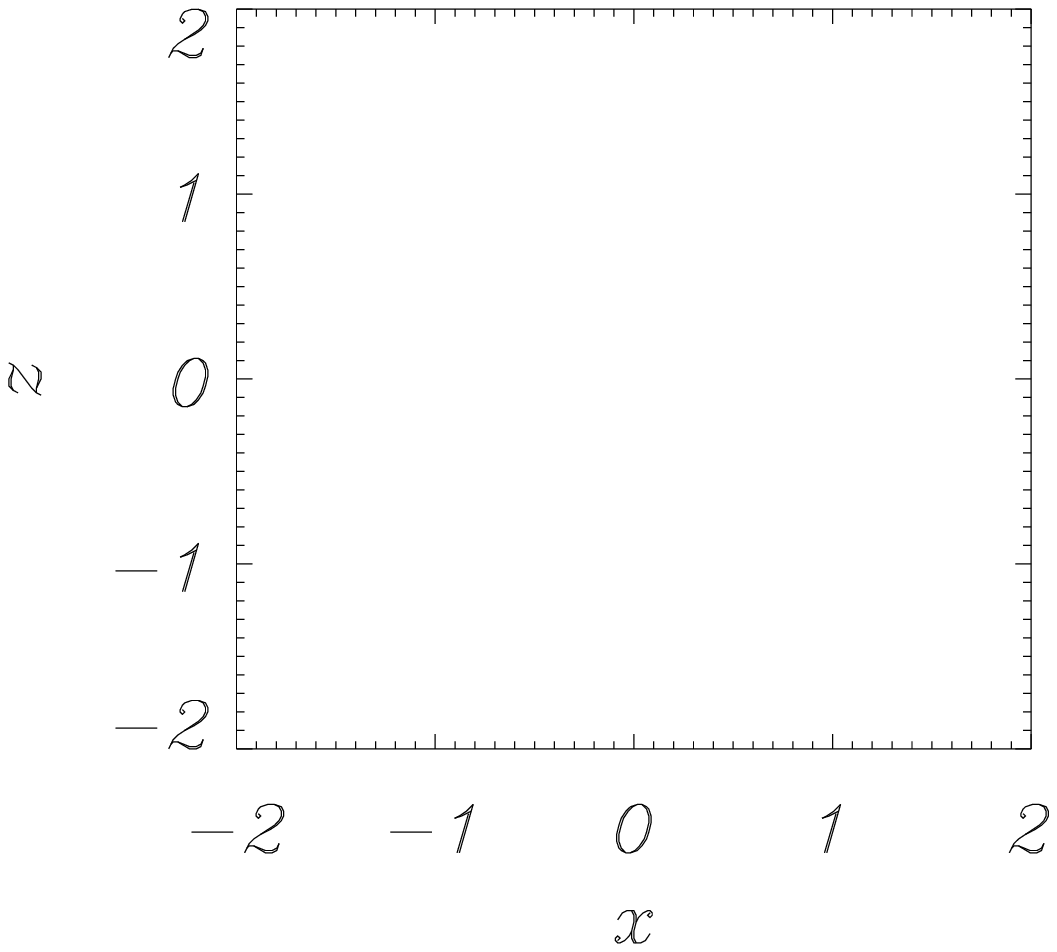}
\hspace{0.15in}
\includegraphics[width=2.0in]{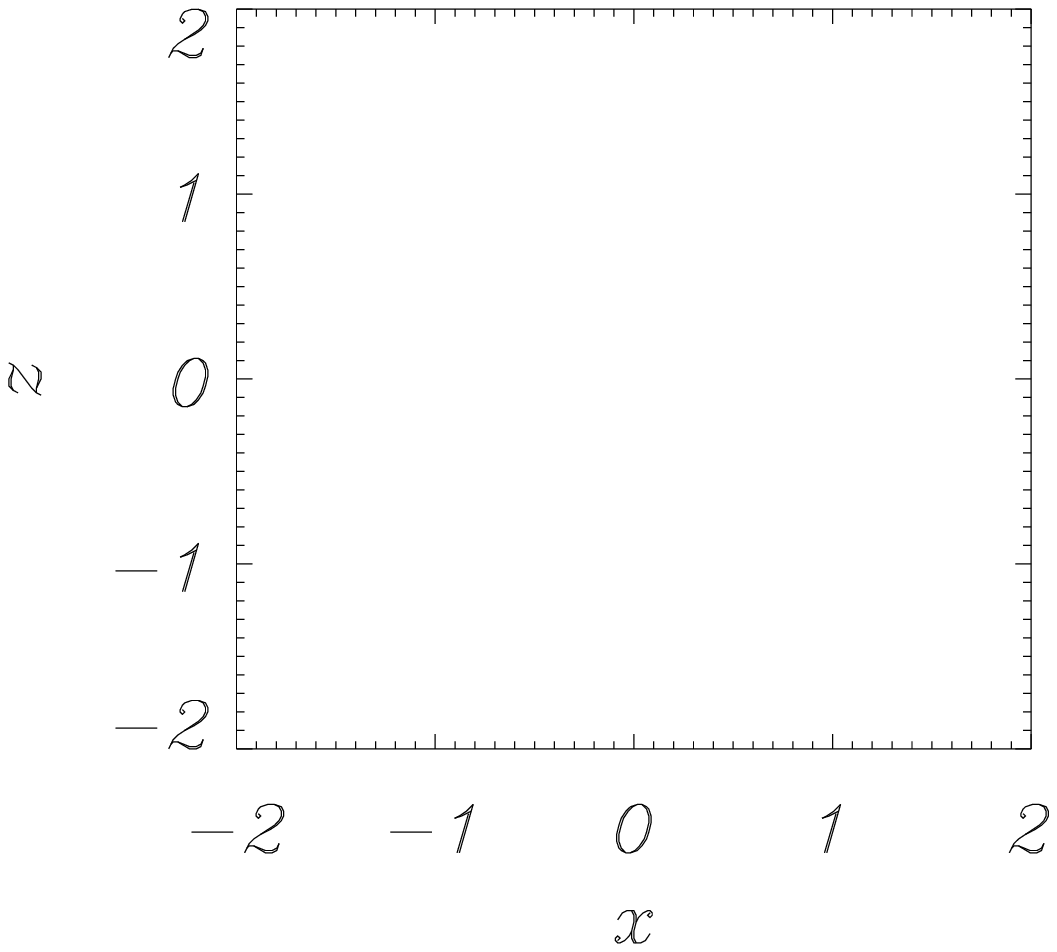}
\hspace{0.7666in}
\includegraphics[width=1.93in]{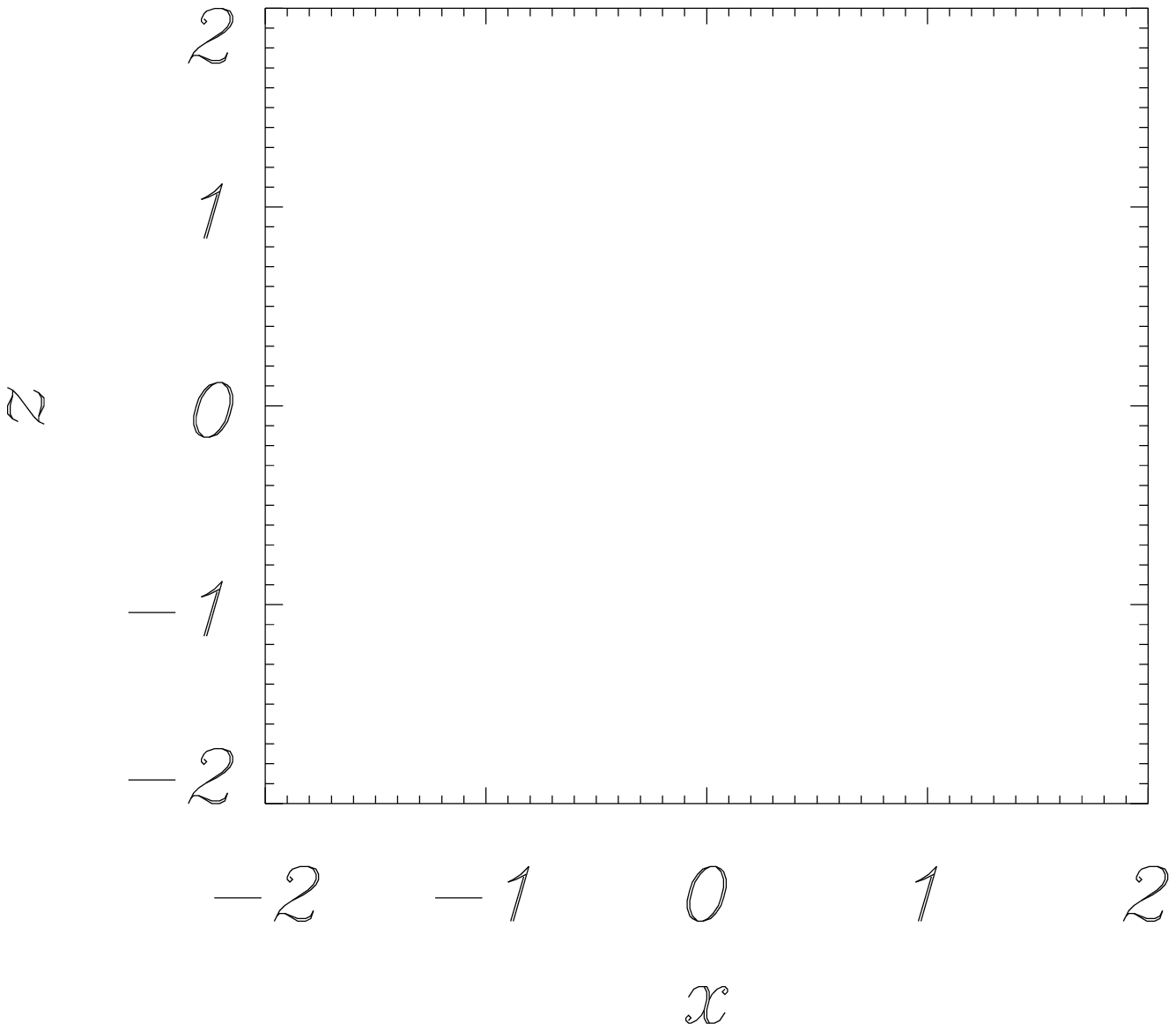}
\caption{Comparison of numerical simulation (shaded area) and analytical solution at times $(a)$ $t$=0.25, $(c)$ $t$=0.75, $(e)$ $t$=1.25, $(g)$ $t$=1.75, $(i)$ $t$=2.25 and $(k)$ $t$=2.75, labelling from top left to bottom right. The lines represent the front, middle and back edges of the wave, where the pulse enters from the top of the box.}
\label{figure30}
\end{figure*}

Again, we have to solve equations (\ref{alphaone}) to gain an analytical solution for our Alfv\'en wave. Here, $B_x = 2xz$ and $B_z = x^2-z^2-1$. We substitute $v_y = e^{i \phi (x,z) } \cdot e^{-i \omega t}$ into (\ref{alphaone}) and assume that $\omega \gg 1 $ (WKB approximation). This leads to a first order PDE of the form $\mathcal{F} \left( x,z,\phi,\frac {\partial \phi } {\partial x}, \frac {\partial \phi } {\partial z} \right)=0$. Applying the method of characteristics to this magnetic setup, we generate the equations:
\begin{eqnarray}
\qquad \frac {d \phi }{ds} &=& - \omega ^2  \nonumber \\
\qquad \frac {dp}{ds} &=& -\left( 2zp + 2xq  \right) \zeta \nonumber \\
\qquad \frac {dq}{ds} &=& -\left( 2xp - 2zq   \right) \zeta \nonumber \\
\qquad \frac {dx}{ds} &=& \left( 2xz \right) \zeta \nonumber \\
\qquad \frac {dz}{ds} &=& \left(   x^2-z^2-1   \right) \zeta \label{C2_characteristics}
\end{eqnarray}
where $\zeta = \left[ 2xzp + \left( x^2-z^2-1 \right) q \right]$, $p=\frac {\partial \phi } {\partial x}$, $q=\frac {\partial \phi } {\partial z}$, $\omega$ is the frequency of our wave and $s$ is some parameter along the characteristic. 

These are solved using a fourth-order Runge-Kutta method and the results can be seen in Figure {\ref{figure30}}.

\subsubsection{side boundary}

We now consider a box ($-2\le x\le 2$, $-2\le z\le 2$) with a single wave pulse coming in across the side boundary ($x=-2$, $0\le z\le 1$). The final, high resolution run was performed in a box  ($-2 \le x \le 0.5$, $-2.5 \le z \le 2.5$). The full boundary conditions were;
\begin{eqnarray*}
v_y(-2,z) = \left\{\begin{array}{cl} {2\sin { \omega t } \sin \frac {\pi \left( z+1 \right) }{2} } & {\mathrm{for} \; \; \left\{\begin{array}{c}  {-1 \leq z \leq 1} \\ {0 \leq t \leq \frac {\pi}{\omega} } \end{array}\right. } \\
{0} & { \mathrm{otherwise} }
\end{array} \right. \; ,  \\
\left. \frac {\partial v_y } {\partial x } \right| _{x=0.5} =0 \; , \qquad \left. \frac {\partial v_y } {\partial z }  \right| _{z=-2.5} = 0 \; , \qquad \left. \frac {\partial v_y } {\partial z }  \right| _{z=2.5}  = 0 \; .
\end{eqnarray*}

We found that the linear Alfv\'en wave travels in from the side boundary and begins to spread out, following the field lines. As the wave approaches the separatix (that passes through $x=-1$, $z=\sqrt{2/3}$), the wave thins (but keeps its original amplitude). The wave eventually accumulates very near this separatrix. This can be seen in Figure \ref{figure37}.

\subsubsection{analytical results}

\begin{figure*}[t]
\includegraphics[width=2.0in]{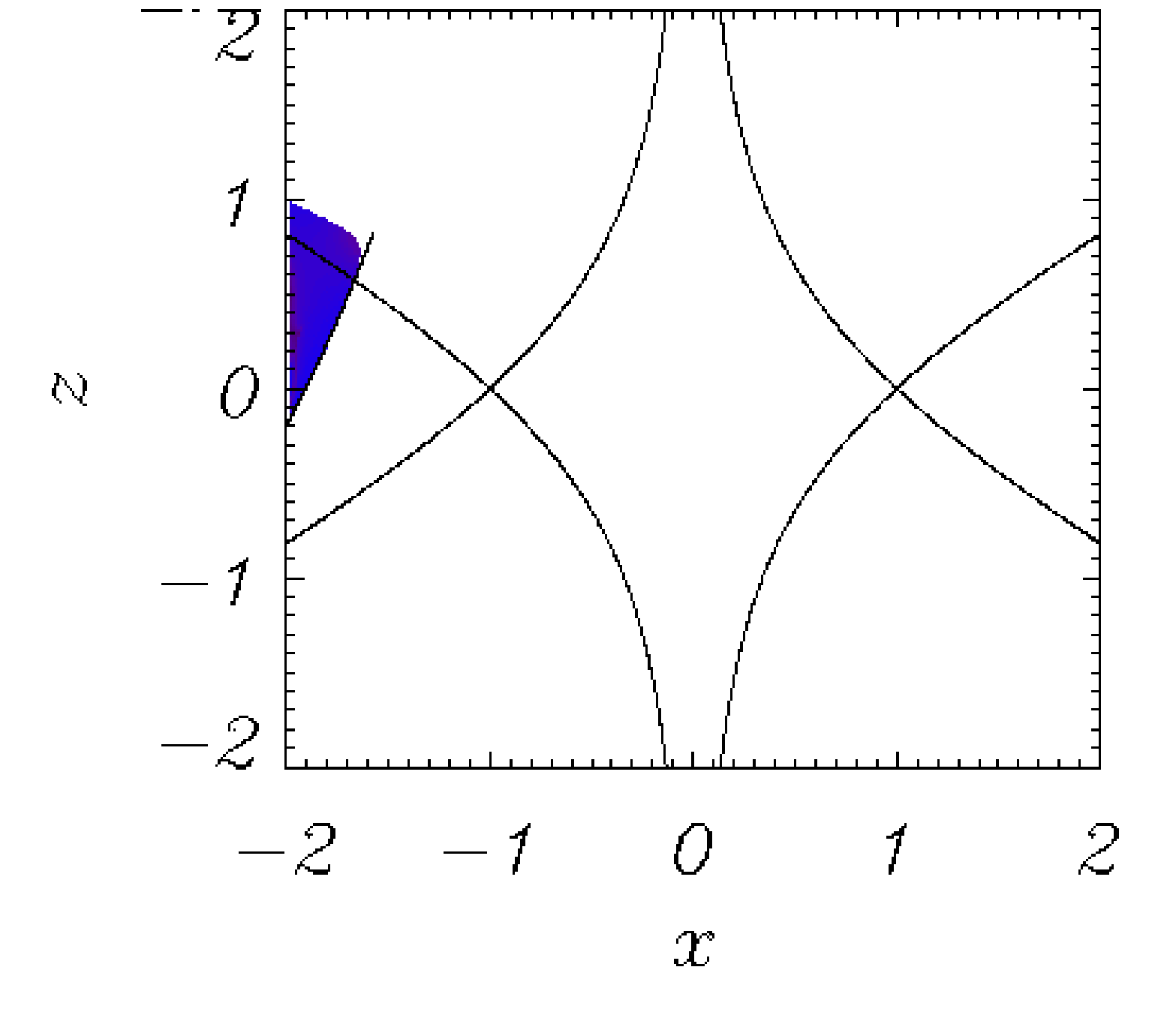}
\hspace{0.15in}
\includegraphics[width=2.0in]{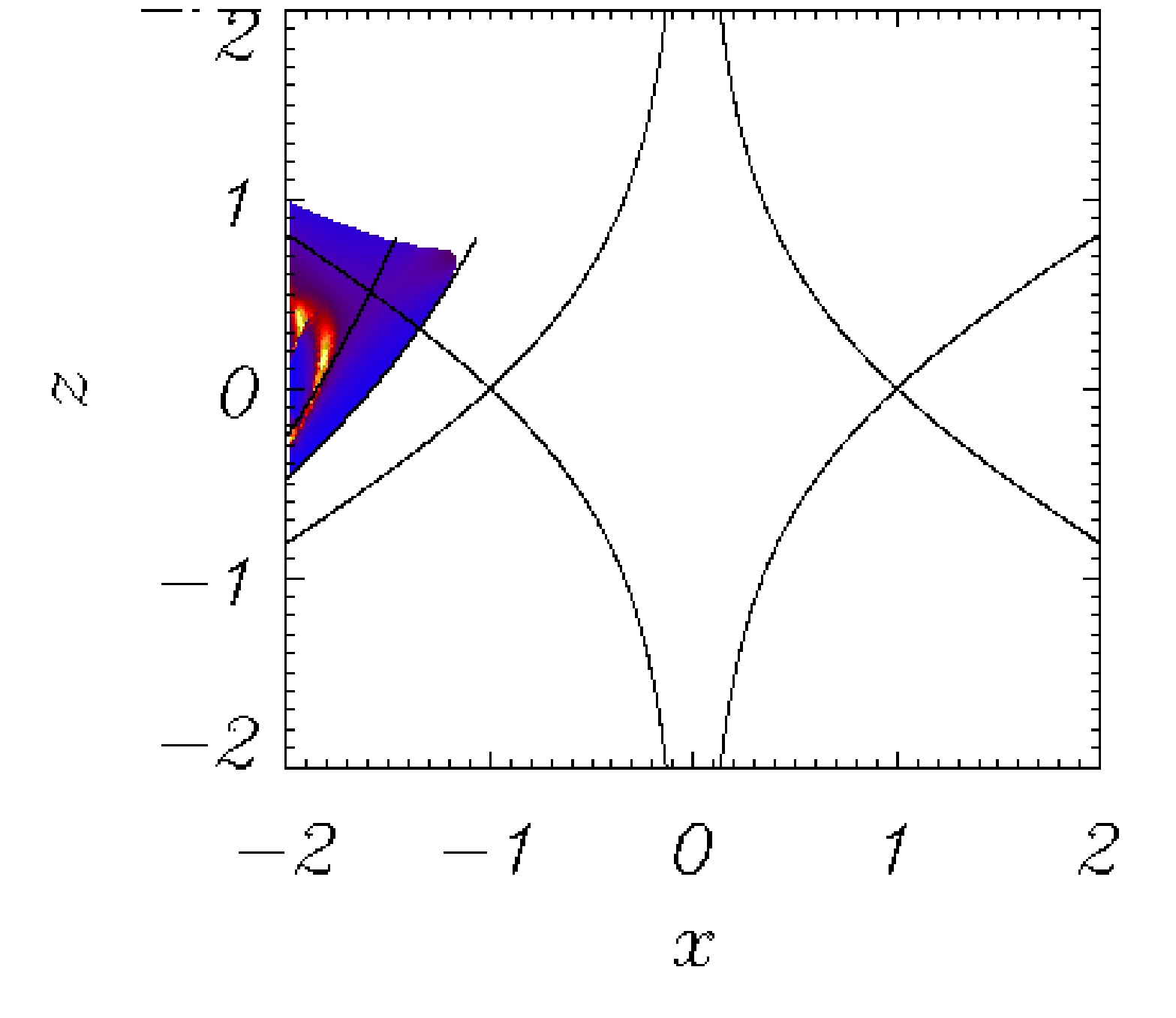}
\hspace{0.15in}
\includegraphics[width=2.0in]{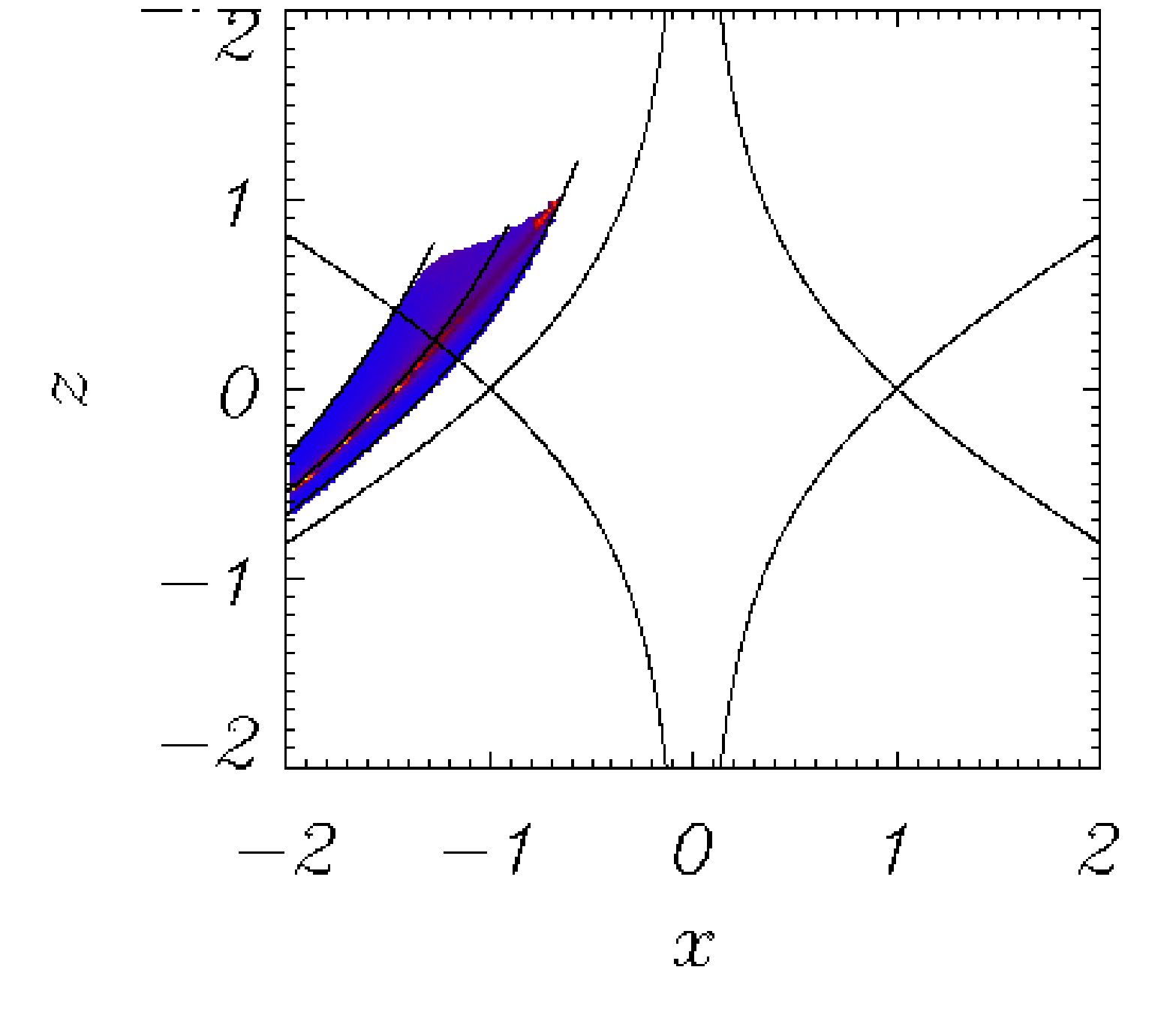}\\
\includegraphics[width=2.0in]{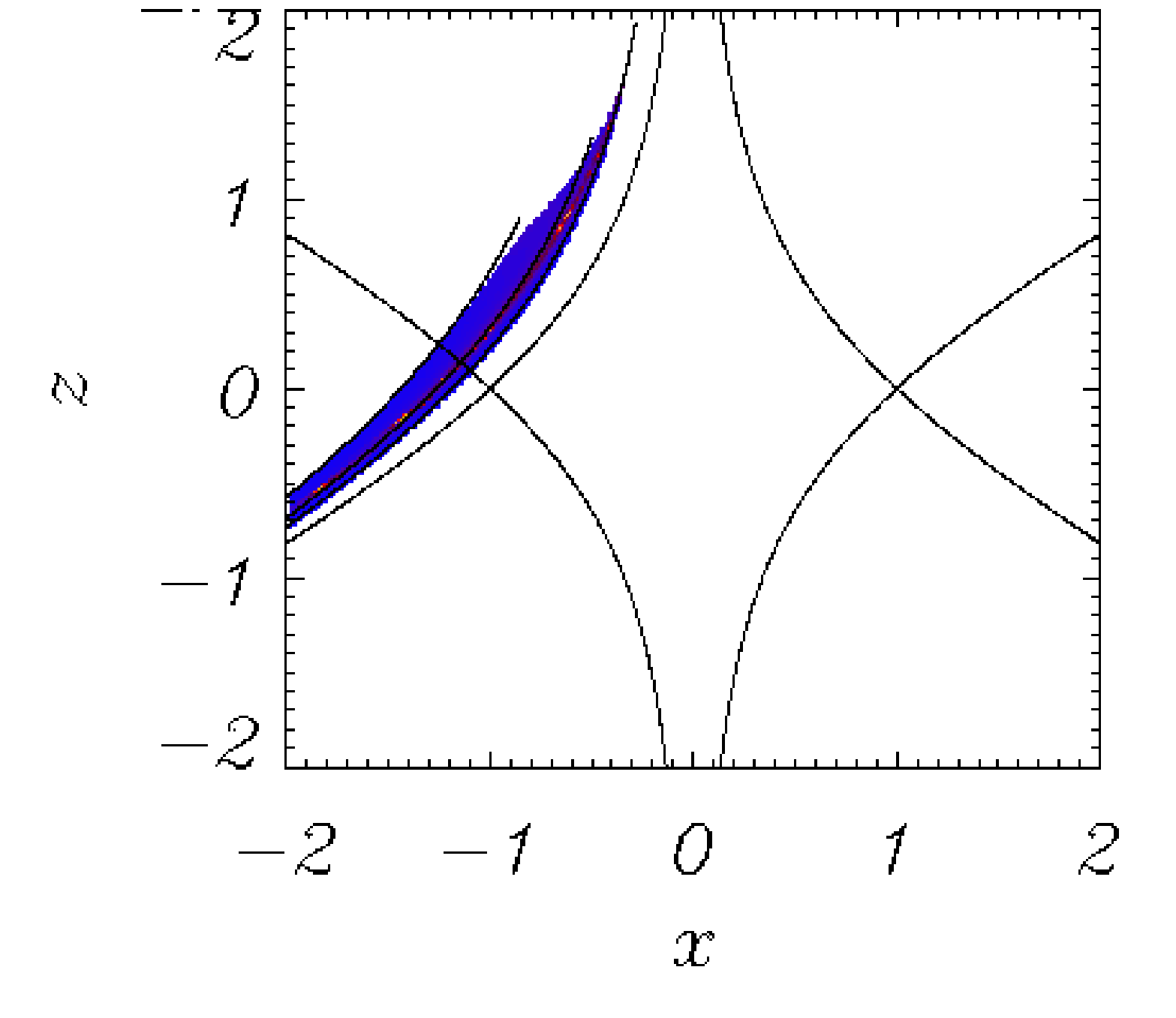}
\hspace{0.15in}
\includegraphics[width=2.0in]{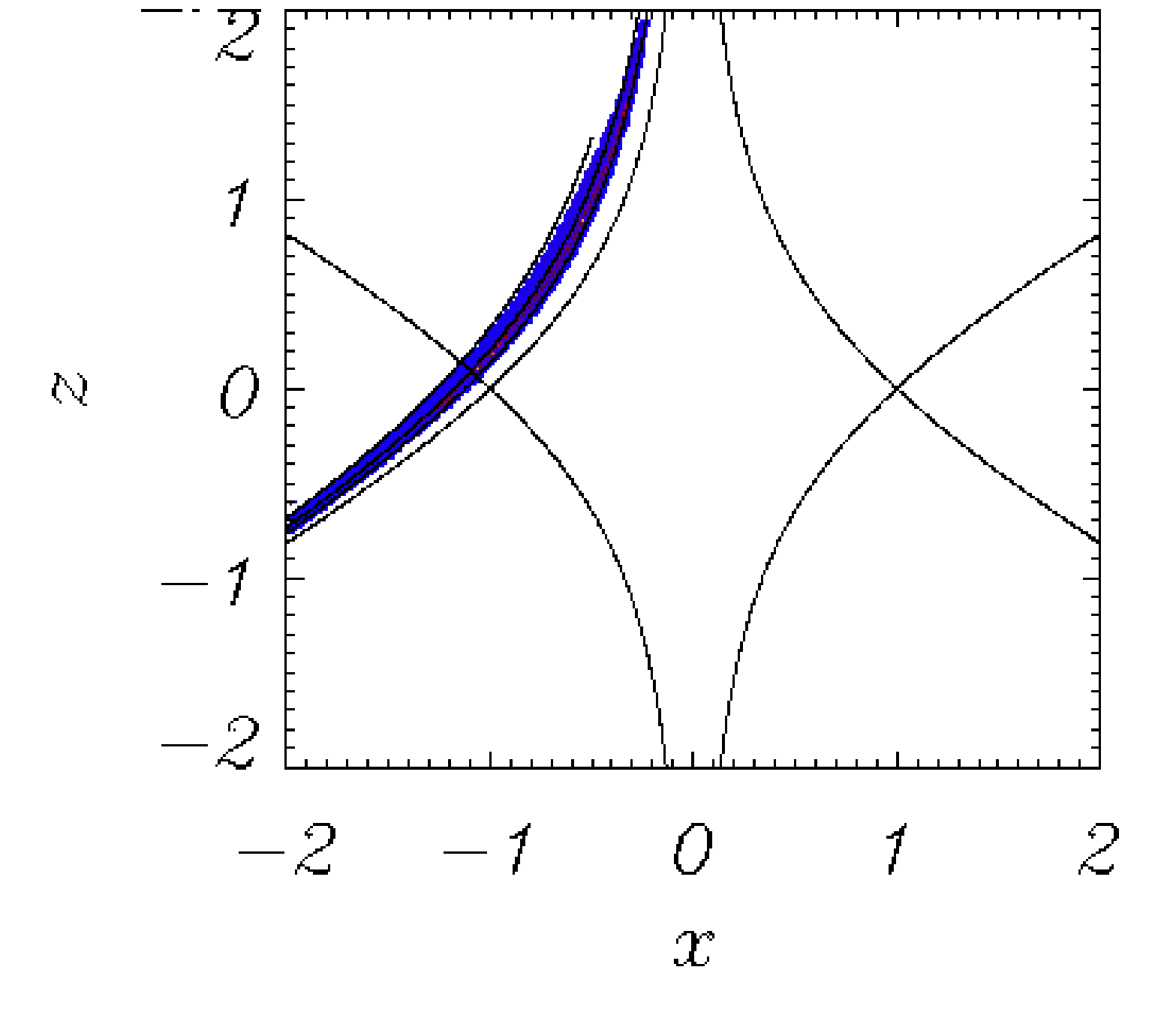}
\hspace{0.7666in}
\includegraphics[width=1.93in]{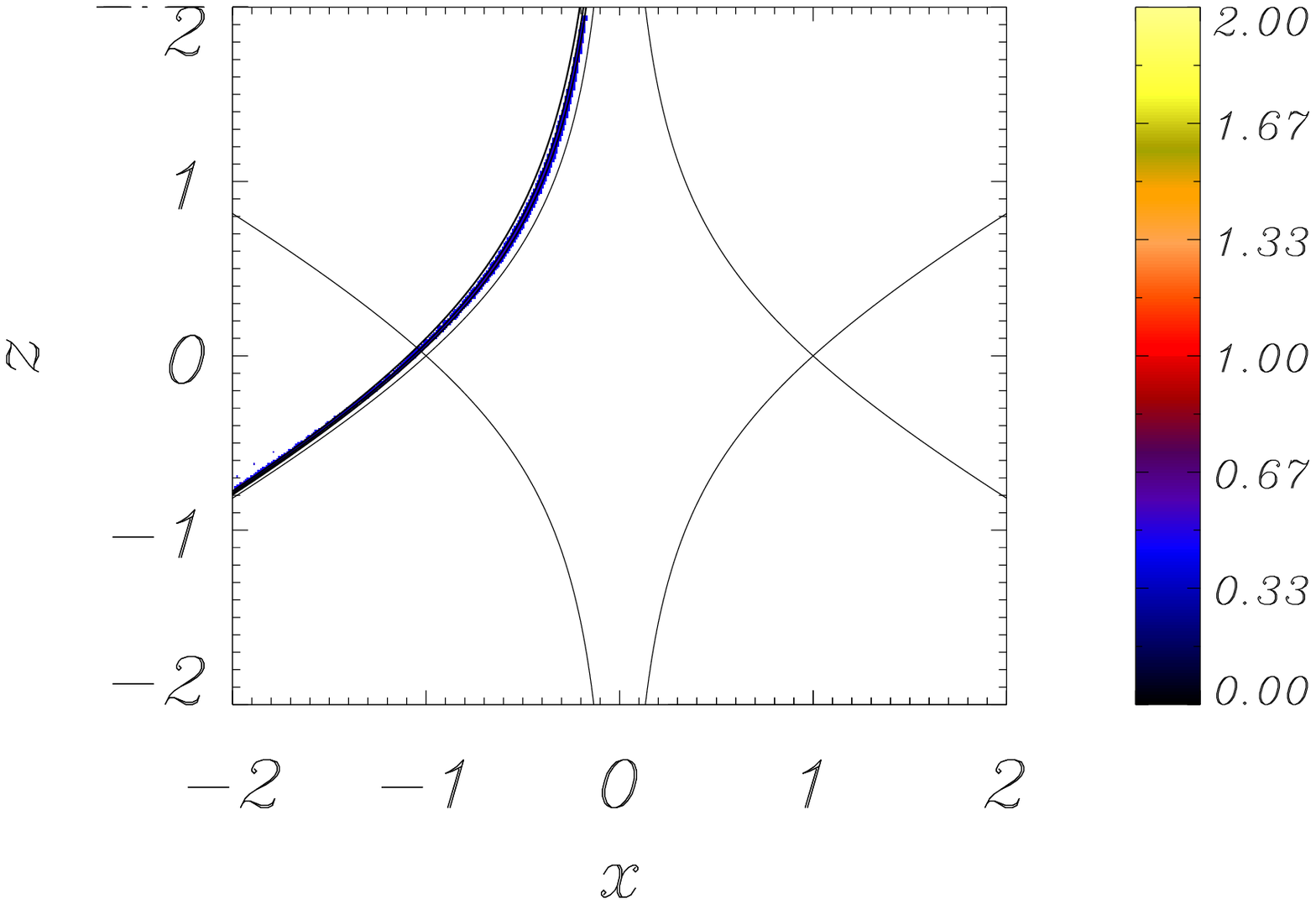}
\caption{Comparison of numerical simulation (shaded area) and analytical solution at times $(a)$ $t$=0.14, $(b)$ $t$=0.39, $(c)$ $t$=0.75, $(d)$ $t$=1.0, $(e)$ $t$=1.25 and $(f)$ $t$=1.75, labelling from top left to bottom right. The lines represent the front, middle and back edges of the wave, where the pulse enters from the side of the box.}
\label{figure37}
\end{figure*}

We use the same characteristic equations, (equations \ref{C2_characteristics}), with different choices of initial condition (i.e. modelling the wave from the side boundary this time) to obtain an analytical result to compare our numerical simulation. The results can be seen in Figure {\ref{figure37}}.

\section{Conclusions}\label{sec:4}

This paper describes an investigation into the nature of MHD waves in the neighbourhood of two null points. From the work explained above, it has been seen that when a fast magnetoacoustic wave propagates near the two null point configuration, the wave bends due to refraction and begins to wrap itself around the null points. The wave 'breaks' and part of it travels (and wraps) around one of the null points and the rest wraps around the other null point. In the case of the fast wave approaching the two null points from above, the wave travels down towards the null points and begins to wrap around them. In addition, since the Alfv\'{e}n speed is non-zero between the null points, the wave can travel through this area (into the lower half plane). This part of the wave is also affected by the refraction effect and the wave continues to wrap around its closest null point. The wave breaks into two along the line $x=0$ (due to symmetry), with each half of the wave going to its closest null point. Each part of the wave then continues to wrap around its respective null point repeatedly, eventually accumulating at that point. In the case of the fast wave pulse travelling in from the side boundary, we see a similar effect (i.e. a refraction effect, wave breakage and accumulation at the nulls), but in this case the wave is \emph{not} equally shared between the null points. For a fast wave travelling in from the left boundary, initially the pulse thins and begins to feel the effect of the left hand-side null point. The wave begins to wrap around this null point (due to refraction). As the ends of the wave wrap around behind the left null point, they then become influenced by the second, right hand side null point. These arms of the wave then proceed to wrap around the right null point, flattening the wave.  Furthermore, the two parts of the wave now travelling through the area between the null points have non-zero  Alfv\'{e}n speed, and so pop through. These parts of the wave break along $x=0$ and then proceed to wrap around the null point closest to them.

So it is clear the refraction effect focusses all the energy of the incident wave towards the null points. The physical significance of this is that any fast magnetoacoustic disturbance in the neighbourhood of a null point pair will be drawn towards the regions of zero magnetic field strength and focus all of its energy at that point.

From this work, it has been seen that when a  linear, fast magnetoacoustic wave propagates near a two null point configuration,  the wave bends due to refraction and begins to wrap itself around the null points (at least in two dimensions). The angle that the fast wave approaches the null points from will determine what proportion of wave ends up at each null point (when the wave 'breaks'). Furthermore, since we have a changing perturbed magnetic field with increasing gradients, we will have a build up of current density. In the case of the fast wave, all the wave is accumulating at the null points. This means that the perturbed magnetic field ($\mathbf{B}_1$) will have large gradients at those points and that is where current will accumulate. This is in good agreeement with similar phenomena noted in \cite{McLaughlin2004}. With a large current accumulation at the nulls, this is where energy will  be dissipated. Therefore, wave heating will naturally occur at these null points. Further experiments are being carried out to confirm this.

In the cases of the Alfv\'en wave, the results show that the wave propagates along the field lines,  thins but keeps its original amplitude,  and eventually accumulates very near to the separatrices. This is again in agreement with work carried out for a single null by \cite{McLaughlin2004}. As hypothesised above, since we have wave accumulation along the separatrices, the gradients of the perturbed magnetic field ($j_x$ and $j_z$ in this case) will build up along these as well. Hence we will have current build up along the separatrices.  Therefore, if all the current is accumulating along the separatrices, then this is where dissipation will naturally occur. Our early experiments  into current accumulation for two null points confirm this. In view of the fact that the waves eventually dissipate  due to resistive damping (for example), it would be useful to have an estimate of the nature of the wave damping. Dissipation enters our model through the addition of a $\eta \nabla^2 \mathbf{B}_1$ term to equation (\ref{eq:2.5}) and the WKB model can be extended to include the effects of resistivity. For the model describing the Alfv\'en wave pulse entering the top boundary (Section \ref{sec:4.1.1}), it was found that along the separatrices the diffusion term become important in a time that depends on $\log {\eta}$ (see Appendix); as found by Craig \& McClymont (1991, 1993) and \cite{Hassam1992}. This means that the linear wave dissipation will be very efficient.  This method can also be applied to the other cases investigated in this paper.


D\'emoulin \emph{et al.} (1994b) investigated the magnetic field topology of a flare event. They found that the high intensity regions of $H_{\alpha}$ were located on or close to the separator lines. Their findings show a link to ours and support the possibility of transferring our results to 3D. We can also easily extend the model to 2.5D with the addition of a third spatial coordinate; by taking into account an extra Fourier component of the form $e^{imy}$, where $m$ is the azimuthal mode number. This would lead to coupling of all the wave modes, and probably result in energy accumulating at \emph{both} the separatrices and the null points.

Note that all the experiments conducted here that send in an Alfv\'en wave pulse from the boundary do so \emph{across} a separatrix. If the initial wave pulse does not cross a separatrix, it will follow the field lines and leave the box (there will be no wave accumulation at any separatrices).

\section*{Appendix}

\begin{figure}[htb]
\begin{center}
\includegraphics[width=2.5in]{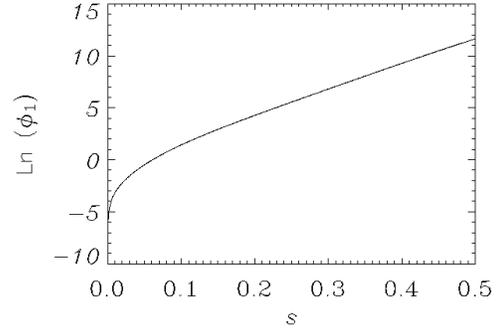}
\caption{ Behaviour of log ( $\phi_1$) against time elapsed. The slope of the line between $s=0.1$ and $s=0.5$ is $4\omega$.} 
\label{figureninety}
\end{center}
\end{figure}

We can learn something about the nature of the wave damping by considering an extension to our WKB expansion for the Alfv\'en wave investigated in Section (\ref{sec:4.1.1}). We assume solutions of the form $e^{i\left( \omega \phi - \omega t \right) }$, where $\omega \gg 1$ and $\phi$ is (now) of order unity. We expand in terms of $1/\omega$, such that $\phi = \phi_0 + {1 \over \omega} \phi_1 + {1 \over \omega}^2 \phi_2 + ...$ and consider $\eta = \eta_0 / \omega^2$, where $\eta_0$ is also of order unity. Here we have assumed a specific scaling for the resistivity. This is for illustration only; alternative forms are possible and they modify the series expansion for $\phi$.

Using the same method with which we reached equation (\ref{alfvenalpha}), the linearised equations for the Alfv\'en wave including  resistivity can be combined to form a single wave equation for $b_y$:
\begin{eqnarray}
\qquad \frac {\partial^2 b_y }{\partial t^2} = \left( \mathbf{B}_0 \cdot \nabla \right) ^2 b_y + \frac{\eta_0}{\omega^2} \frac {\partial}{\partial t} \nabla^2 b_y \nonumber \;.
\end{eqnarray}
Substituting in $b_y =e^{i\left( \omega \phi - \omega t \right) }$ gives:
\begin{eqnarray}
-\omega^2 b_y &=& -\omega^2 \left( {\bf{B}}_0 \cdot \nabla \phi \right) ^2 b_y + i \omega b_y \left( {\bf{B}}_0 \cdot \nabla \right) ^2 \phi \nonumber\\
 &+& {\eta_0} \left( i \omega \left| \nabla \phi \right|^2 b_y + b_y \nabla^2 \phi \right) \; \nonumber .
\end{eqnarray}
Now let $\phi = \phi_0 + i {{\eta_0} \over {\omega }} \phi_1$, where we assume $\phi_1$ has a complex form (for consistency later) and we have taken out a factor of $\eta_0$ (for simplification).
\begin{eqnarray}
-\omega^2 &=& -\omega^2 \left( {\bf{B}}_0 \cdot \nabla \phi_0 \right) ^2 -2 i \omega \eta_0 \left( {\bf{B}}_0 \cdot \nabla \phi_0 \right) \left( {\bf{B}}_0 \cdot \nabla \phi_1 \right)  \nonumber\\
&+& \eta_0^2\left( {\bf{B}}_0 \cdot \nabla \phi_1 \right)^2 + i \omega  \left( {\bf{B}}_0 \cdot \nabla \right)^2 \phi_0 - \eta_0\left( {\bf{B}}_0 \cdot \nabla  \right)^2 \phi_1   \nonumber\\
&+&  i \eta_0 \omega \left| \nabla \phi_0 \right| ^2 + \left[\textrm{other terms}\right] \; \nonumber . 
\end{eqnarray}
Comparing terms of order of $\omega^2$ and $\omega$:
\begin{eqnarray}
\left( {\bf{B}}_0 \cdot \nabla \phi_0 \right) ^2  &=& 1 \Rightarrow \left( {\bf{B}}_0 \cdot \nabla \phi_0 \right) = 1 \;, \nonumber\\
-2 \eta_0 \left( {\bf{B}}_0 \cdot \nabla \phi_0 \right)  \left( {\bf{B}}_0 \cdot \nabla \phi_1 \right)  &+&   \left( {\bf{B}}_0 \cdot \nabla \right)^2 \phi_0 + \eta_0 \left| \nabla \phi_0 \right| ^2 = 0 \nonumber\\ 
 \Rightarrow    \left( {\bf{B}}_0 \cdot \nabla \right) \phi_1 &=& {1 \over 2} \left| \nabla \phi_0 \right| ^2 \;  \nonumber .
\end{eqnarray}
Now:
\begin{eqnarray}
\nabla \phi_0 &=& \left( \frac{\partial \phi_0}{\partial x}, 0, \frac{\partial \phi_0}{\partial z}\right) \nonumber\\ 
\Rightarrow \left| \nabla \phi_0  \right|^2 &=& \left(  \frac{\partial \phi_0}{\partial x} \right)^2 + \left( \frac{\partial \phi_0}{\partial z} \right)^2  = p^2 + q^2 \; \nonumber, 
\end{eqnarray}
where $p=  \frac{\partial \phi_0}{\partial x}$ and $q= \frac{\partial \phi_0}{\partial z} $ as in section (\ref{sec:WKB}). Hence:
 \begin{eqnarray}
\xi\left( B_x\frac {\partial \phi_1}{\partial x} +B_z \frac {\partial \phi_1}{\partial z } \right) &=& {1 \over 2} \xi \left( p^2+q^2 \right) \;,\nonumber\\
\Rightarrow \frac { d \phi_1 }{d s} = \frac {dx}{ds}\frac {\partial \phi_1}{\partial x}+\frac {dz}{ds}\frac {\partial \phi_1}{\partial z} &=& {1 \over 2} \xi \left( p^2+q^2 \right) \label{appendix} \;, 
\end{eqnarray}
where $\xi = \left[ \left( x^2-z^2-1 \right) p -2xzq \right]$, and we have used the characteristic equations from section (\ref{sec:WKB}). Hence, we can solve for $\phi_1$ by adding equation (\ref{appendix}) to  (\ref{C1_characteristics}) and utilising the same Runge-Kutta method. The behaviour of $\phi_1$ can be seen in Figure \ref{figureninety}, where $\log {\phi_1} $ along the separatrix is plotted against $s$. The behaviour is described by $\phi_1 = {1 \over 2}e^{4 \omega s}$, where $\omega = 2\pi$.

Hence:
\begin{eqnarray}
 b_y &=& \exp{\left[i\left(\omega \phi_0 - \omega t \right) \right]} \exp{\left(- \omega^2 \eta \phi_1\right)} \; , \nonumber \\
&=& \exp{\left[i\left(\omega \phi_0 - \omega t \right)\right] }  \exp{ \left\{  - {1 \over 2} \omega^2 \eta \left[  \exp{ \left( 4 \omega s \right) } \right]     \right\} } \; .
\end{eqnarray}
Therefore, the diffusive term will become important when $ {1 \over 2}{\omega^2 \eta} {e^{4 \omega s}} $ becomes of order unity, i.e. on a timescale of $s \approx - {1 \over {4 \omega} } \log { \eta}$, under the above assumptions.

\section*{Acknowledgements}

James McLaughlin acknowledges financial assistance from the Particle Physics and Astronomy Research Council (PPARC). He also wishes to thank the referee for constructive comments, and Valery Nakariakov and Robert Kevis for helpful discussions.


%

%

\end{document}